\documentclass[11pt]{article}
\usepackage[letterpaper,margin=1in]{geometry}
\usepackage[numbers,sort]{natbib}
\usepackage[utf8]{inputenc}
\usepackage{amsmath, amsthm, amssymb}
\usepackage{thmtools}
\usepackage{thm-restate}
\usepackage{amsfonts, bm, bbold, dsfont}
\usepackage{enumitem}
\usepackage[normalem]{ulem}
\usepackage{xspace}
\usepackage[usenames, dvipsnames]{xcolor}
\usepackage[colorlinks=true, citecolor=ForestGreen, linkcolor=ForestGreen, urlcolor=NavyBlue]{hyperref}
\usepackage{booktabs}
\usepackage{multirow}
\usepackage{relsize}
\usepackage{float}
\usepackage[nameinlink, capitalize]{cleveref}

\usepackage{comment}
\usepackage{subcaption}
\usepackage{multirow}
\usepackage{bbold}
\usepackage{pgf, tikz, pgfplots}
\usepackage{nicefrac}
\usepackage{bbm} 
\usepackage[ruled,noend]{algorithm2e}
\crefname{algocf}{algorithm}{algorithms}
\Crefname{algocf}{Algorithm}{Algorithms}

\usetikzlibrary{matrix}
\usetikzlibrary{calc, patterns, intersections, pgfplots.fillbetween}
\pgfplotsset{compat=1.18}
\usetikzlibrary{arrows.meta,fillbetween} 

\usepackage[suppress]{color-edits}

\DeclareMathOperator{\E}{\mathbb{E}}

\newtheorem{theorem}{Theorem}[section]
\newtheorem{proposition}[theorem]{Proposition}
\newtheorem{lemma}[theorem]{Lemma}

\theoremstyle{definition}
\newtheorem{definition}{Definition}[section]

\newcommand{\eps}{\epsilon}

\allowdisplaybreaks
\SetAlFnt{\normalsize}
\SetAlCapFnt{\normalsize}
\SetAlCapNameFnt{\normalsize}
\SetAlCapHSkip{0pt}
\IncMargin{-\parindent}

\addauthor{AE}{magenta}
\addauthor{AB}{blue}
\addauthor{Omri}{red}
\addauthor{ITC}{purple}
\addauthor{IRC}{orange}
\addauthor{Zohar}{red}

\usepackage{mathtools} 
\newcommand{\N}{\mathbb{N}} 
\DeclarePairedDelimiter{\brk}{[}{]}
\DeclarePairedDelimiter{\crl}{\{}{\}}
\DeclarePairedDelimiter{\prn}{(}{)}
\newcommand\numberthis{\addtocounter{equation}{1}\tag{\theequation}}
\newcommand{\indic}[1]{\mathbbm{1}_{\crl*{#1}}}  
\DeclareMathOperator{\I}{\mathcal{I}}

\newcommand{\stadpt}{\mathrm{STOP\_ADAPT}}
\newcommand{\skadpt}{\mathrm{SKIP\_ADAPT}}

\newcommand{\nonadpt}{\mathrm{NON\_ADAPT}}
\newcommand{\adapt}{\mathrm{ADAPT}}
\newcommand{\batch}{\mathrm{BATCH}}

\newcommand{\bF}{\mathbf{F}}
\newcommand{\supp}{\mathrm{supp}}

\newcommand{\utility}{\mathrm{utility}}

\newcommand{\batchpk}{\textbf{BATCH-PK}}
\newcommand{\sapk}{\textbf{SA-PK}}

\newcommand{\pkof}{\textbf{PK-OF}}
\newcommand{\pkbound}{\textbf{PK-BOUND}}

\newcommand{\mpk}{\textsf{MPK}}
\newcommand{\pbatch}{\textsf{PB-BATCH}}
\title{Pandora's Knapsack via Simple Policies}

\newif\ifanonymous
\anonymousfalse

\ifanonymous
\author{Anonymous Authors}
\hypersetup{pdfauthor={Anonymous Authors}}
\else
\author{Zohar Barak \thanks{Tel Aviv University (\texttt{zoharbarak@mail.tau.ac.il; omriporat@gmail.com; inbaltalgam@gmail.com}).} \and
Asnat Berlin \thanks{Hebrew University (\texttt{asnat.berlin@mail.huji.ac.il,
alon.eden@mail.huji.ac.il}).} \and
Ilan Reuven Cohen \thanks{Bar-Ilan University (\texttt{ilan-reuven.cohen@biu.ac.il}).} \and
Alon Eden$^{~\dagger}$ \and
Omri Porat$^{~*}$ 
\and Inbal Talgam-Cohen$^{~*}$ }
\fi

\date{}

\begin{document}
\maketitle
    \setcounter{page}{1}
\thispagestyle{empty}

\pagestyle{empty}
\begin{abstract}
\normalsize

We introduce \emph{Pandora's Knapsack}: a hybrid of the classic stochastic knapsack [Dean, Goemans, and Vondr\'ak, 2008] and Pandora's box [Weitzman, 1979]. As in stochastic knapsack, items have sizes drawn from known distributions, and items that fit within a knapsack contribute to the total value. As in Pandora, every item $i$ comes in a box that costs $c_i$ to open. What distinguishes our problem is that the size is revealed only after opening the box and paying the cost. A motivating application from contract design is \emph{knapsack contracts}, where a principal sequentially delegates jobs with stochastic processing times to agents before a deadline, and must pay each agent to incentivize effort, even if the job ends up not completing in time.

We study the power of simple decision-making policies to approximate Pandora's Knapsack, along two complexity axes: (i)~knowledge of size distributions, and (ii)~adaptivity. We show that, compared to the classic stochastic knapsack and Pandora's box, policies in our model must be significantly more complex along both axes to be approximately optimal. 
Conversely, we show that with full distributional information and by allowing adaptive skipping of items, a constant approximation can be recovered. 

Our analysis reveals an economic quantity, namely ROI (return-on-investment, defined as the jobs' minimum utility over cost), which smoothly characterizes the performance of simple policies. Across a hierarchy of increasingly adaptive policies, we establish tight adaptivity gaps all governed by ROI. To demonstrate the importance of the ROI parameter in characterizing simple policies, we revisit Pandora's box, and show that return-on-investment 
exactly captures the adaptivity gap in this classic problem as well.

\end{abstract}
\newpage
    \setcounter{tocdepth}{2} 
    \tableofcontents

    \newpage

    \pagestyle{plain}
    \pagenumbering{arabic}

\section{Introduction}
\label{sec:intro}

\paragraph{Stochastic Optimization with Costs.}
In many task-scheduling and job-processing settings, executing a job carries an inherent cost. This cost may represent the effort required to perform the job, the energy consumed while running it, or the payment made to an executing agent.  
From an 
optimization perspective, such costs can often be absorbed directly into the objective: one simply replaces the value of each completed job by its net value, obtained by subtracting the cost from the reward. In \emph{stochastic} settings, however, this reduction is no longer innocuous. When job outcomes or processing times are uncertain, the decision-maker may have to incur costs before knowing whether the corresponding value will actually be realized. This creates a fundamental asymmetry between costs and rewards, and makes 
stochastic scheduling with costs qualitatively different from its deterministic counterpart. 

The following toy example illustrates job-processing with costs under uncertainty. Consider a busy PI with a to-do list of tasks (``jobs'') to carry out before the end of the day, when a submission deadline expires.
Each job has value for the PI if completed on time, but each job requires costly effort and takes time. The PI has only a rough idea of how much time each job will take until it is actually executed---this is modeled as having distributional knowledge about the stochastic job sizes. If a job runs over the deadline, the PI incurs its cost without receiving its value.

In which order should the PI execute jobs to maximize the total utility obtained before the deadline? Should the PI prepare a schedule at the beginning of the day and stick to the decided order? Or should the PI decide after each completed job what the next one will be depending on how much time the previous job actually required? 

\paragraph{Adaptivity Gaps in Stochastic Knapsack.} 
When jobs do not carry inherent execution costs, these questions have been addressed by Dean, Goemans, and Vondr\'ak~\cite{dean2008approximating} in their seminal study of the \emph{Stochastic Knapsack} problem, where jobs have known values and stochastic processing times, but no execution costs. A \emph{policy}  
chooses which jobs to run subject to a time budget; each chosen job reveals its processing time only after execution, and the policy collects value from the jobs that complete before the budget is exhausted. 

Dean, Goemans, and Vondr\'ak~\cite{dean2008approximating} introduced the notion of \emph{adaptivity gap}, now a central way to quantify the simplicity-vs.-optimality tradeoff between fully-adaptive and non-adaptive policies. A fully-adaptive policy may choose the next job based on the realized processing times of previously executed jobs, while a non-adaptive policy commits to an execution order in advance and executes sequentially until the deadline. Adaptivity can potentially be powerful, but it also comes with significant complexity. E.g., representing a fully-adaptive policy as a decision tree may require space exponential in the number of jobs. This motivates the search for simpler policies that are easier to describe, plan and implement while retaining most of the value of adaptivity. 

Dean, Goemans, and Vondr\'ak~\cite{dean2008approximating} showed that non-adaptive policies, 
despite their simplicity, achieve a constant-factor approximation of the optimal fully-adaptive policy, implying that the adaptivity gap is bounded by a constant.
The policies they design are simple along another axis as well: they require limited distributional information about the job sizes. 
If the guarantees of Dean, Goemans, and Vondr\'ak~\cite{dean2008approximating} were to apply to the PI scenario above, then the PI could fix the schedule at the start of the day using only limited information about the job-size distributions, while still achieving near-optimal utility. 
Interestingly, in this work we show their conclusion no longer holds once jobs carry execution costs.

\paragraph{Introducing Pandora's Knapsack, and the Role of ROI.}
We model this problem as \emph{Pandora's Knapsack}, where each of $n$ jobs has a value,
\footnote{Our results extend in a straightforward manner to stochastic values, by letting the expected value play the role of the deterministic one.}
a stochastic size, and an execution cost. The cost of a job is incurred when the policy decides to execute it, but the size is revealed only after this decision; the value is gained only if the realized size fits within the remaining knapsack capacity. Thus, Pandora's Knapsack combines the feasibility constraints of stochastic knapsack, with the defining feature of \emph{Pandora's Box} problems---paying for a box before seeing what's in it. 

In the classic stochastic knapsack model, an unsuccessful job may fail to contribute value, but it does not directly decrease the policy's utility. 
In Pandora's Knapsack, by contrast, a policy may pay for a job before knowing whether it will complete on time. This changes the objective from reward-only to mixed-sign, and makes it substantially harder for simple policies to approximate the optimal adaptive policy. In our first set of results, we show that neither knowing only the first $k$ moments of the size distributions, nor fixing the order of jobs in advance while allowing adaptive stopping, is sufficient for a constant approximation.

Our impossibility results do not apply in full force to all problem instances; rather they are governed by an important feature of the instance's jobs. A job's \emph{return-on-investment (ROI)} is the utility obtained from a successful execution (i.e., that finishes before the deadline), relative to the cost paid to execute the job. Low-ROI jobs are precisely those for which the reward only slightly exceeds the cost, so paying for such a job without receiving its value can be especially damaging. We show that these are essentially the only jobs that create hard instances. For non-adaptive policies and policies with limited distributional information, we prove matching lower and upper bounds, up to constant factors, which are characterized exactly by the ROI. 
We also revisit the classic Pandora's Box problem and show that ROI characterizes the adaptivity gap between non-adaptive policies and policies that adaptively decide when to stop.

\paragraph{How Much Adaptivity is Enough.}

How much adaptivity is needed to recover a constant approximation---independent of ROI---for Pandora's Knapsack? A natural first attempt is to keep a fixed execution order, but allow the policy to stop when continuing the execution gives negative expected contribution given the remaining time. This type of adaptivity allows one to obtain optimal utility in the classic Pandora's box problem~\cite{weitzman1978optimal}. 

We show that such stopping-adaptive policies still do not suffice: their gap relative to the optimal adaptive policy can remain unbounded, growing as the square root of the inverse ROI, and we show that this rate is tight. However, a slightly richer form of adaptivity does suffice: We consider 
\emph{skipping-adaptive} policies, which follow a fixed order but assign each job a threshold, and skip that job whenever the remaining time is below the threshold. This ability to skip unpromising jobs, while limited, is enough to obtain a constant-factor approximation to the optimal fully-adaptive policy. Thus, while costs rule out the simplest policies, they do not require full adaptivity; the right level of limited adaptivity restores the guarantee for a simple near-optimal policy.

\subsection{Our Model} 
\label{sub:intro-our-model}

In Section~\ref{sub:MSKC} we formalize a \emph{multi-choice} version 
of the Pandora's Knapsack problem (\mpk). The multi-choice feature captures settings in which each job can be executed in one of several possible modes, with corresponding execution costs and size distributions. In our example above, the PI can choose to spend different levels of effort on a job, where more effort is more costly, but it results in a distribution concentrated over shorter execution times (job sizes). The multi-choice generality is useful for applications such as \emph{knapsack contracts} (see below), and we are able to handle it without loss, albeit with added complication to the design of policies and algorithms.
In what follows, we sometimes refer to jobs with the understanding that they can be replaced with job-mode pairs.

\paragraph{Adaptivity Hierarchy.}

The classic adaptivity gap framework of \cite{2004DeanGoemansVon} compares two extremes: non-adaptive and fully-adaptive policies. In the presence of costs, these endpoints no longer tell the entire story, and we must consider 
intermediate policy classes that allow for more nuanced adaptivity with respect to the remaining capacity.
The policies are ordered from least to most adaptive, according to the extent to which they can react to realized job sizes (see Section~\ref{sub:hierarchy}): 

\begin{itemize}[itemsep=0pt]
    \item \textit{Batch policies.} The policy chooses a batch of jobs before execution begins, pays all their costs upfront, and receives value only if the entire batch fits within the time budget. 

    \item \textit{Non-adaptive policies.} The policy fixes an order in advance and then attempts jobs in that order, paying only for jobs it tries to execute, and stopping with the first overflow of capacity.

    \item \textit{Stopping-adaptive policies.} The policy fixes an order in advance, and may decide to stop ahead of any overflow, depending on the remaining capacity.
    
    \item \textit{Skipping-adaptive policies.} The policy fixes an order along with job-specific thresholds in advance, and may either stop or \emph{skip} jobs whose threshold exceeds the remaining capacity. 

    \item \textit{Fully-adaptive policies.} The policy may sequentially choose the next job based on the full realized history of job sizes up to that point.
\end{itemize}

\subsection{Our Results}

We quantify the guarantees of the different policy classes in terms of ROI. The ROI of a job-mode pair is (roughly) its net utility from successful execution, divided by the cost paid to execute it. Let the parameter \(\alpha\) be the largest inverse ROI over all job-mode pairs of the instance (equivalently, the largest ratio between execution cost and utility). Thus, large~\(\alpha\) corresponds to the presence of low-ROI pairs 
(pairs with high cost relative to the utility gained if the job succeeds). Figure~\ref{fig:adaptivity_levels} summarizes our results: The adaptivity gaps among the different levels are all tight up to constant factors, and all grow polynomially with $\alpha$, except for the constant gap between the second-to-last and last levels. We interpret these results as saying that ROI characterizes the power of limited adaptivity for Pandora's Knapsack, and that skipping-adaptivity is the appropriate level of adaptivity for this problem. 
\begin{figure}[ht]
\centering

\begin{tikzpicture}[
    scale = 0.75,
    axis/.style={->, >=Latex, thick},
    tick/.style={thick},
    jump/.style={->, >=Latex, semithick},
    label text/.style={font=\sffamily\footnotesize, align=center},
    circle label/.style={font=\sffamily\bfseries}
]

    \begin{scope}[yshift=-6cm, xshift=-7cm, font=\footnotesize]
        
        \draw[axis] (0,0) -- (15.3,0) node[right, font=\sffamily\footnotesize] {more adaptivity};

        \coordinate (tB) at (0,0);
        \coordinate (tNA) at (3.2,0);
        \coordinate (tStA) at (6.8,0);
        \coordinate (tSkA) at (10.4,0);
        \coordinate (tFA) at (13.5,0);

        \foreach \pos/\name in {
            (tB)/Batch,
            (tNA)/Non-\\Adaptive, 
            (tStA)/Stopping-\\Adaptive, 
            (tSkA)/Skipping-\\Adaptive, 
            (tFA)/Fully-\\Adaptive} {
            \draw[tick] \pos -- ++(0, 0.2) -- ++(0, -0.4);
            \node[below=4pt, font=\sffamily\footnotesize, align=center] at \pos {\name};
        }

        \draw[jump] (tB) to[bend left=40] node[midway, above] {Thm~\ref{thm:batch}: $\Omega(\alpha)$} (tNA);
        \draw[jump] (tNA) to[bend left=40] node[midway, above] {Thm~\ref{thm:alpha-gap}: $\Omega(\alpha)$} (tStA);
        \draw[jump] (tStA) to[bend left=40] node[midway, above] {Thm~\ref{thm:adapt_stadapt_gap}: $\Omega(\sqrt{\alpha})$} (tSkA);
        \draw[jump] (tSkA) to[bend left=40] node[midway, above] {Thm~\ref{thm:a}: $O(1)$} (tFA);

        \coordinate (bB) at ($(tB)+(0,-1.3)$);
        \coordinate (bStA) at ($(tStA)+(0,-1.3)$);
        \coordinate (bFA) at ($(tFA)+(0,-1.3)$);
        \draw[jump] (bB) to[bend right=30] node[midway, below] {Thm~\ref{thm:batch-approx}: $O(\alpha)$} (bFA);
        \draw[jump] (bStA) to[bend right=12] node[pos=0.38, below] {Thm~\ref{thm:stop-sqrt-ub}: $O(\sqrt{\alpha})$} (bFA);

    \end{scope}
\end{tikzpicture}

\caption{Upper ($O$) and lower ($\Omega$) bounds on the adaptivity gaps among different policy classes.} \label{fig:adaptivity_levels}
\end{figure}

\paragraph{Information Gap.}

    In classic stochastic knapsack, constant approximations can be obtained from limited information about the size distributions, such as expected sizes and overflow probabilities. We show that this is no longer true in Pandora's Knapsack. We prove in \cref{thm:beyond-const-moments} that any policy, adaptive or not, whose access to each size distribution is limited to finitely many moments and  
    overflow probabilities, can be an \(\Omega(\alpha)\)-factor worse than the optimal policy with full distributional information. 
    The proof requires interesting techniques---see Sections \ref{subsec:info_gap} and~\ref{subsec:info-gap}.
    This dependence is tight up to constants: our \(O(\alpha)\)-approximate batch policy uses only first-moment and overflow-probability information (Theorem~\ref{thm:batch-approx}).

\paragraph{Adaptivity Gap Characterized by ROI.}
   
    We show that the ROI dependence is tight for non-adaptive policies. On the upper-bound side, we give an efficient algorithm that computes a fixed batch of job-mode pairs, whose expected utility is within an \(O(\alpha)\) factor of the optimal fully-adaptive policy (\cref{thm:batch-approx}). The algorithm is based on an LP upper bound for adaptive policies, together with a rounding argument that selects either a structured ordered set or a single high-utility job. On the lower-bound side, we construct instances in which every non-adaptive policy is an \(\Omega(\alpha)\)-factor worse than the optimal adaptive policy (\cref{thm:alpha-gap}). Thus the adaptivity gap for non-adaptive policies is \(\Theta(\alpha)\).
    
\paragraph{Constant Approximation with Limited Adaptivity.}
    
    We then ask which limited forms of adaptivity suffice to remove the ROI dependence. We first show that stopping-adaptive policies, which are optimal for the Pandora's box problem, are not enough in our context: even though they may halt when continuing has negative expected utility, their gap from the optimal skipping-adaptive (and hence fully-adaptive) policy can be \(\Omega(\sqrt{\alpha})\) (Theorem~\ref{thm:adapt_stadapt_gap}). We show that this is tight: every instance admits a stopping-adaptive policy whose expected utility is within an \(O(\sqrt{\alpha})\) factor of the optimal fully-adaptive utility (Theorem~\ref{thm:stop-sqrt-ub}). Thus the adaptivity gap of stopping-adaptive policies is \(\Theta(\sqrt{\alpha})\), strictly between the \(\Theta(\alpha)\) gap of non-adaptive policies and the constant gap of skipping-adaptive policies. Finally, we give an efficient algorithm that computes a skipping-adaptive policy with a constant-factor approximation to the optimal fully-adaptive policy (Theorem~\ref{thm:a}). The algorithm is based on a global time-indexed LP~\cite{akker1996time,gupta2011approximation,ma2014improvements} to handle costs and multiple execution modes.

    In \Cref{sec:relaxations}, we identify mild structural assumptions under which simpler policies recover guarantees independent of the ROI. In particular, if every job has a small probability of consuming almost the entire budget (where ``almost'' is measured by a parameter $\delta$), we give a stopping-adaptive policy with an \(O(1/\delta)\)-approximation to the optimal fully-adaptive utility. We also show that some restriction of this kind is necessary: non-adaptive policies can still fail to obtain an approximation independent of \(\alpha\) even when all job sizes are bounded by half the total budget.
    
\paragraph{ROI and Pandora's Box.}
To test whether the importance of ROI extends beyond Pandora's Knapsack, we revisit the adaptivity gap of the classic Pandora's Box problem~\cite{weitzman1978optimal} through the lens of ROI. Weitzman's optimal policy opens boxes in a fixed order, but stops once the expected marginal gain from opening the next box is non-positive; in this sense, it is a stopping-adaptive policy. A non-adaptive policy, by contrast, chooses in advance a set of boxes to open and pays their costs. Both policies obtain the maximum realized reward (as opposed to the sum of realized rewards in our Pandora's Knapsack problem). The gap between non-adaptive and stopping-adaptive policies for Pandora's Box was shown to be arbitrarily large by \cite{boodaghians2020pandora,singla2018price}.

In \Cref{sec:pandora} we show that, as in Pandora's Knapsack, the problematic boxes in Pandora's Box for non-adaptive policies are precisely those with low return-on-investment; this holds for an ROI notion that is adapted to the max-value objective. We denote by \(\tilde{\alpha}\) the inverse of the adapted \emph{Tail-ROI} parameter, 
and design an \(O(\tilde{\alpha})\) non-adaptive policy. Our policy uses the explicit box-opening probabilities of Weitzman's optimal policy to down-sample a set of boxes that gives good utility in expectation. In Section~\ref{app:pandora-lower-bound}, we show this dependence on \(O(\tilde{\alpha})\) is essentially tight by giving, for every \(\tilde{\alpha}\), an instance with an \(\Omega(\tilde{\alpha})\) gap between the optimal adaptive and non-adaptive policies. 

\paragraph{Application to Contract Design---Knapsack Contracts.} 
The multi-choice aspect of \mpk\ is motivated by settings in which the decision-maker can execute the same job in different ways, e.g., by choosing a faster machine. Such choices influence the execution cost as well as the distribution of the job's processing times. 
One such setting is \textit{Knapsack Contracts}, a principal-agent model in which a principal must incentivize agents to complete jobs before a deadline (in our example, this models a scenario where the PI delegates her tasks to students). 
The agents choose which level of effort to apply to their assigned jobs and these choices are not directly observable by the principal.
Principal-agent models are challenging due to the misalignment of interests among the parties, which necessitates the design of payment mechanisms known as contracts. Despite this complexity,
in \Cref{sec:contract-application}, we show that multi-choice Pandora's Knapsack is general enough to capture Knapsack Contracts. We design a reduction from Knapsack Contracts to \mpk\ such that our approximation guarantees transfer directly to this contract-design setting.

\subsection{Our Techniques} \label{sec:techniques}

\paragraph{Lower Bounds.} Our lower bounds isolate the two sources of difficulty: limited information and limited adaptivity. For the information gap, we construct two types of items, ``good'' and ``bad'', with identical values and costs and size distributions that agree on all the first $k$ moments. Good items have size zero with high probability, so a full-information policy can repeatedly attempt them and obtain constant expected utility. Bad items instead consume a small positive amount of capacity even in their smallest realization; after attempting one bad item, every subsequent attempt has negative expected marginal utility. The main technical step is making these distributions indistinguishable. The moment-matching constraints form a transposed Vandermonde system, whose unique solution yields valid probabilities when the smallest positive size is sufficiently small. Finally, we create many more bad items than good ones. A limited-information policy then selects a bad item first with high probability and obtains only $O(\epsilon)$ utility, whereas a full-information policy identifies the good items and obtains $\Omega(1)$. Since $\alpha=\Theta(1/\epsilon)$, this gives the $\Omega(\alpha)$ information gap.
 
To show a gap between stopping-adaptive and non-adaptive policies, we construct an instance where each job $i$ has two possible sizes in the support, $a_i\ll b_i=1-\sum_{j<i}a_j$. $a_i$ and $b_i$ are set such that if jobs are executed in order, then job $i$ always fits the knapsack if previous jobs $j<i$ are realized at small size $a_j$. We show that an optimal stopping-adaptive policy executes jobs in order, and halts once a job is realized at a large size. A non-adaptive policy, on the other hand, has to execute jobs until all jobs are executed or until a job overflows the knapsack. We set the parameters in a way that once a job is realized at a large size, future jobs give a large negative utility. As the policy cannot stop, the only non-risky non-adaptive policy schedules a single job. This enables the gap.

A more involved construction is needed for the gap between stopping-adaptive and skipping-adaptive policies, as the instance needs to give extra power to policies that skip over jobs. This is achieved by constructing an instance with $k=\Theta(\sqrt{\alpha})$ classes of $k^2$ jobs each, where every job has utility $k$ and cost $\alpha k$. Each job in class $i$ has four size realizations: a tiny ``dust'' size, a moderate size $\approx a_i$, a large size $\approx b_i$ with $a_i+b_i>1$, and the full capacity. The parameters are set such that a skipping-adaptive policy executes jobs of class $i$ until one is realized at a moderate size, in which case it skips the rest of the class and continues to the next class, or at a large size, in which case it stops; it thus executes $\Omega(k^2)$ jobs in expectation. The full-size outcome, which overflows once the knapsack is non-empty, caps the expected utility of every job after the first at $1$. A stopping-adaptive policy, in contrast, cannot skip: once a job of class $i$ is realized at a moderate or large size, every later job of class $i$ overflows whenever it is realized at a large size, which makes attempting it so costly that the policy must stop there. Hence, in any fixed order, all but the last executed job of each class must be realized at dust size, so the policy executes only $O(k)$ jobs in expectation and obtains utility $O(k)$. This gives a gap of $\Omega(k)=\Omega(\sqrt{\alpha})$.

\paragraph{Upper Bounds.}
Our upper bounds are based on LP relaxations that upper-bound the utility of fully-adaptive policies, and guide us in choosing which items the simple policy should pick in order to obtain high expected utility. For the \(O(\alpha)\)-approximation, we adapt the stochastic-knapsack LP~\cite{goemans2006stochastic}  to handle costs and multiple choices per item. The main technical challenge is that the LP does not consider the high cost that the policy might incur in case of an overflow. We address this by carefully selecting either a sparse high value-over-cost set of items or a single high-utility item, balancing the value obtained against the cost incurred in overflow events. The ROI parameter is crucially used in choosing the right balancing point.

As we have shown that our policy cannot use a fixed execution order and limited distributional knowledge, our policy uses richer distributional information and allows deviating from the fixed execution order. We adapt the global time-indexed LP framework~\cite{gupta2011approximation, ma2014improvements,akker1996time}, and design an algorithm that extracts a constant fraction of its value. The LP incorporates the overflow probability of each item at each discretized time step, allowing the algorithm to use precisely the information that is missing from moment-based policies. The algorithm then assigns randomized maximum starting times, or deadlines, to item-choice pairs. During execution, when an item-choice pair is considered, the policy checks whether it is still worthwhile to pay its cost, taking into account the overflow probability at the current remaining capacity.
    
The proof has two parts. First, we show that the LP upper-bounds $\adapt$. Second, we show that the algorithm obtains a constant fraction of the LP value. For the rounding, we first randomly select items so that each item is chosen at most once, and then randomly select a single choice for each selected item. The LP upper bound is obtained by setting each variable to the probability that the optimal adaptive policy attempts the corresponding item-choice pair at the corresponding time. The approximation guarantee follows by showing that, for every item-choice pair separately, the algorithm captures a constant fraction of its expected contribution in the LP.

For the matching $O(\sqrt{\alpha})$ upper bound for stopping-adaptive policies, we use the same LP. The algorithm randomly selects item-choice pairs and start times as before, but scales down the selection probabilities by a factor of $\Theta(1/\sqrt{\alpha})$. It then attempts the selected pairs in order of their start times, and stops once half of the capacity is used. Alternatively, if a single item-choice pair already has a large expected utility, the algorithm attempts only this pair. Since a stopping-adaptive policy cannot skip, a pair may be attempted later than its start time, and lose part of its value. The key observation is that this loss is controlled by the best single pair: by the definition of the ROI parameter, the value a pair can lose is at most a $(1+\alpha)$ factor larger than its expected utility when attempted alone. Scaling down the selection probabilities makes such delays rare, and balancing the two cases yields the $O(\sqrt{\alpha})$ guarantee.

\paragraph{Pandora's Box.}
We then consider the classic Pandora's Box model, where the ROI parameter is adjusted to be compatible with the objective, which keeps the highest realized value: When computing the return-on-investment of a box, we consider a value distribution that keeps only the values above Weitzman's fair cap for the box. Our batch policy chooses a random set of boxes to simultaneously open as follows. Let $p_i$ be the probability that Weitzman's optimal policy opens box $i$. Our policy picks each box $i$ independently with probability proportional to $p_i$ and to the box's return-on-investment. That is, it gives priority to boxes with high ROI that are likely to get picked by the optimal policy. Choosing such a probability ensures that, on the one hand, the value obtained is not too small compared to the value obtained by the optimal policy, and on the other hand, the additional cost of opening all boxes simultaneously is not too high. For the matching lower bound, we consider an instance with identical boxes with a Bernoulli value distribution that gives high value with small probability. While the optimal policy opens boxes until it gets the value, a non-adaptive policy has to commit to the number of boxes in advance, which creates the gap.

\subsection{Related Work}
\label{appx:related-work}

\paragraph{The benefit of adaptivity.}
Our work is most closely related to 
Dean, Goemans, and Vondr\'ak~\cite{dean2008approximating}, who initiated a broader line of research on 
stochastic packing/bandit models 
\citep[e.g.][]{bhalgatImprovedApproximationResults2011,gupta2011approximation,ma2014improvements}.
Singla~\cite{singla2018price} develops a general framework for stochastic combinatorial optimization with costly information acquisition. This work already observes that non-adaptive policies cannot give constant approximations in such mixed-sign settings, including Pandora's Box, and therefore focuses on adaptive policies. Our work takes a complementary perspective: we ask which simple policy classes remain useful, and identify ROI as the parameter governing when non-adaptive or limited-adaptivity policies can approximate the optimum.

The role of adaptivity has also received substantial attention in stochastic probing, where a policy probes elements subject to feasibility constraints, and observes random outcomes only after probing~\cite{gupta2016algorithms,gupta2017adaptivity,DBLP:conf/approx/Bradac0Z19}. Related adaptivity-gap questions arise in stochastic orienteering, stochastic graph exploration, and adaptive routing~\cite{GuptaKNR15,bansal2015adaptivity,anagnostopoulos2019stochastic,DBLP:conf/innovations/JiangLL020}, as well as in optimal decision-tree problems and adaptive TSP~\cite{gupta2017approximation}. These works share our interest in comparing simple policies with fully-adaptive decision trees, but the technical setting is different: in our model, the main obstruction is not only feasibility or routing uncertainty, but the mixed-sign effect created by paying costs before knowing whether value will be realized.

\paragraph{Notions of limited adaptivity.}
Different limited-adaptivity notions have been studied in the literature. Instead of studying the size of the decision tree, several works study the number of adaptive rounds or choices. Recent works in stochastic optimization explore the power of $k$ adaptive queries compared to full adaptivity: \cite{DBLP:conf/icml/GhugeGN21,agarwal2019stochastic} study Stochastic Submodular Cover, while \cite{DBLP:conf/aistats/TanGN24} study Informative Path Planning from an information-theoretic perspective.
Esfandiari, Karbasi, and Mirrokni~\cite{esfandiari2021adaptivity} consider both Stochastic Submodular Maximization and Stochastic Minimum Cost Coverage.
The studied problems and techniques are distinct from ours, and focus on the number of adaptive decision rounds or choices; to our knowledge, our policy hierarchy has not been previously studied.

The power of limited adaptivity has also been studied in the context of non-stochastic optimization, in particular, in maximization of submodular and other complement-free set functions: see e.g.~the works of \cite{balkanski2018adaptive,balkanski2018non,DBLP:conf/soda/BalkanskiRS19,chekuri2019parallelizing,kupfer2020adaptive}.

\paragraph{Pandora's Box and prophet inequalities.}
Our work is related to Pandora's Box problems and free-order prophet inequalities~\citep[e.g.][]{hill1983prophet,arsenis2021constrained}, which study adaptive ordering and stopping under uncertainty. However, the Pandora's Knapsack model studied here appears to be new. 
\cite{EsfandiariHLM19} introduce variations to online Pandora's Box including the use of stochastic types for boxes, which can model stochastic sizes. They allow the collection of multiple prizes subject to constraints, including a knapsack constraint on the opened boxes' realized sizes. However, in their model a box's type is revealed upon arrival and before deciding whether to open it, whereas in our model learning the size requires paying a cost to open the box. This creates a fundamental difference and indeed in their model the adaptivity gap is constant.  
We further discuss the relationship to classic Pandora's Box
in Appendix~\ref{app:pandora-related-work}.

\paragraph{LP-based approach.} Our approach builds on LP-based upper bounds for adaptive policies introduced in Dean, Goemans, and Vondr\'ak~\cite{dean2008approximating}, extending them to settings with costs and mixed-sign objectives. We use convexity and dominance arguments to obtain well-structured LP solutions. In the context of limited adaptivity, we adapt global time-indexed LP techniques from \cite{akker1996time,gupta2011approximation,ma2014improvements} to design constant-approximation skipping-adaptive policies. Unlike \cite{gupta2011approximation,ma2014improvements}, our setting also captures multiple choices for each item and negative item utility.

    \paragraph{Contract design.}

    In recent years, a series of studies has emerged that study contracts through the lens of the theory of computation, initiated by \cite{BabaioffFN06,DuttingRT19}; a recent survey paper \cite{surveyContract} provides an overview of developments in algorithmic contract theory. The multi-agent setting was initially explored by Babaioff, Feldman, and Nisan~\cite{BabaioffFN06} and further developed by~\cite{dutting2023multi}, but in their setting the agents only choose between effort and no-effort, and the outcome is only success or failure of the project; we allow multiple actions and outcomes.
    The study of sequential contracts in \cite{SequentialContracts24} investigates the problem of incentivizing a single agent to perform multiple jobs in a sequential manner. 
    In contrast, our work considers a sequence of contracts in which the jobs are executed sequentially by different agents, with no interdependence between the jobs.
    The works of \cite{SaigET24,Budget-Feasible25,AharoniHT25,budget-contracts} consider contract design subject to a limited monetary budget constraining the transfers from principal to agent. 
    In our work, we study a multi-agent setting under a time budget constraint on the agents' actions, where time is limited and the principal must incentivize the agents to complete their jobs on time.
Several works consider agent decisions under costs and incentives, including sequential exploration settings studied in Hoefer, Schecker, and Schewior~\cite{HoeferSS25}, where optimal contracts are designed for a principal–agent exploration task. The work of Hoefer, Schecker, and Schewior~\cite{HoeferSS25} is similar in flavor to ours as it combines stochastic optimization with contract design and delegation. 

    \section{Model: The Pandora's Knapsack Problem} 
    \label{sec:model} 
    \label{sub:MSKC}

    We generalize Stochastic Knapsack 
    in two ways: (i)~items can have multiple tiers, called \emph{choices}, and (ii) each choice is associated with a \emph{cost}. 
    An instance $I=\{v_{i}, \{p_{i}^{j},F_{i}^{j}\}_{j\in [m]}\}_{i\in[n]}$ of the Multi-choice Pandora's Knapsack problem (or \mpk\ for short) consists of $n$ items. Each item $i\in[n]$ has a \emph{value} $v_i$ and $m$ choices. Each choice $j\in[m]$ has a (deterministic) cost $p_{i}^{j}$, and a \emph{size distribution} $F_{i}^{j}$. There is also a knapsack capacity of size $S$. Without loss of generality, $S=1$. The assumption of deterministic values is also without loss, as we can equivalently allow stochastic values drawn independently from known distributions, by replacing each random value with its expectation (\cite{2004DeanGoemansVon} note this reduction in their seminal paper).

    When item $i$ is selected for the knapsack, a single choice $j$ must be specified. We refer to this as the \emph{item-choice pair} (sometimes referring just to the item when the choice is clear from context). In this case, the cost $p_{i}^{j}$ is incurred up-front, and the item's size $s_i$ is realized from distribution $F_{i}^{j}$. Item $i$'s realized \emph{utility} is then $v_i-p_{i}^{j}$ (more precisely, this is the ``potential'' realized utility, i.e., the utility that will be gained if the item fits in the knapsack). 
    
    For a given instance $I$, a \emph{policy} $\varphi$ possibly inspects the knapsack, then either selects an item-choice pair, or terminates. Item-choice pairs are selected according to the policy sequentially, until the total size of selected items exceeds the capacity, or until the policy decides to terminate. In the former case, we refer to the last selected item as the \emph{overflow item}. 
    
    The realized \emph{utility} of the policy is the total realized utility of items selected before the cumulative size exceeded the capacity, minus the cost of the overflow item, if one exists. Let $\utility(\varphi)$ be a random variable representing the utility obtained by the policy $\varphi$ on instance $I$. We also design algorithms that return policies. When referring to an algorithm $\mathcal{A}$ that is given an instance $I$ and returns a policy $\varphi_I$, we denote by $\mathcal{A}(I)=\utility(\varphi_I)$ the random variable representing the utility obtained by the policy on instance~$I$.

    We are also interested in the \emph{expected} utility of an item-choice pair $(i,j)$. This depends on the remaining capacity $t\in[0,1]$, and is equal to $v_i\cdot \Pr_{s_i\sim F_i^j}[s_i\le t] \; - \; p_i^j.$
    Let 
    $w_{i,j}:=v_i\cdot \Pr_{s_i\sim F_i^j}[s_i\le 1] - p_i^j$
    be the expected utility of inserting $(i,j)$ when the knapsack is empty ($t=1$). Without loss of generality, we assume $w_{i,j}>0$ for every item-choice pair $(i,j)$, as otherwise, inserting $(i,j)$ is never profitable.

    \paragraph{Return-on-Investment.} 
    \label{sub:hierarchy}
    
    The lower bounds in this paper critically exploit instances with item-choice pairs $(i,j)$ that have \emph{high} costs, but only \emph{low} expected utilities; that is, $0 < w_{i,j} \ll p_i^j.$    
    In such cases, an item-choice pair may yield a \emph{small positive} expected utility when the knapsack is empty, yet incur a \emph{large negative} expected utility, of up to $-p_i^j$, when the capacity becomes limited. We show that this drastic instability property is an inherent obstacle in our model, via the following definitions: 

    \begin{definition}[ROI and Inverse ROI Parameters] \label{def:RoI}
        Let
        $\mathrm{ROI}(i,j)\ :=\ w_{i,j}/p_{i}^{j}$ and $\alpha \ :=\ \mathrm{IOR} \ :=\ \max_{i,j}\{1/\mathrm{ROI}(i,j)\}.$
        
    \end{definition}

    Intuitively, a large value of $\alpha$ signifies the presence of at least one item-choice pair where the upfront cost of the item greatly exceeds the expected increase in utility from successfully inserting the item. Our results show that $\alpha$ is the key parameter governing the adaptivity gaps between different levels of adaptivity, which we now specify.

\paragraph{Adaptivity Hierarchy.}
The policy classes considered in this paper appear in Section~\ref{sub:intro-our-model}. 
They differ in the amount of adaptivity they allow with respect to the realized remaining capacity.
Except for fully-adaptive policies, all of them are based on a fixed priority order over jobs. The following definition extends this notion to item-choice pairs:

\begin{definition}[Priority Order]\label{def:priority-order}
A \emph{priority order} is an ordered list
$\sigma=\big((i_1,j_1),\ldots,(i_\ell,j_\ell)\big)$
of item-choice pairs, fixed before any item sizes are realized. We require that every item appears at most once in \(\sigma\), i.e.,
\(i_r\neq i_{r'}\) for every \(r\neq r'\). Thus, committing to \(\sigma\) fixes both the order in which items may be considered and the choice used for each such item.
\end{definition}

\noindent Using \Cref{def:priority-order}, we provide here for completeness a formal definition of the policy classes we consider. 

\begin{definition}
~
\begin{itemize}[itemsep=0pt]
    \item A \emph{batch policy} returns a priority order $\sigma$. The policy pays the cost of the entire batch, but gains the batch's value only if the entire batch fits in the knapsack without overflowing. 
    
    \item A \emph{non-adaptive policy} returns a priority order \(\sigma\). The policy then attempts to insert the item-choice pairs in \(\sigma\) according to this order, until either all pairs in \(\sigma\) have been attempted, or the knapsack overflows.

    \item A \emph{stopping-adaptive policy} returns a priority order $\sigma$. The policy considers the pairs in the order \(\sigma\). When pair \((i,j)\) is considered, the policy observes the remaining capacity, and accordingly decides whether it's beneficial to continue the execution of the policy, or stop the execution. 
    
    \item A \emph{skipping-adaptive policy} returns a priority order $\sigma$, and a skipping threshold \(\tau_{i,j}\in[0,1]\) for each item-choice pair \((i,j)\in\sigma\). The policy considers the pairs in the order \(\sigma\). When pair \((i,j)\) is considered, the policy observes the current remaining capacity \(B\). If $B < \tau_{i,j}$, then the policy skips \((i,j)\) and continues to the next pair in the order. Otherwise, it attempts to insert \((i,j)\). The policy continues in this way until all pairs in \(\sigma\) have been considered, or until the knapsack overflows.

    \item A \emph{fully-adaptive policy} may choose the next item-choice pair as an arbitrary function of the realized history. In particular, after each attempted insertion, it may observe the remaining capacity and the set of previously attempted items, and then either select any not-yet-attempted item-choice pair, or terminate.
\end{itemize}
\end{definition}

For an instance \(I\), let \(\batch(I)\), \(\nonadpt(I)\), \(\stadpt(I)\), \(\skadpt(I)\), and \(\adapt(I)\) denote the maximum expected utilities obtained by the corresponding policies. We remark that each policy can be represented by a \emph{decision tree}; a fully-adaptive policy needs an exponentially sized tree, but the other policies require only polynomial-size trees (in fact, stopping-adaptive and non-adaptive policies correspond to linear-size trees).

\section{Roadmap}
\label{sec:roadmap} 
The next two sections give our main results for Pandora's Knapsack: \Cref{sec:lower-bounds} presents the lower bound constructions, showing why limited information, non-adaptivity, and stopping-adaptivity are insufficient in the presence of costs; \Cref{sec:upper-bounds} presents the complementary positive results, showing that slightly richer policy classes recover meaningful guarantees, and that ROI governs the performance of batch policies.
Our extension of the ROI perspective to the classic Pandora's Box model appears in \Cref{sec:pandora}.
Our application to contract design appears in \Cref{sec:contract-application}.
Some details are deferred to Appendices \ref{app:pandora-related-work}-\ref{sec:bad-roi}. Model relaxations appear in \Cref{sec:relaxations}.

    \section{Tight Lower Bounds on Simple Policies}
    \label{sec:lower-bounds}
    In this section we establish lower bounds on the performance of simple policies, where simplicity is measured along two axes: $(i)$~the amount of distributional information and $(ii)$~the level of adaptivity. We show that while simple non-adaptive policies with limited information give a constant approximation in the regular Stochastic Knapsack problem (without costs), they fail to do so in the presence of costs. In the Pandora's Knapsack model, any policy (even a fully adaptive one) with access only to a constant number of moments is at least $\Omega(\alpha)$ away from the optimal fully adaptive policy with full information, and any non-adaptive policy (even with full information) is at least $\Omega(\alpha)$ away from the optimal fully adaptive policy. We further show that even stopping-adaptive policies (with full information) are at least $\Omega(\sqrt{\alpha})$ away from the optimal skipping-adaptive policy, and hence from the optimal fully adaptive policy; by \Cref{thm:stop-sqrt-ub}, this is tight. This suggests that an $O(1)$-approximation policy must use extensive amounts of information and adaptivity, and that the IOR parameter $\alpha$ is a fundamental barrier to simple policies.

    \subsection{\texorpdfstring{$\Omega(\alpha)$}{Omega(alpha)} Informational Lower Bound} 
    \label{subsec:info_gap}
    We show that any policy (even a fully adaptive one) with access only to a constant number of moments of the size distribution (as well as the first $k$ truncated moments) is at least $\Omega(\alpha)$ away from the optimal fully adaptive policy with full information. Moreover, we show that limited access to the CDF is also insufficient, even if the algorithm has access to the probability that an item fits in the \emph{current} remaining capacity. Indeed, the skipping-adaptive algorithm presented in \Cref{sub:adaptive-const-approx} uses the entire size distribution of each item to achieve a constant approximation guarantee. Our main result is as follows.

    \begin{theorem}
    \label{thm:beyond-const-moments}
        Fix $k\ge 1$.
        Let $\mathcal{A}$ be any algorithm whose knowledge of the item size distributions is restricted to the moment $\E[s_i^j]$, and truncated moment $\mu_i^j=\E[(\min\{s_{i},1\})^{j}]$ for every item $i\in[n]$ and moment $j\in[k]$, and oracle access to the function $q_{i}(B)=\Pr[s_{i}\le B]$, where $B$ is the current remaining knapsack capacity. For every $\alpha$, there exists an instance $I$ of \mpk\ with IOR $\alpha$ such that
        \[
            \frac{\adapt(I)}{\E[\mathcal{A}(I)]} = \Omega(\alpha).
        \]
    \end{theorem}
    We begin with a warm-up lower bound that shows that any (possibly fully adaptive) algorithm whose knowledge of the item size distributions is restricted to the expected sizes (and the expected truncated sizes) and oracle access to the probability of overflowing the current knapsack capacity, is at least $\Omega(\alpha)$ away from the optimal adaptive policy. We then generalize this lower bound to an algorithm with access to the first $k$ moments of the item size distributions. 
    \subsubsection{Warm-Up: First Moment}
    
    \begin{proposition} 
    \label{pro:info_gap}
        Let $\mathcal{A}$ be any algorithm whose knowledge of the item size distributions is restricted to the expected values $\E[s_i]$, the truncated expected values $\mu_i = \E[\min\{s_i,1\}]$, and oracle access to the function $q_{i}(B)= \Pr[s_{i}\le B]$ for each $i\in[n]$, where $B$ is the current remaining knapsack capacity. For every $\alpha$, there exists an instance $I$ of \mpk\ with IOR $\alpha$ such that
        \[
            \frac{\adapt(I)}{\E[\mathcal{A}(I)]} = \Omega(\alpha).
        \]
    \end{proposition}

    The proof relies on constructing two types of items, one ``good'' and one ``bad''. The optimal strategy would be to insert only the ``good'' items, and any algorithm that starts by inserting a ``bad'' item will result in a value that is only a $\frac{1}{\alpha}$-fraction of OPT. We design the size distributions to be indistinguishable to algorithm $\mathcal{A}$, forcing $\mathcal{A}$ to arbitrarily pick an item to insert into the knapsack. By creating an instance where the number of ``bad'' items far exceeds the number of ``good'' items, algorithm $\mathcal{A}$ will almost surely pick a ``bad'' item first, resulting in our lower bound.

    \begin{proof}[Proof of Proposition~\ref{pro:info_gap}]
        Fix $\epsilon\in(0,1)$.
        The two item types are defined as follows:
        \begin{center}
            \begin{tabular}{l@{\quad}|l@{\quad}|l}
                \toprule Property                 & Type 1 Item (``good'')                                                                       & Type 2 Item (``bad'')                                                                                                                         \\
                \hline
                \midrule Value $v_{i}$           & $2$                                                                                        & $2$                                                                                                                                         \\
                Cost $p_{i}$                      & $2 - \epsilon$                                                                             & $2 - \epsilon$                                                                                                                              \\
                \cmidrule{1-3} Size $s_{i}$ dist. & $s_{1}= \begin{cases}0, & \text{w.p. }1-\epsilon,\\ 1, & \text{w.p. }\epsilon;\end{cases}$ & $s_{2}= \begin{cases}\epsilon/2, & \text{w.p. }1 - \frac{\epsilon}{2-\epsilon},\\ 1, & \text{w.p. }\frac{\epsilon}{2-\epsilon}.\end{cases}$ \\
                \bottomrule
            \end{tabular}
        \end{center}

        Note that in this instance the expected values and the truncated expected values are identical: $\E[s_1] = \E[s_2] = \mu_{1}=\mu_{2}=\epsilon$ and $q_{1}(1)=q_{2}(1)=1$, hence, algorithm $\mathcal{A}$ cannot distinguish between the two items. Inserting one item of type 2 results in a net utility of $\epsilon$, and reduces the remaining capacity $B$ by at least $\epsilon/2$. Inserting any further items (of either type) results in a negative expected utility:
        \begin{align*}
            \E[\text{inserting item of type $i$}\;|\; B \le 1-\epsilon/2] & \le \Pr[s_{i}\ne 1]\cdot v_{i}-p_{i}                    \\
                                                                        & \le (1-\frac{\epsilon}{2-\epsilon})\cdot 2-(2-\epsilon) \\
                                                                        & = \epsilon(1-\frac{2}{2-\epsilon})\; <\; 0.
        \end{align*}
        Now, consider an input instance $I$ with $n >\frac{1}{\epsilon}$ items of type 1 and $n^{2}-n$ items of type 2. The optimal policy will attempt to insert items of type 1, and stop if an item's size is realized as $s_{i}=1$, or if no more items of type 1 remain. The expected number of items successfully inserted by OPT is $\frac{1-(1-\epsilon)^n}{\epsilon}$, each item contributing $\epsilon$ net utility, resulting in
        \[
            \adapt(I) \ge \frac{1-(1-\epsilon)^n}{\epsilon}\cdot \epsilon \xrightarrow{n\to\infty} 1.
        \]
        Since $\mathcal{A}$'s only strategy is to randomly select an item to insert, with probability $1-\frac{1}{n}$ it begins by inserting a ``bad'' item. By taking $n\to\infty$, the expected utility of $\mathcal{A}$ approaches $\epsilon$:
        \[
            \lim_{n\to\infty}\E[\mathcal{A}(I)] = \epsilon.
        \]
        Thus for a large enough $n$, the performance ratio of the algorithm can be bounded:
        \[
            \frac{\adapt(I)}{\E[\mathcal{A}(I)]}\ge \frac{1}{2\epsilon}\ge \frac{1}{4}\alpha.
        \]
    \end{proof}

    \subsubsection{Main Lower Bound: $k$ Moments}
    \label{subsec:info-gap}

    We generalize the distributional informational gap from the warm-up and show that, even an algorithm with knowledge of the first $k$ (or truncated) moments of the distribution, has an $\Omega(\alpha)$ gap compared to an algorithm with knowledge of the exact distribution.

    \begin{proof}[Proof of Theorem~\ref{thm:beyond-const-moments}]
        Fix $\epsilon\in(0,1/2)$.
        Fix distinct sizes
        \begin{align*}
             & t_{j} := 1 - 2^{-j},\quad j=1,2,\dots,k, \\
             & t_{k+1} := 1.
        \end{align*}

        The two item types are defined as follows.
        For sufficiently small $\delta>0$ and for $p_{j}\in [0,1]$ to be determined later: 
        \begin{center}
            \begin{tabular}{l@{\quad}|l@{\quad}|l}
                \toprule Property                 & Type 1 Item (``good'')                                                                                                                               & Type 2 Item (``bad'')                                                                                                                     \\
                \hline
                \midrule Value $v_{i}$           & $2$                                                                                                                                                & $2$                                                                                                                                     \\
                Cost $p_{i}$                      & $2 - \epsilon$                                                                                                                                     & $2 - \epsilon$                                                                                                                          \\
                \cmidrule{1-3} Size $s_{i}$ dist. & $s_{1}= \begin{cases}0,&\text{w.p. }1-2\epsilon\\ t_{j},&\text{w.p. }\tfrac{\epsilon}{k+1},\; j=1,\dots,k+1 \\ 1,&\text{w.p. }\epsilon\end{cases}$ & $s_{2}= \begin{cases}\delta,&\text{w.p. }1-2\epsilon\\ t_{j},&\text{w.p. }p_{j},\; j=1,\dots,k+1 \\ 1,&\text{w.p. }\epsilon\end{cases}$ \\
                \bottomrule
            \end{tabular}
        \end{center}

        Note that since $\Pr[s_i \le 1] = 1$, the truncated moments are identical to the original moments. We now focus on finding the correct parameters such that the two distributions have the same moments; we seek $p_{j}$ with $\sum_{j}p_{j}=\epsilon$, satisfying for each $r=1,\dots,k$:
        \[
            \E[s_{1}^{r}] = \E[s_{2}^{r}].
        \]
        That is, for $r=1,\dots,k$:
        \begin{align*}
            \sum_{j=1}^{k+1}\frac{\epsilon}{k+1}\,t_{j}^{r}+\epsilon                & = (1-2\epsilon)\,\delta^{r}+ \sum_{j=1}^{k+1}p_{j}\,t_{j}^{r}+ \epsilon \\
            \iff \sum_{j=1}^{k+1}t_{j}^{r}\,\left(\frac{\epsilon}{k+1}-p_{j}\right) & = (1-2\epsilon)\,\delta^{r}.
        \end{align*}
        Denote $x_{j}:= \frac{\epsilon}{k+1}- p_{j}$. Then $\sum_{j}x_{j}=0$ (since $\sum p_{j}= \epsilon$ and $\sum \frac{\epsilon}{k+1}= \epsilon$). We get the following system of equations:
        \begin{align}
            \sum_{j=1}^{k+1}x_{j}            & = 0; \label{eq:zero_sum}                                                       \\
            \sum_{j=1}^{k+1}t_{j}^{r}\,x_{j} & = (1-2\epsilon)\,\delta^{r}, \quad \text{for }r=1,\dots,k. \label{eq:moments}
        \end{align}
        We now rewrite \eqref{eq:zero_sum} and \eqref{eq:moments} in matrix form $Ax = b$, where
        \[
            A =
            \begin{pmatrix}
                1         & 1         & \cdots & 1           \\
                t_{1}     & t_{2}     & \cdots & t_{k+1}     \\
                t_{1}^{2} & t_{2}^{2} & \cdots & t_{k+1}^{2} \\
                \vdots    & \vdots    &        & \vdots      \\
                t_{1}^{k} & t_{2}^{k} & \cdots & t_{k+1}^{k}
            \end{pmatrix}, x =
            \begin{pmatrix}
                x_{1}   \\
                \vdots  \\
                x_{k+1}
            \end{pmatrix}, b =
            \begin{pmatrix}
                0                       \\
                (1-2\epsilon)\delta     \\
                \vdots                  \\
                (1-2\epsilon)\delta^{k}
            \end{pmatrix}.
        \]
        Note that $A$ is a $(k+1) \times (k+1)$ matrix where the first row corresponds to the zero-sum constraint \eqref{eq:zero_sum}, and the subsequent $k$ rows correspond to the moment-matching conditions \eqref{eq:moments}. This is a transposed Vandermonde matrix \cite{golub2013matrix}. Since the $t_{j}$ are distinct, it yields a unique solution $x = A^{-1}b$. Hence
        \[
            p_{j}= \frac{\epsilon}{k+1}- x_{j}.
        \]
        It is left to prove that $p_{j}\ge 0$ to conclude that $s_{2}$ is indeed a valid distribution. Note that since $A$ and $A^{-1}$ are functions of $k$ (and the fixed $t_{j}$ values), then after fixing $k$, $x_{j}$ is a function of $\epsilon$ and $\delta$:
        \[
            x_{j}(\epsilon, \delta)=\sum_{r=1}^{k}(A^{-1})_{j,r+1}[(1-2\epsilon)\delta^{r}].
        \]
        For a fixed $\epsilon$, we have that $|x_{j}(\delta)| = O(\delta)$. Thus, for $p_{j}= \frac{\epsilon}{k+1}- x_{j}$ to be non-negative, we can choose $\delta$ sufficiently small such that $\frac{\epsilon}{k+1}$ dominates the $x_{j}$ terms, ensuring $p_{j}\ge 0$. Hence, the two distributions are indistinguishable to $\mathcal{A}$.

        Inserting a single item of type 2 results in a net utility of $\epsilon$, and reduces the remaining capacity $B$ by at least $\delta$. Inserting any further items (of either type) results in a negative expected utility:

        \begin{align*}
            \E[\text{inserting item of type $i$}\;|\; B \le 1-\delta] & \le \Pr[s_{i}\ne 1]\cdot v_{i}-p_{i} \\
                                                                      & \le (1-\epsilon)\cdot 2-(2-\epsilon)   \\
                                                                      & = -\epsilon < 0.
        \end{align*}

        We conclude the proof by considering an input instance $I$ with $n>\frac{1}{\epsilon}$ items of type 1 and $n^{2}-n$ items of type 2. The optimal policy does as well as the following policy: it will attempt to insert items of type 1, and stop if an item's size is realized as $s_{i}\ne0$, or if no more items of type 1 remain. The expected number of items successfully inserted by OPT is $\frac{1-(1-2\epsilon)^n}{2\epsilon}$, each item contributing $\epsilon$ net utility, resulting in
        \[
            \adapt(I) \ge \frac{1-(1-2\epsilon)^n}{2\epsilon}\cdot \epsilon \xrightarrow{n\to\infty} \frac{1}{2}.
        \]
        Since $\mathcal{A}$'s only strategy is to randomly select an item to insert, with probability $1-\frac{1}{n}$ it begins by inserting a ``bad'' item. By taking $n\to\infty$, the expected utility of $\mathcal{A}$ approaches $\epsilon$:
        \[
            \lim_{n\to\infty}\E[\mathcal{A}(I)] = \epsilon.
        \]
        Thus for a large enough $n$, the performance ratio of the algorithm can be bounded:
        \[
            \frac{\adapt(I)}{\E[\mathcal{A}(I)]}\ge \frac{\frac{1}{2}}{2\epsilon}\ge \frac{1}{8}\alpha.
        \]
    \end{proof}

    \noindent\textbf{Remark:} In concurrent and independent work, \cite{CorreaCL+26} study the tradeoff between information and approximation for prophet inequality problems. There are similarities between their work and ours, in terms of both the result and the construction.\footnote{The first draft of our paper was uploaded to arXiv in September 2025. \cite{CorreaCL+26} appeared in the STOC proceedings in June 2026 and to our knowledge was not available online beforehand.} We believe these techniques can be useful in studying the information-approximation tradeoff of other stochastic optimization problems.
    

    \subsection{Limited Adaptivity Lower Bounds} 
    \label{sec:limited-adaptivity}
        \subsubsection{Batch vs. Non-Adaptive Policies}
        \label{subsec:batch-vs-nonadaptive}
        We consider the batch model, where the policy must first commit to a bundle of items, pay their costs upfront, and gain their values only if the entire bundle fits in the knapsack. This is clearly a restriction of the non-adaptive model, where the policy stops after the first overflow.

        In the original work on stochastic knapsack without costs by Dean, Goemans, and Vondr\'ak~\cite{dean2008approximating}, both batch policies and non-adaptive policies are shown to achieve a constant approximation of the optimal fully-adaptive policy. However, in our setting, if the IOR is large, neither policy type can achieve a constant approximation of the optimal fully-adaptive policy. Moreover, the best batch policy cannot approximate the best non-adaptive policy. This model describes real-world scenarios where payments for jobs are made in advance, before any jobs have been executed and their durations are realized.
        
        In this section, we show that in our model there is an $\Omega(\alpha)$-gap between the optimal batch policy and the optimal non-adaptive policy. In \Cref{subsec:batch-policy}, we present a batch policy that gives an $O(\alpha)$ approximation to the optimal fully-adaptive policy.

        \begin{theorem}
        \label{thm:batch}
        For every $\alpha$, there exists an instance $I$ of \mpk\ with IOR $\alpha$ such that
        \[
            \frac{\nonadpt(I)}{\batch(I)} = \Omega(\alpha).
        \]
        \end{theorem}

        \begin{proof}
        Consider $n$ items with value $v_i=1$, cost $p_i=1-\epsilon$, and size distribution 
        $$s_i =
        \begin{cases}
        0, & \text{w.p. }1-\epsilon^2,\\
        2, & \text{w.p. }\epsilon^2.
        \end{cases}
        $$
        Note that $\alpha = \frac{1}{\epsilon}$. Denote by $\pi_{NA}$ a non-adaptive policy that inserts all items until the first overflow (i.e., until an item realizes to $s_i=2$). The expected number of attempted insertions is $\frac{1-(1-\epsilon^2)^n}{\epsilon^2}$, each contributing expected utility $\epsilon-\epsilon^2$. Taking $n\ge\lceil1/\epsilon^2\rceil$ gives $$\utility(\pi_{NA}) = (1/\epsilon-1)\bigl(1-(1-\epsilon^2)^n\bigr) \ge \frac{1}{2}(1/\epsilon-1).$$

        For a batch policy, since all items are identical, the only degree of freedom for the algorithm is choosing how many items it would attempt to insert. Consider the batch policy $\pi_{B_{k}}$ that inserts $k$ items, for some $1\le k \le n$. We prove that the expected utility of this policy is at most 2.
         The fixed total cost of inserting $k$ items is $k(1-\epsilon)$, and their total value $k$ is gained only if all $k$ items fit, which happens with probability $(1-\epsilon^2)^k$. Therefore, the expected utility of $\pi_{B_{k}}$ is
        $$
        \utility(\pi_{B_{k}}) = k\left[(1-\epsilon^2)^k - (1-\epsilon)\right].
        $$

        This utility is positive only for small $k$. Solving for $k$, we get

        \begin{align*}
        & \utility(\pi_{B_{k}}) > 0 \\
        \iff & (1-\epsilon^2)^{k} > 1-\epsilon \\
        \iff & k\ln(1-\epsilon^2) > \ln(1-\epsilon) \\
        \iff & k - 1 < \frac{\ln(1-\epsilon)}{\ln(1-\epsilon^2)} - 1
        \end{align*}
        We rewrite the RHS:
        \begin{align*}
        \frac{\ln(1-\epsilon)}{\ln(1-\epsilon^2)} - 1 & = \frac{\ln(1-\epsilon) - \ln(1-\epsilon^2)}{\ln(1-\epsilon^2)} \\
        & = \frac{\ln(1-\epsilon) - \left[\ln(1-\epsilon) + \ln(1+\epsilon)\right]}{\ln(1-\epsilon^2)} \\
        & = \frac{-\ln(1+\epsilon)}{\ln(1-\epsilon^2)} \\
        & \le \frac{-\epsilon}{-\epsilon^2} = \frac{1}{\epsilon},
        \end{align*}

        where the final inequality uses the fact that $\ln(1+x) \le x$ for $x>-1$, for both the numerator and denominator.
        Therefore, if $\utility(\pi_{B_{k}}) > 0$, then $k < \frac{1}{\epsilon}+1$. Even if all $k$ items fit, the algorithm's utility is at most $$\utility(\pi_{B_{k}})\le k\cdot\epsilon < \left(\frac{1}{\epsilon}+1\right)\cdot\epsilon \le 2.$$

        We conclude that for this instance,
        $$
        \frac{\nonadpt(I)}{\batch(I)} \ge \frac{\utility(\pi_{NA})}{\max_k[\utility(\pi_{B_{k}})]} \ge \frac{1/\epsilon - 1}{4} = \Omega(\alpha).
        $$

        \end{proof}  


    \subsubsection{Non-Adaptive vs. Stopping-Adaptive Policies}
    \label{sec:tight-alpha-gap}
    \label{sec:proof-of-thm-stopping-adapt-to-non-adapt-lb}
    
    We show that there exists an $\Omega(\alpha)$ gap between the optimal Non-Adaptive policy and the optimal Stopping-Adaptive policy. Recall that both models require the algorithm to choose an item ordering at the beginning; the stopping-adaptive policy can then adaptively choose when to halt, while the non-adaptive policy must attempt all items in the chosen order, until all items are inserted or an overflow occurs. Paired with the $O(\alpha)$-approximation batch policy (\cref{subsec:batch-policy}), this establishes a tight $\Theta(\alpha)$ adaptivity gap for the Pandora's Knapsack problem, in contrast to the constant adaptivity gap for the Stochastic Knapsack problem.

\begin{theorem}
\label{thm:alpha-gap}
For every $\alpha$, there exists an instance $I$ of \mpk\ with IOR $\alpha$ such that
\[
    \frac{\stadpt(I)}{\nonadpt(I)} = \Omega(\alpha).
\]
\end{theorem}

\begin{proof}
For a constant $\alpha$, the claim is trivially true. We prove the claim for every $\alpha>3$. We build an instance $I$ as follows. Consider some $\epsilon <\frac{1}{2}$ and any $\gamma \in (0,\tfrac12)$, where $\epsilon$ will be used to determine $\alpha$. Our instance has $n$ items for $n = \lceil\frac{1}{\epsilon}\rceil$. 
Each item has value $v = 2$ and cost $p = 2 - \epsilon$.

The size distribution of item $i \in [n]$ is

$s_i \;=\;
\begin{cases}
  a_i, & \text{with probability } 1-\epsilon,\\[4pt]
  b_i, & \text{with probability } \epsilon,
\end{cases}$
where
$
  \qquad a_i = \gamma^{\,n-i+1},
  \qquad
  b_i = 1 - \displaystyle\sum_{j=1}^{i-1} a_j
$

By \Cref{def:RoI}, we have
$
\alpha = \frac{2-\epsilon}{\epsilon} \;=\; \Theta\!\left(\frac{1}{\epsilon}\right).
$
Note that for every $\alpha>3$, we can choose an $\epsilon<1/2$ in order to set the IOR of our instance to be $\alpha$.

Since for every $i$, $\sum_{j=1}^{i-1} a_{j} + b_i = 1,$ a stopping-adaptive policy can safely insert the items sequentially until it encounters the first item \(i\) whose realized size equals \(b_i\), and then halt.

We show that since \(b_i + a_{i'} > 1\) for every pair \(i<i'\), attempting to
insert item \(i'\) without considering the realization of earlier items' sizes yields
negative expected utility.  Consequently, the optimal \(\nonadpt\) policy executes at most a single item. 
Our proof combines these two facts to yield the desired bound.

Formally, consider a stopping-adaptive policy that inserts items in increasing index order
($i=1,2,\ldots$) until it encounters the first item whose size realizes as
$s_i=b_i$, and then stops.
Let $X$ denote the number of items the policy inserts.
We have
$$
\E[X]
  \;=\;
  \sum_{i=1}^{n} \Pr[X \ge i]
  \;=\;
  \sum_{i=1}^{n}(1-\epsilon)^{i-1}
  \;=\;
  \frac{1-(1-\epsilon)^{n}}{\epsilon}.
$$
Since $n \geq 1/\epsilon$, we get that $\E[X] \ge \frac{1-e^{-1}}{\epsilon}$.
Therefore the expected utility is at least
$$
\stadpt(I)
  \;=\;
    \E[X]\,(v-p)
  \;\ge\;
  \frac{1-e^{-1}}{\epsilon}\,\cdot\,\epsilon
  \;=\;
  1-e^{-1}.
$$
  
      Consider now a non‑adaptive policy. This policy fixes the number of items to insert, $k$, and an ordering $\pi=i_1,i_2,\ldots,i_k$. For each $\ell\in[k]$, let $\utility(\ell)$ denote the utility from the $\ell$‑th item in the sequence. Clearly, $\E[\utility(1)]=\epsilon$ for every non‑empty sequence of items. We will show that for every sequence with $k>1$ we have $\E[\utility(k)]<0$, which implies that it is always better to execute at most a single item, since we can iteratively remove items and only improve the expected utility of the policy.

       Given a sequence $\pi=i_1,i_2,\ldots,i_k$ of items, consider first the case in which  
$i_\ell>i_k$ for some earlier item in the sequence ($\ell<k$).
If $i_k$ is reached and its size realizes as $s_{i_k}=b_{i_k}$, the item
overflows, because
$s_{i_\ell}+s_{i_k}\ge a_{i_\ell}+b_{i_k}>1$.
Hence
\begin{align*}
  \E[\utility(k)]
  &= \Pr[i_k\text{ is reached}]
     \left(\begin{aligned}
       &\Pr[i_k\text{ overflows} \mid i_k\text{ is reached}] (-p)\\
       &\quad+\Pr[i_k\text{ fits} \mid i_k\text{ is reached}] (v-p)
     \end{aligned}\right) \\[4pt]
  &\le \Pr[i_k\text{ is reached}]
     \Bigl(
    \Pr[s_{i_k}=b_{i_k}] (-p)
    + \Pr[s_{i_k}=a_{i_k}] (v-p)
     \Bigr) \\[4pt]
  &= \Pr[i_k\text{ is reached}]
     \bigl(
    \epsilon(-p) + (1-\epsilon)(v-p)
     \bigr) \\[4pt]
  &= \Pr[i_k\text{ is reached}]
     \bigl(-2\epsilon+\epsilon^2+\epsilon-\epsilon^2\bigr) \\[4pt]
  &= -\,\epsilon\,
     \Pr[i_k\text{ is reached}]
     \;<\; 0.
\end{align*}

        Now consider the case where $i_\ell<i_k$ for all $\ell<k$. Let $j=\arg\max_{\ell<k}i_\ell$, the largest‑index item among the first 
$k-1$ inserted. Note that $i_j<i_k$ by our assumption. We have that
        \begin{eqnarray*}
            \E[\utility(k)]  &= & \E[\utility(k)|i_k\mbox{ is reached}]\cdot \Pr[i_k\mbox{ is reached}]\\
            & = & \E[\utility(k)|s_{i_\ell}=a_{i_\ell}\quad \forall \ell<k]\cdot\Pr[s_{i_\ell}=a_{i_\ell}\quad \forall \ell<k]\\
            & & + \E[\utility(k)|i_k\mbox{ is reached}\ \wedge\ \exists\ell<k:\ s_{i_\ell}=b_{i_\ell}]\\
            & &\ \ \cdot\Pr[i_k\mbox{ is reached}\ \wedge\ \exists\ell<k:\ s_{i_\ell}=b_{i_\ell}]\\
            & = & (v-p)\cdot (1-\epsilon)^{k-1} + (-p)\cdot \Pr[i_k\mbox{ is reached}\ \wedge\ \exists\ell<k:\ s_{i_\ell}=b_{i_\ell}] \\
            & = & \epsilon\cdot (1-\epsilon)^{k-1} +(-p)\cdot \Pr[i_k\mbox{ is reached}\ \wedge\ \exists\ell<k:\ s_{i_\ell}=b_{i_\ell}] \\
            &\le& \epsilon \cdot(1-\epsilon)^{k-1}+ (-p)\cdot \Pr[s_{i_j}=b_{i_j} \ \wedge\ (s_{i_\ell}=a_{i_\ell} \quad \forall \ell <k , \ell \neq j)]\\
            & = & \epsilon \cdot(1-\epsilon)^{k-1}+(-2+\epsilon)\cdot(1-\epsilon)^{k-2}\cdot\epsilon\\
            & = & \epsilon \cdot(1-\epsilon)^{k-2}(1-\epsilon-2+\epsilon) \\             
            & = & -\epsilon \cdot(1-\epsilon)^{k-2}<0.         
        \end{eqnarray*}
        In the above, the third equality follows since $i_k$ always overflows in case some $s_{i_\ell}=b_{i_\ell}$ for some $i_\ell<i_k$, as $b_{i_\ell}+a_{i_k}>1$. The first inequality follows from the fact that $-p$ is negative, 
        and $
        \Pr[i_k\mbox{ is reached}\ \wedge\ \exists\ell<k:\ s_{i_\ell}=b_{i_\ell}] \geq 
        \Pr[s_{i_j}=b_{i_j} \ \wedge\ (s_{i_\ell}=a_{i_\ell} \quad \forall \ell <k , \ell \neq j)]$ since for the event $s_{i_j}=b_{i_j}$ and for all other $\ell< k$, $s_{i_\ell}=a_{i_\ell},$ we have that $i_k$ is reached.
        Since $\sum_{\ell<k, \ell \neq j}a_{i_\ell}\le \sum_{\ell=1}^{i_j-1}a_{\ell}=1-b_{i_j},$  there will be no overflowing item before $i_k$ is executed.

        We conclude that the optimal non-adaptive policy only inserts a single item, and obtains a utility of $\epsilon$, while the optimal stopping-adaptive policy obtains a utility of at least $1-e^{-1}$, as argued above. Therefore,  
        \[
        \frac{\stadpt(I)}{\nonadpt(I)}\ge\frac{1-e^{-1}}{\epsilon}=\Omega(\alpha).
        \]
    \end{proof}

    \subsubsection{Stopping-Adaptive vs. Skipping-Adaptive Policies}
    \label{sec:proof-of-thm-stopping-adapt-to-skipping-adapt-lb}

    We show that there exists an $\Omega(\sqrt{\alpha})$ gap between the optimal stopping-adaptive policy and the optimal skipping-adaptive policy. Paired with the $O(1)$-approximation skipping-adaptive policy (\Cref{sub:adaptive-const-approx}), this result highlights the power of skipping-adaptive policies, and identifies the power of skipping items in the fixed order as the minimal level of adaptivity required to overcome the $\alpha$-dependent adaptivity gap. This gap is tight: by \Cref{thm:stop-sqrt-ub} in \Cref{sub:stopping-ub}, every instance admits a stopping-adaptive policy that is an $O(\sqrt{\alpha})$-approximation to the optimal skipping-adaptive policy.

\begin{theorem}
\label{thm:adapt_stadapt_gap}
For every $\alpha\ge59$, there exists an instance $I$ of \mpk\ with IOR $\alpha$ such that
\[
    \frac{\skadpt(I)}{\stadpt(I)} = \Omega(\sqrt{\alpha}).
\]
\end{theorem}

\begin{proof}
Let $k:=\left\lfloor\sqrt{\frac{\alpha+1}{15}}\right\rfloor$; the assumption $\alpha\ge59$ guarantees $k\ge2$. The instance consists of $k$ types $L_1,\ldots,L_k$, each containing $k^2$ items. Every item has a single choice, with value $v:=(\alpha+1)k$ and cost $p:=\alpha k$, and an item of type $L_i$ has the size distribution
\begin{center}
\begin{tabular}{c|c}
    \toprule
    \textbf{Size} & \textbf{Probability} \\
    \midrule
    $\eta$ & $z$ \\
    $a_i$ & $1/k$ \\
    $1-a_i/2$ & $1/k^2$ \\
    $1$ & $\beta$ \\
    \bottomrule
\end{tabular}
\end{center}
where
\[
a_i:=3^{-(k-i+1)},\qquad \eta:=\frac{a_1}{2k^3},\qquad \beta:=\frac{k-1}{v},\qquad z:=1-\frac1k-\frac1{k^2}-\beta.
\]
Since $k\ge2$ and $\beta<k/v=1/(\alpha+1)$, we have $z>0$. All sizes lie in $(0,1]$, so every item fits in the empty knapsack, and by \Cref{def:RoI}
\[
w=v-p=k,\qquad \mathrm{IOR}=\frac{p}{w}=\alpha.
\]
The probability $\beta$ is chosen so that an item that fits unless its size is $1$ has expected utility exactly
\begin{equation}\label{eq:sqrt-lb-later}
v(1-\beta)-p=1.
\end{equation}

An item of type $L_i$ with size $1-a_i/2$ does not fit together with an item of the same type with size $a_i$, since $a_i+(1-a_i/2)>1$, but it does fit together with one item of size $a_j$ from each lower type $j<i$. A skipping-adaptive policy that processes the types in increasing order can therefore skip the remaining items of $L_i$ once one of them has size $a_i$, and it attempts $\Omega(k)$ items of each type (\Cref{lem:E_ADAPT}). A stopping-adaptive policy cannot skip: once an item of type $L_i$ has size $a_i$, it must either stop before the next item of $L_i$ or attempt that item at a large expected loss. We show that, as a result, its expected utility is only $O(k)$ (\Cref{lem:adapt-lb}), which yields the $\Omega(k)=\Omega(\sqrt{\alpha})$ gap.

\begin{lemma}
   \label{lem:E_ADAPT}
   For the instance given above, $\skadpt(I)\ge\frac12e^{-16/15}k^2$.
\end{lemma}

\begin{proof}
We construct a skipping-adaptive policy $\varphi$ and lower-bound its expected utility.

The policy $\varphi$ lists the items of $L_1$, then the items of $L_2$, and so on in its priority order, and assigns the threshold $\tau_i:=1-a_i/2$ to every item of type $L_i$. Thus an item of type $L_i$ is attempted if and only if the used capacity is at most $a_i/2$.

The key property of these thresholds is that one size-$a_j$ outcome from each earlier type $L_j$, together with all preceding size-$\eta$ outcomes, leaves the used capacity strictly below $a_i/2$. Indeed, there are $k^3$ items in total, so before any item is considered there can have been at most $k^3-1$ size-$\eta$ outcomes. Since $a_{j+1}=3a_j$ and $k^3\eta=a_1/2$, we have for every $i\in[k]$,
\begin{equation}\label{eq:sqrt-lb-moderate}
\sum_{j<i}a_j+(k^3-1)\eta\;<\;\sum_{j=1}^{i-1}3^{-(k-j+1)}+k^3\eta\;=\;\frac{a_i-a_1}{2}+\frac{a_1}{2}\;=\;\frac{a_i}{2}.
\end{equation}
Under this policy, within type $L_i$, items are attempted as long as their size is $\eta$; after an item of size $a_i$, the policy skips the remaining items of $L_i$ and moves on to $L_{i+1}$; and after an item of size $1-a_i/2$ or $1$, it attempts no further items.

Indeed, as long as no item has had size $1-a_j/2$ or $1$, every earlier type $L_j$ has contributed at most one item of size $a_j$, and all other attempted items have size $\eta$. Hence, when an item of $L_i$ is considered and no earlier item of $L_i$ has had size $a_i$, the used capacity is at most $\sum_{j<i}a_j+(k^3-1)\eta<a_i/2$ by~\eqref{eq:sqrt-lb-moderate}, and the item is attempted. After an item of $L_i$ has size $a_i$, the used capacity is at least $a_i>a_i/2$, so the remaining items of $L_i$ are skipped. After an item of $L_i$ has size $1-a_i/2\ge5/6$, the used capacity exceeds $a_{i'}/2\le1/6$ for every $i'$, so all remaining items are skipped. Finally, an item of size $1$ either overflows or, if it is the first attempt, fills the knapsack, after which all remaining items are skipped.

Moreover, since the used capacity before an attempt of an item of $L_i$ is less than $a_i/2$, every attempted item fits unless it has size $1$ and is not the first attempt. Consequently, the first attempted item contributes $v-p=k$ in expectation, and by~\eqref{eq:sqrt-lb-later} every later attempted item contributes exactly $1$ in expectation. Since $k\ge1$, the expected utility of the policy is at least its expected number of attempts.

It remains to bound the expected number of attempts. By the choice of $k$, we have $k^2\le(\alpha+1)/15$, and hence
\begin{equation}\label{eq:sqrt-lb-v}
v=(\alpha+1)k\ge15k^3,\qquad\text{so}\qquad\beta<\frac{k}{v}\le\frac{1}{15k^2}.
\end{equation}
Together with $k\ge2$, this gives
\[
1-z=\frac1k+\frac1{k^2}+\beta\le\frac1k\Bigl(1+\frac{16}{15k}\Bigr)\le\frac{8}{5k},
\qquad
z^{k^2}\le\Bigl(1-\frac1k\Bigr)^{k^2}\le e^{-k}\le\frac15.
\]
Conditional on reaching type $L_i$, the number of attempts in $L_i$ is the minimum of $k^2$ and a geometric random variable with success probability $1-z$, so its expectation is
\[
\frac{1-z^{k^2}}{1-z}\;\ge\;\frac{4/5}{8/(5k)}\;=\;\frac{k}{2}.
\]
The policy moves from $L_i$ to $L_{i+1}$ if all $k^2$ items of $L_i$ have size $\eta$, or if the first item of $L_i$ whose size is not $\eta$ has size $a_i$. This happens with probability
\[
z^{k^2}+\frac{1/k}{1-z}\bigl(1-z^{k^2}\bigr)\;\ge\;\frac{1/k}{1-z}\;\ge\;\Bigl(1+\frac{16}{15k}\Bigr)^{-1}.
\]
Therefore every type is reached with probability at least $(1+\frac{16}{15k})^{-k}\ge e^{-16/15}$, and the expected number of attempts is at least
\[
k\cdot e^{-16/15}\cdot\frac{k}{2}\;=\;\frac12e^{-16/15}k^2.
\]
Hence $\E[\utility(\varphi)]\ge\frac12e^{-16/15}k^2$, and the lemma follows.
\end{proof}

\begin{lemma}
\label{lem:adapt-lb}
   For the instance given above, $\stadpt(I)\le3k$.
\end{lemma}
\begin{proof}
Fix a stopping-adaptive policy with priority order $\sigma=(x_1,x_2,\ldots)$. Let $H_\ell$ denote the realized sizes of $x_1,\ldots,x_\ell$. Since the policy cannot skip, the attempted items are $x_1,\ldots,x_T$ for a random $T$. Whether $x_\ell$ is attempted is determined by $H_{\ell-1}$, and the size of $x_\ell$ is independent of $H_{\ell-1}$. Hence the expected utility of the policy is the sum over $\ell$ of $\E\bigl[\mathbf{1}\{T\ge\ell\}\cdot u_\ell\bigr]$, where $u_\ell$ is the expected utility of attempting $x_\ell$ conditional on $H_{\ell-1}$.

Since all sizes are at most $1$, the first attempt always fits and contributes $u_1=v-p=k$. We show that the attempts at positions $\ell\ge2$ contribute at most $2k$ in total, which proves the lemma. Consider a position $\ell\ge2$ on the event $T\ge\ell$. The item $x_1$ fit and has size at least $\eta>0$, so the used capacity is positive and size $1$ overflows; by~\eqref{eq:sqrt-lb-later}, $u_\ell\le v(1-\beta)-p=1$. We call position $\ell\ge2$ \emph{unsafe} if $x_\ell$ is of some type $L_i$ and an earlier item $x_j$, $j<\ell$, of the same type has size $a_i$ or $1-a_i/2$; otherwise it is \emph{safe}. Whether position $\ell$ is unsafe is determined by $H_{\ell-1}$. If position $\ell$ is unsafe, then $x_j$ fit, so the used capacity is at least $a_i$, and since $a_i+(1-a_i/2)>1$, size $1-a_i/2$ overflows as well. In this case, by~\eqref{eq:sqrt-lb-v},
\begin{equation}\label{eq:sqrt-lb-unsafe}
u_\ell\;\le\;v\Bigl(1-\beta-\frac1{k^2}\Bigr)-p\;=\;1-\frac{v}{k^2}\;\le\;1-15k.
\end{equation}

Let $S$ and $U$ be the numbers of attempted safe and attempted unsafe positions, respectively. Splitting the positions into safe and unsafe ones and applying the corresponding bound on $u_\ell$, the attempts at positions $\ell\ge2$ contribute
\begin{align}
\sum_{\ell\ge2}\E\bigl[\mathbf{1}\{T\ge\ell\}\cdot u_\ell\bigr]
&=\sum_{\ell\ge2}\E\bigl[\mathbf{1}\{T\ge\ell,\ \ell\text{ safe}\}\cdot u_\ell\bigr]
+\sum_{\ell\ge2}\E\bigl[\mathbf{1}\{T\ge\ell,\ \ell\text{ unsafe}\}\cdot u_\ell\bigr]\notag\\
&\le\sum_{\ell\ge2}\Pr[T\ge\ell,\ \ell\text{ safe}]
+(1-15k)\sum_{\ell\ge2}\Pr[T\ge\ell,\ \ell\text{ unsafe}]\notag\\
&=\E[S]+(1-15k)\,\E[U].\label{eq:sqrt-lb-stop-total}
\end{align}

It remains to bound $\E[S]$ in terms of $\E[U]$. Let $R_0$ be the number of consecutive safe attempts immediately after the first attempt, and for $j\ge1$ let $R_j$ be the number immediately after the $j$-th unsafe attempt, assigning length $0$ to missing or empty runs. Then $S=\sum_{j\ge0}R_j$.

Fix a history through the first attempt or through an unsafe attempt. If the ensuing safe run has length at least $t$, every one of the next $t$ positions except the last of each type must have size $\eta$: a size of $a_i$ or $1-a_i/2$ would make the later occurrence of that type unsafe, while size $1$ would overflow before that later occurrence could be attempted. There are at least $\max\{0,t-k\}$ such positions, fixed by the priority order and the starting position. By independence, the conditional probability that the run has length at least $t$ is therefore at most $z^{\max\{0,t-k\}}$ (and is zero if fewer than $t$ positions remain). Its conditional expected length is thus at most
\[
\sum_{t\ge1}z^{\max\{0,t-k\}}\;=\;k+\frac{z}{1-z}\;\le\;2k,
\]
where we used $1-z\ge1/k$. For $j\ge1$, the $j$-th unsafe attempt occurs exactly on $\{U\ge j\}$. Conditioning on the corresponding histories gives $\E[R_j]\le2k\Pr[U\ge j]$, and also $\E[R_0]\le2k$. Hence
\[
\E[S]\;=\;\sum_{j\ge0}\E[R_j]\;\le\;2k\sum_{j\ge0}\Pr[U\ge j]\;=\;2k\bigl(1+\E[U]\bigr).
\]
Plugging this into~\eqref{eq:sqrt-lb-stop-total}, the attempts at positions $\ell\ge2$ contribute at most
\[
2k\bigl(1+\E[U]\bigr)+(1-15k)\,\E[U]\;=\;2k+(1-13k)\,\E[U]\;\le\;2k.
\]
Together with the first attempt, $\stadpt(I)\le k+2k=3k$.
\end{proof}

Combining \Cref{lem:E_ADAPT,lem:adapt-lb}, we get
\[
\frac{\skadpt(I)}{\stadpt(I)}\ge\frac{e^{-16/15}}{6}\,k\ge\frac{e^{-16/15}}{12\sqrt{15}}\sqrt{1+\alpha},
\]
as claimed.
\end{proof}


\section{Tight Upper Bounds with Limited Adaptivity}
\label{sec:upper-bounds}
In \Cref{sub:adaptive-const-approx,sub:stopping-ub}, we assume all item sizes (or times) are in $\crl*{0,1,\ldots,T+1}$ and that the knapsack size is $T$ (equivalently, that all items have rational sizes), where size $T+1$ represents any size larger than the knapsack (all such sizes are equivalent, as the item overflows).\footnote{In the original problem we assumed that the knapsack size is $1$. Here we rescale all sizes by $T$ since it is more convenient. In \cref{sec:continuous-distributions} we discuss the handling of continuous distributions, using $\eps$-resource augmentation.} Both sections use the time-indexed LP $\Psi(T)$, defined in \Cref{sub:adaptive-const-approx}. While the runtime of computing its optimal solution is pseudo-polynomial in $T$, it is possible to write a compact (approximate) LP and then use its solution to get a $4$-approximate solution for $\Psi(T)$ in polynomial time. All details of the compact LP appear in \cref{sec:round-lp-for-adaptive-policy}.

\subsection{\texorpdfstring{$O(1)$}{O(1)} Skipping-Adaptive Policy}
\label{sub:adaptive-const-approx}
\label{sec:skipping-ub}
    Our lower bounds imply that without skipping items and without using sufficient distributional information, we may not achieve an $\alpha$-independent approximation ratio. We present an algorithm that uses full knowledge of the size distributions and finds a skipping-adaptive policy with a constant approximation guarantee.
    \begin{theorem}\label{thm:a}
        There exists a polynomial-time algorithm for \mpk\ that computes, for every instance $I$, a skipping-adaptive policy $\pi_I$ such that
        \[
            \frac{\adapt(I)}{\E[\utility(\pi_I)]} = O(1).
        \]
    \end{theorem}
The idea is to use a global time-indexed LP as in \cite{gupta2011approximation, ma2014improvements,akker1996time}, and design an algorithm that aims to get a constant fraction of its value.
    In the global time-indexed LP, there is a variable not only for each item-choice $i,j$, but also for each time step $t \in \crl*{0, \ldots T}$ (each size is interpreted as a job duration). Thus the decision to insert any item depends also on the time of the decision.
    For every $i\in[n],\ j\in[m],\ t\in \crl*{0, \ldots T}$ let $w_{i,j,t} := v_{i}\cdot\Pr_{{s_{i}^j\sim F_i^j}}[s_{i}^j\le T-t]-p_i^j$ be the effective value of choice $i,j$ given remaining capacity of $T-t$ and let $\mu_{i,j,t} =\E_{s_{i}^j\sim F_i^j}[\min \{s_i^j,t\}]$ be the $t$-truncated expected size of choice $i,j$.
    
    For any $b \in \N$, let $\Psi(b)$ be the following linear program:
    \begin{align*}
    \Psi(b) &:= \max_{x} \quad 
      \sum_{i\in[n]} \sum_{j\in[m]} \sum_{t \in \crl*{0, \ldots b}} w_{i,j,t} \, x_{i,j,t} 
      \\[2mm]\notag
    \text{s.t.}\quad 
    & \sum_{i\in[n]} \sum_{j\in[m]} \sum_{t' \le t} x_{i,j,t'} \, \mu_{i,j,t} \le 2t,
    && \forall\, t \in \crl*{0, \ldots b}, 
    \numberthis \label{constraint:capacity}
    \\[1mm]
    & \sum_{j \in [m]} \sum_{t \in \crl*{0, \ldots b}} x_{i,j,t} \le 1,
    && \forall\, i \in [n],
    \numberthis \label{constraint:take-each-item-time-at-most-once}
    \\[2mm]
    & 0 \le x_{i,j,t} \le 1,
    && \forall\, i \in [n],\, j \in [m],\, t \in \crl*{0, \ldots b}.
    \numberthis \label{constraint:x-is-prob}
    \end{align*}

    This LP bounds the expected gain of the optimal adaptive policy:
    \begin{lemma}\label{lem:adapt-ub-of-Psi-T}
        $\E[\adapt] \le \Psi(T)$ where $\adapt$ is the utility of the optimal adaptive policy.
    \end{lemma}

    \begin{proof}
    Let $\sigma_A$ be an optimal adaptive policy. Let $x_{i,j,t}$ be the probability that $\sigma_A$ attempts to insert $(i,j)$ at time $t$. We claim that $x$ is a feasible solution to the LP $\Psi(T)$.
    It is immediate that \cref{constraint:x-is-prob}, \cref{constraint:take-each-item-time-at-most-once} hold (by the definition of $x_{i,j,t}$, each item and choice are inserted at most once in the run of $\sigma_A$). It remains to show that \cref{constraint:capacity} holds. For any $i \in [n]$, $j \in [m]$, $t' \in \crl*{0,\ldots,T}$: let $\indic{i,j,t'}^{ins}$ be the indicator for the event where item-choice $(i,j)$ is inserted at time $t'$ and let $\indic{i,j,s}^{size}$ be the indicator for the event where item-choice $(i,j)$ realizes to size $s$; that is: $s_{i}^j = s$.
    
    For any time $t$, let $(i_t, j_t)$ be the last item-choice pair that $\sigma_A$ inserts up until or at time $t$, and let $P_t$ be the set of all item-choices successfully inserted by $\sigma_A$ before inserting $(i_t, j_t)$. So the sum of all inserted item sizes is at most $t$. 
    \[
        \sum_{i, j \in P_t} \sum_{s \le T+1} \indic{i,j,s}^{size} \cdot s \le t.
    \]
    Let $P'_t = P_t \cup \crl*{(i_t,j_t)}$. We get $\sum_{i,j \in P'_t} s_i^j = \sum_{i,j \in P_t} s_i^j + s_{i_t}^{j_t}$ and thus:
    \[
        \sum_{i, j \in P'_t} \sum_{s \le T+1} \indic{i,j,s}^{size} \cdot \min(s,t) \le 2t.
    \]
    For any $i \in [n], j \in [m]$: $\sum_{t' \le t} \indic{i,j,t'}^{ins} = 1$ if $(i,j) \in P'_t$, and $0$ otherwise, and thus we have:
    \[
        \sum_{i, j \in [n] \times [m]} \  \sum_{t' \le t} \ \sum_{s \le T+1} \indic{i,j,t'}^{ins} \cdot \indic{i,j,s}^{size} \cdot \min(s,t) \le 2t.
    \]
    By taking expectations and using linearity of expectation we have:
    \begin{align*}
        & \sum_{i, j \in [n] \times [m]} \ \sum_{t' \le t} \ \sum_{s \le T+1} \E\brk*{\indic{i,j,t'}^{ins} \cdot \indic{i,j,s}^{size}} \cdot \min(s,t) = \\
        & \qquad = \sum_{i, j \in [n] \times [m]} \ \sum_{t' \le t} \Pr \prn*{\indic{i,j,t'}^{ins}} \prn*{\sum_{s \le T+1} \Pr \prn*{\indic{i,j,s}^{size}} \min(s,t)} \\
        & \qquad = \sum_{i, j \in [n] \times [m]} \ \sum_{t' \le t} x_{i,j,t'} \ \mu_{i,j,t},
    \end{align*}
    where the first equality holds due to the independence between $\indic{i,j,t'}^{ins}$ and $\indic{i,j,s}^{size}$ (the events are independent as the policy inserts item-choice $(i,j)$ before seeing the realized value of $s_{i}^j$).
    Hence \[\sum_{i, j \in [n] \times [m]} \ \sum_{t' \le t} x_{i,j,t'} \ \mu_{i,j,t} \le 2t.\]
    
    We have shown that $x$ is indeed a feasible solution to the LP.

    We now bound $E[\adapt]$. Let $W_{i,j}$ be the utility $\sigma_A$ gets from item-choice $i,j$ (if $\sigma_A$ did not attempt to insert the item-choice then the corresponding utility is 0). From linearity of expectation we get:
    \begin{align*}
        \E[\adapt] &= \sum_{i,j} \E[W_{i,j}] = \sum_{i,j} \sum_{t \in \crl*{0,\ldots,T}} \E[W_{i,j} \mid \sigma_A \text{ inserts } (i,j) \text{ at time } t ] \cdot \Pr \prn*{\indic{i,j,t}^{ins}} \\
        & = \sum_{i,j} \sum_{t \in \crl*{0,\ldots,T}} \E[W_{i,j} \mid \sigma_A \text{ inserts } (i,j) \text{ at time } t ] \cdot x_{i,j,t} \\
        & \le \sum_{i,j} \sum_{t \in \crl*{0,\ldots,T}}  x_{i,j,t} \cdot w_{i,j,t} \le \sum_{i,j} \sum_{t \in \crl*{0,\ldots,T}}  x^*_{i,j,t} \cdot w_{i,j,t} = \Psi(T),
    \end{align*}
    
    where the last inequality holds due to the fact that $x$ is a feasible solution to the LP.
    \end{proof}

    We now give the algorithm that finds a skipping-adaptive policy that achieves a constant approximation of the LP $\Psi(T)$. Since we must account for the multi-choice setting, we use per-item randomized rounding so that at most one choice is selected for each item.
    
    \begin{algorithm}[H]
    \caption{SA-PK (Skipping-Adaptive Pandora's Knapsack)}
    \label{alg:adaptive-skc}
    \KwIn{Instance $I$ with items $i\in[n]$, choices $j\in[m]$, values $v_i$, costs $p_i^j$, distributions $F_i^j$}
    \KwOut{Adaptive policy $\pi$}
    
    \SetKwFunction{Bern}{Bernoulli}
    \SetKwFunction{Cat}{Categorical}
    \SetKwInput{KwInit}{Initialize}
    
    \tcp{Step 1: Solve LP}
    Compute an optimal solution $x^*$ to $\Psi(T)$
    
    \tcp{Step 2: Per-item randomized start time (at most one $(j,t)$ per item)}
    \KwInit{$\I \gets \emptyset$}
    \For{$i\in[n]$}{
      $\alpha_i \gets \sum_{j\in[m],\, t \in \crl*{0,\ldots,T}} x^*_{i,j,t}$
    
      Sample $A_i \sim \Bern\!\left(\frac{\alpha_i}{4}\right)$\\
      \lIf{$A_i = 0$}{\textbf{continue}}
      Draw $(j_i,t_i) \sim \Cat\!\left(\frac{x^*_{i,j,t}}{\alpha_i}\right)$, $\quad$
      $b_{i,j_i} \gets t_i$ \tcp*[f]{Implicitly $b_{i,j'} \gets \infty$ for $j'\neq j_i$}\\
      
      Append $(i,j_i,b_{i,j_i})$ to $\I$
    }
    
    \tcp{Step 3: Adaptive execution in the random order}
    Sort $\I$ by non-decreasing $b_{i,j}$ (break ties by increasing $(i,j)$)
    \Return the policy that works as follows
    $S \gets 0$\;
    \ForEach{$(i,j,b_{i,j})\in \I$ \textbf{in order}}{
      \If{$S \le b_{i,j}$}{
        Insert $(i,j)$; Observe $s_i^j$
        $S \gets S + s_i^j$
      }
    }
    \end{algorithm}
    
    The next lemma tells us that given that the randomly assigned start time of item-choice $i,j$ is $b_{i,j} = t$, the probability that all previous item-choices took at most $t$ time is at least half.
    
    \begin{lemma}
    \label{lem:adaptive-skc-prob-of-insert-ub}
        For any $i,j \in [n] \times [m]$, $t \in \crl*{0,\ldots,T}$, let $\I_{i,j}$ be the multiset of item-choices inserted by Algorithm \ref{alg:adaptive-skc} in the sorted order of $\I$ before inserting item-choice $(i,j)$. Let $s(\I_{i,j}) = \sum_{(i',j') \in \I_{i,j}} s_{i'}^{j'}$. Then $\Pr\prn*{s(\I_{i,j}) \le t \ \middle| \ b_{i,j} = t} \ge \frac{1}{2}.$
    \end{lemma}
    \begin{proof}
    Let $S = \min\crl*{\sum_{(i',j') \in \I_{i,j}} s_{i'}^{j'}, t}$ denote how much of the $[0,t]$ interval is occupied by items inserted before time $t$. Let $x^*$ be the optimal solution to $\Psi(T)$.
    Note that if $(i',j') \in \I_{i,j}$ it must be that $b_{i',j'} \le t$. Thus:
    \begin{align*}
        \E[S \mid b_{i,j} = t] & \le \sum_{i', j' \in \I_{i,j}} \E\brk*{\min\crl*{s_{i'}^{j'}, t} \ \middle| \ b_{i,j} = t} \\
        & = \sum_{i', j' \in \I_{i,j}} \E\brk*{\min\crl*{s_{i'}^{j'}, t} \ \middle| \ b_{i,j} = t, b_{i',j'} \le t} \Pr\prn*{b_{i',j'} \le t} \\
        & \le \sum_{i',j' \in [n] \times [m]} \mu_{i',j',t} \cdot \prn*{\frac{\alpha_{i'}}{4} \cdot \sum_{t' \le t} \frac{x^*_{i',j',t'}}{\alpha_{i'}}} \\
        & = \frac{1}{4} \sum_{i',j' \in [n] \times [m]} \sum_{t' \le t} x^*_{i',j',t'}  \cdot \mu_{i',j',t} \le \frac{t}{2},
    \end{align*}
    where the first inequality follows from the definition of $S$. The second inequality follows from $\I_{i,j} \subseteq [n] \times [m]$, the independence between the item size $s_{i'}^{j'}$ and the assigned starting times $b_{i,j}, b_{i',j'}$, and the starting-time assignment probability. The last inequality holds due to constraint
    \cref{constraint:capacity} of the LP (as $x^*$ is a solution to $\Psi(T)$).
    
    By Markov's inequality: $\Pr \prn*{S \ge t \mid b_{i,j} = t} \le \frac{\E[S \mid b_{i,j} = t]}{t} \le \frac{1}{2}$, implying the desired result.
    
    \end{proof}
    We can now prove the constant-approximation guarantee for the adaptive policy:
    
    \begin{proof}[Proof of Theorem~\ref{thm:a}]
    Let $\sigma_{ALG}$ be the policy returned by Algorithm \ref{alg:adaptive-skc}, and let $ALG$ be its total obtained utility.  For any $i \in [n]$, $j \in [m]$, $t \in \crl*{0,\ldots,T}$: let $\indic{i,j,t}^{ins}$ be the indicator for the event where item-choice $(i,j)$ is inserted and $b_{i,j} = t$. Let $W_{i,j}$ be the (random) utility the policy gets from item $i$ choice $j$, and let $W_{i,j,t}$ be the (random) utility the policy gets from inserting item $i$ choice $j$ with $b_{i,j} = t$. Let $\I_{i,j}$ be the multiset of item-choices inserted by Algorithm \ref{alg:adaptive-skc} before item-choice $(i,j)$ according to the sorted order of $\I$, and let $s(\I_{i,j}) = \sum_{(i',j') \in \I_{i,j}} s_{i'}^{j'}$ be the total size of $\I_{i,j}$ items.
    We get:
    \begin{equation}\label{eq:prob-of-insert-i-j-at-time-t-lower-bound}
        \Pr \prn*{\indic{i,j,t}^{ins}} \ge \Pr\prn*{b_{i,j} = t, s(\I_{i,j}) \le t} = \Pr \prn*{s(\I_{i,j}) \le t \mid b_{i,j} = t} \cdot \Pr \prn*{b_{i,j} = t} \ge \frac{1}{2} \cdot \frac{x^*_{i,j,t}}{4} = \frac{x^*_{i,j,t}}{8},
    \end{equation}
    where the last inequality is due to \cref{lem:adaptive-skc-prob-of-insert-ub} and the fact that \[\Pr \prn*{b_{i,j} = t} = \Pr \prn*{A_i = 1} \cdot \Pr \prn*{(j_i, t_i) = (j,t) \mid A_i = 1} = \frac{\alpha_i}{4} \cdot \frac{x^*_{i,j,t}}{\alpha_i} = \frac{x^*_{i,j,t}}{4}.\]
    
    For any $i \in [n], j \in [m]$:
    \begin{align*}
        \E[W_{i,j}] & = \sum_{t \in \crl*{0,\ldots,T}} \E[W_{i,j,t}] = \sum_{t \in \crl*{0,\ldots,T}} \E[W_{i,j,t} \mid \indic{i,j,t}^{ins} ] \cdot \Pr \prn*{\indic{i,j,t}^{ins}} \ge \sum_{t \in \crl*{0,\ldots,T}} w_{i,j,t} \cdot \Pr \prn*{\indic{i,j,t}^{ins}} \\
        & \ge \frac{1}{8} \sum_{t \in \crl*{0,\ldots,T}} x^*_{i,j,t} \cdot w_{i,j,t}, \numberthis \label{eq:W-i-j-bound}
    \end{align*}
    where the first inequality holds since on the event $\indic{i,j,t}^{ins}$, $(i,j)$ is inserted with remaining capacity $T - s(\I_{i,j}) \ge T - t$, and the last inequality holds due to \cref{eq:prob-of-insert-i-j-at-time-t-lower-bound} and since $w_{i,j,t} \ge 0$ whenever $x^*_{i,j,t} > 0$ (otherwise, setting $x^*_{i,j,t}$ to $0$ would preserve feasibility and increase the objective).
    
    From linearity of expectation we get:
    \begin{align*}
        \E[\sapk] &= \sum_{i,j} \E[W_{i,j}] \ge  \sum_{i,j} \frac{1}{8} \sum_{t \in \crl*{0,\ldots,T}}  x^*_{i,j,t} \cdot w_{i,j,t} = \frac{1}{8} \Psi(T) \ge \frac{1}{8} \adapt,
    \end{align*}
    
    where the first inequality follows from \cref{eq:prob-of-insert-i-j-at-time-t-lower-bound} and the second inequality is due to \cref{lem:adapt-ub-of-Psi-T}.
    \end{proof}

    \subsection{\texorpdfstring{$O(\sqrt{\alpha})$}{O(sqrt(alpha))} Stopping-Adaptive Policy}
    \label{sub:stopping-ub}
    \label{sec:stopping-ub}
    We now show that the lower bound of \Cref{thm:adapt_stadapt_gap} is tight: we give an algorithm that computes a stopping-adaptive policy whose expected utility is within a factor of $O(\sqrt{\alpha})$ of the optimal fully-adaptive policy. Hence the adaptivity gap of stopping-adaptive policies is $\Theta(\sqrt{\alpha})$, strictly between the $\Theta(\alpha)$ gap of non-adaptive policies and the constant gap of skipping-adaptive policies. We use the same LP $\Psi(T)$ as in \Cref{sub:adaptive-const-approx}.

    \begin{theorem}\label{thm:stop-sqrt-ub}
        There exists a polynomial-time algorithm for \mpk\ that computes, for every instance $I$, a stopping-adaptive policy $\pi_I$ such that
        \[
            \frac{\adapt(I)}{\E[\utility(\pi_I)]} = O\bigl(\sqrt{\alpha}\bigr).
        \]
        
    \end{theorem}

    The algorithm is similar to \Cref{alg:adaptive-skc}, with three differences. First, each item is selected with probability proportional to $\lambda:=1/(8\sqrt{1+\alpha})$ instead of $1/4$. Second, the policy cannot skip; instead, it stops once more than half of the capacity is used. Third, if a single item-choice pair already achieves a large fraction of the LP value, the algorithm returns the policy that attempts only this pair.

    \begin{algorithm}[H]
    \caption{StA-PK (Stopping-Adaptive Pandora's Knapsack)}
    \label{alg:stopping-pk}
    \KwIn{Instance $I$ with items $i\in[n]$, choices $j\in[m]$, values $v_i$, costs $p_i^j$, distributions $F_i^j$}
    \KwOut{Stopping-adaptive policy $\pi$}

    \SetKwFunction{Bern}{Bernoulli}
    \SetKwFunction{Cat}{Categorical}
    \SetKwInput{KwInit}{Initialize}

    \tcp{Step 1: Solve LP}
    Compute an optimal solution $x^*$ to $\Psi(T)$\;

    \tcp{Step 2: Best single pair}
    $(i^*,j^*)\gets\arg\max_{i,j}w_{i,j}$\;
    \lIf{$w_{i^*,j^*}>\Psi(T)/(18\sqrt{1+\alpha})$}{\Return the policy that attempts only $(i^*,j^*)$}

    \tcp{Step 3: Per-item randomized start time (at most one $(j,t)$ per item)}
    $\lambda\gets1/(8\sqrt{1+\alpha})$\;
    \KwInit{$\I\gets\emptyset$}
    \For{$i\in[n]$}{
      $q_i\gets\sum_{j\in[m],\,t\in\crl*{0,\ldots,T}}x^*_{i,j,t}$\;
      Sample $A_i\sim\Bern\!\left(\lambda q_i\right)$\;
      \lIf{$A_i=0$}{\textbf{continue}}
      Draw $(j_i,t_i)\sim\Cat\!\left(\frac{x^*_{i,j,t}}{q_i}\right)$\;
      Append $(i,j_i,t_i)$ to $\I$\;
    }

    \tcp{Step 4: Execution in start-time order with a stopping rule}
    Sort $\I$ by non-decreasing $t$ (break ties by increasing $i$)\;
    \Return the policy that works as follows\;
    $S\gets0$\;
    \ForEach{$(i,j,t)\in\I$ \textbf{in order}}{
      \lIf{$S>T/2$}{\textbf{stop}}
      Insert $(i,j)$; observe $s_i^j$; $S\gets S+s_i^j$\;
    }
    \end{algorithm}

   The returned policy attempts the selected pairs in a fixed order and only decides when to stop, so it is a stopping-adaptive policy. Each item $i$ is assigned at most one choice $j$ and start time $t$, and it is assigned $(j,t)$ with probability $\lambda q_i\cdot x^*_{i,j,t}/q_i=\lambda x^*_{i,j,t}$, independently of the other items.

    The next lemma bounds the total size of the items preceding a selected item, analogously to \Cref{lem:adaptive-skc-prob-of-insert-ub}.

    \begin{lemma}\label{lem:stop-ub-load}
        Fix an item $i$ and a pair $(j,t)$ with $x^*_{i,j,t}>0$, and condition on the event that $i$ is assigned $(j,t)$ in Step~3. Let $S_i$ be the total size of the selected items that precede $i$ in the sorted order. Then
        \[
            \Pr[t<S_i\le T/2]\le2\lambda
            \qquad\text{and}\qquad
            \Pr[S_i>T/2]\le4\lambda.
        \]
    \end{lemma}
    \begin{proof}
    Conditioned on the assignment of $i$, the assignments of the other items remain independent, with the same distribution, and all sizes are independent of all assignments. Every item preceding $i$ has start time at most $t$.

    For the first bound, we may assume $t<T/2$, as otherwise the event is empty; in particular, $t+1\le T$. If $S_i>t$, then $S_i\ge t+1$, and the sum of the $(t+1)$-truncated sizes of the items preceding $i$ is at least $t+1$: either one of them has size at least $t+1$, or none of the truncations is active. Hence, by Markov's inequality and constraint~\eqref{constraint:capacity} for $t+1$,
    \[
        \Pr[S_i>t]
        \le\frac{1}{t+1}\sum_{i'\ne i}\sum_{j'}\sum_{t'\le t}\lambda x^*_{i',j',t'}\,\mu_{i',j',t+1}
        \le\frac{\lambda}{t+1}\sum_{i',j'}\sum_{t'\le t+1}x^*_{i',j',t'}\,\mu_{i',j',t+1}
        \le2\lambda.
    \]
    For the second bound, $S_i$ is at most the total size of all selected items other than $i$. If $S_i>T/2$, then the sum of the $T$-truncated sizes of these items is larger than $T/2$, by the same argument. Hence, by Markov's inequality and constraint~\eqref{constraint:capacity} for $T$,
    \[
        \Pr[S_i>T/2]\le\frac{2}{T}\sum_{i',j'}\sum_{t'\le T}\lambda x^*_{i',j',t'}\,\mu_{i',j',T}\le\frac2T\cdot2\lambda T=4\lambda.
    \]
    \end{proof}

    We now bound the contribution of a single selected item. Unlike in \Cref{alg:adaptive-skc}, an item may be attempted at a larger used capacity than its start time, which incurs a loss.

    \begin{lemma}\label{lem:stop-ub-contribution}
        Fix an item $i$ and a pair $(j,t)$ with $x^*_{i,j,t}>0$, and condition on the event that $i$ is assigned $(j,t)$ in Step~3. Then the expected utility that the policy of Step~4 obtains from item $i$ is at least
        \[
            (1-4\lambda)\,w_{i,j,t}-2\lambda(1+\alpha)\,w_{i^*,j^*}\Pr\bigl[s_i^j>T/2\bigr].
        \]
    \end{lemma}
    \begin{proof}
    Let $S_i$ be as in \Cref{lem:stop-ub-load}, where we pre-sample the sizes of all selected items, including those that are not attempted. Item $i$ is attempted if and only if $S_i\le T/2$, in which case it is attempted at used capacity $S_i$. Indeed, if $S_i\le T/2$, then the used capacity before each item preceding $i$ is at most $S_i\le T/2$, so the policy neither stops nor overflows before reaching $i$. If $S_i>T/2$, then the policy either stops or overflows before reaching $i$, or reaches $i$ at used capacity $S_i>T/2$ and stops. Since $s_i^j$ is independent of $S_i$, the expected utility obtained from item $i$ is
    \[
        \E\bigl[w_{i,j,S_i}\cdot\indic{S_i\le T/2}\bigr].
    \]
    If $S_i\le t$, then $w_{i,j,S_i}\ge w_{i,j,t}$. If $t<S_i\le T/2$, then $T-S_i\ge T/2$ and $T-t\le T$, and hence
    \begin{align*}
        w_{i,j,t}-w_{i,j,S_i}
        =v_i\Pr\bigl[T-S_i<s_i^j\le T-t\bigr]
        &\le v_i\Pr\bigl[T/2<s_i^j\le T\bigr]\\
        &\le v_i\Pr\bigl[s_i^j\le T\bigr]\Pr\bigl[s_i^j>T/2\bigr]\\
        &\le(1+\alpha)\,w_{i^*,j^*}\Pr\bigl[s_i^j>T/2\bigr],
    \end{align*}
    where the second inequality holds since $\Pr[T/2<s_i^j\le T]=\Pr[s_i^j>T/2]-\Pr[s_i^j>T]\le\Pr[s_i^j>T/2]\bigl(1-\Pr[s_i^j>T]\bigr)$, and the last inequality holds since $v_i\Pr[s_i^j\le T]=w_{i,j}+p_i^j\le(1+\alpha)\,w_{i,j}$ by the definition of $\alpha$. Finally, $w_{i,j,t}\ge0$, as in the proof of \Cref{thm:a}. Hence, by \Cref{lem:stop-ub-load},
    \[
        \E\bigl[w_{i,j,S_i}\cdot\indic{S_i\le T/2}\bigr]
        \ge w_{i,j,t}\Pr[S_i\le T/2]-(1+\alpha)\,w_{i^*,j^*}\Pr\bigl[s_i^j>T/2\bigr]\Pr[t<S_i\le T/2],
    \]
    which gives the claim.
    \end{proof}

    Summing over all items, choices and start times, we bound the expected utility of the policy returned in Step~4.

    \begin{lemma}\label{lem:stop-ub-random-policy}
        The policy returned in Step~4 has expected utility at least
        \[
            \frac{\Psi(T)}{16\sqrt{1+\alpha}}-\frac{w_{i^*,j^*}}{8}.
        \]
    \end{lemma}
    \begin{proof}
    Item $i$ is assigned $(j,t)$ with probability $\lambda x^*_{i,j,t}$. Summing the bound of \Cref{lem:stop-ub-contribution} over all items, choices and start times, the expected utility is at least
    \[
        \lambda(1-4\lambda)\,\Psi(T)-2\lambda^2(1+\alpha)\,w_{i^*,j^*}\sum_{i,j,t}x^*_{i,j,t}\Pr\bigl[s_i^j>T/2\bigr].
    \]
    Since $\min\crl*{s_i^j,T}>T/2$ whenever $s_i^j>T/2$, we have $\mu_{i,j,T}\ge\frac{T}{2}\Pr[s_i^j>T/2]$, and thus, by constraint~\eqref{constraint:capacity} for $T$, the last sum is at most $\frac{2}{T}\cdot2T=4$. Since $\lambda=1/(8\sqrt{1+\alpha})\le1/8$, we have $1-4\lambda\ge1/2$ and $8\lambda^2(1+\alpha)=1/8$, which gives the claim.
    \end{proof}

    \begin{proof}[Proof of \Cref{thm:stop-sqrt-ub}]
    By \Cref{lem:adapt-ub-of-Psi-T}, $\Psi(T)\ge\adapt(I)$. Suppose first that the algorithm returns in Step~2. Attempting $(i^*,j^*)$ alone has expected utility exactly $w_{i^*,j^*}$, so
    \[
        \E[\utility(\pi_I)]=w_{i^*,j^*}>\frac{\Psi(T)}{18\sqrt{1+\alpha}}\ge\frac{\adapt(I)}{18\sqrt{1+\alpha}}.
    \]
    Otherwise, $w_{i^*,j^*}\le\Psi(T)/(18\sqrt{1+\alpha})$, and by \Cref{lem:stop-ub-random-policy},
    \[
        \E[\utility(\pi_I)]
        \ge\frac{\Psi(T)}{16\sqrt{1+\alpha}}-\frac{\Psi(T)}{144\sqrt{1+\alpha}}
        =\frac{\Psi(T)}{18\sqrt{1+\alpha}}
        \ge\frac{\adapt(I)}{18\sqrt{1+\alpha}}.
    \]
    Finally, $\pi_I$ is a random stopping-adaptive policy, so $\E[\utility(\pi_I)]\le\stadpt(I)$.
    \end{proof}
    \subsection{$O(\alpha)$ Batch Policy}
    \label{subsec:batch-policy}
    \label{sec:batch-ub}
        In this section, we present an $O(\alpha)$-approximate batch algorithm for the $\mpk$ problem. We recall that a batch policy must choose a subset of item-choice pairs, pay their full cost upfront, and collect their values \emph{only if} all of the items fit the knapsack. If items overflow, the utility may be extremely negative; therefore, there is a subtle tradeoff between selecting a valuable batch and keeping the probability of overflow low. Together with Section~\ref{sec:limited-adaptivity}, this establishes a tight $\Theta(\alpha)$ adaptivity gap between the best batch policy and the optimal fully adaptive policy, as $\alpha$ grows. The following explicit guarantee also covers $\alpha<1$, including $\alpha=0$.

        \begin{theorem}
        \label{thm:batch-approx}
            There exists a polynomial-time algorithm for \mpk\ that computes, for every instance $I$, a batch policy $\pi_I$ such that
            \[
                \frac{\adapt(I)}{\E[\utility(\pi_I)]} = O(\alpha).
            \]
            
        \end{theorem}

\paragraph{An overview of the algorithm.} First, we solve the LP relaxation $\Phi(1)=\Phi(1;1)$ defined in \Cref{sec:alg} to obtain an optimal solution $x$ of item choices. As shown there, every item has one choice with $x_{ij}=1$, except for at most one item with two choices $x_{ij}+x_{ij'}=1$, which can be converted into a single virtual choice. Null choices are omitted from the candidate set $\mathcal C$. If $\mathcal C=\emptyset$, the algorithm returns the empty policy. Since each remaining item has a single (possibly virtual) choice, we refer to each item-choice pair $(i,j)$ by the item $i$ for convenience. In this subsection, $\mu_i:=\mu_i(1)$ and $\mu(S):=\mu_1(S)$.

        We sort the candidates by non-increasing density $d_i=w_i/\mu_i$, assigning density $+\infty$ when $\mu_i=0$. Ties are broken by placing all integral candidates before the possibly virtual item. In particular, as shown in \Cref{sec:alg}, a virtual item that mixes a non-null choice with the null choice is never included in the prefix used below. Thus every possibly virtual item in this prefix retains its deterministic item value $v_i$.

        Define the threshold $U=1/[3(1+\alpha)]$. If $\sum_{i\in\mathcal C}\mu_i\le U$, let $S=\mathcal C$. Otherwise, let $S=\{1,\ldots,k\}$ be the maximal prefix such that
        \[
            \mu(S)=\sum_{i=1}^k\mu_i\le U<\sum_{i=1}^{k+1}\mu_i.
        \]
        The batch algorithm either inserts the items in $S$, or the single most profitable candidate $i_{\max}$, with $w_{\max}=\max_{i\in\mathcal C}w_i$. Specifically, it inserts $S$ if $\frac23\sum_{i=1}^k w_i\ge w_{\max}$, and inserts $i_{\max}$ otherwise. Denote these policies by $\pi_B(S)$ and $\pi_B(\{i_{\max}\})$, respectively.

        \begin{algorithm}[H]
        \caption{$O(\alpha)$ Batch Policy}
        \label{alg:batch-policy}
        \KwIn{Instance $I$ of \mpk}
        \KwOut{Batch policy $\pi_B$}
        $\mathcal C\gets\text{\textbf{Virtual-Items}}(I;1,1)$\; \tcp{Algorithm~\ref{alg:virtual-items}}
        \If{$\mathcal C=\emptyset$}{\Return the empty policy\;}
        Sort $\mathcal C$ by non-increasing density $d_i=w_i/\mu_i(1)$, with density $+\infty$ when $\mu_i(1)=0$ and integral candidates first in ties\;
        $U\gets 1/[3(1+\alpha)]$\;
        $S\gets\{1,\ldots,k\}$, the maximal prefix satisfying $\sum_{i=1}^k\mu_i(1)\le U$; set $S=\mathcal C$ if all candidates fit this bound\;
        $i_{\max}\gets\arg\max_{i\in\mathcal C}w_i$ and $w_{\max}\gets w_{i_{\max}}$\;
        \eIf{$\frac23\sum_{i\in S}w_i\ge w_{\max}$}{
            \Return $\pi_B(S)$\;
        }{
            \Return $\pi_B(\{i_{\max}\})$\;
        }
        \end{algorithm}

        The following lemma lower bounds the expected utility obtained when inserting $S$.
        \begin{lemma}
        \label{lem:batch-utility-S}
            \[
                \E[\utility(\pi_B(S))]\ge\frac23\sum_{i=1}^k w_i.
            \]
        \end{lemma}
        \begin{proof}
        Denote by $E_S$ the event that all items in $S$ fit in the knapsack. For a virtual item, $p_i$ denotes its expected cost; its value $v_i$ is deterministic because it mixes only non-null choices. The expected utility of the policy is
        \begin{align*}
        \E[\utility(\pi_B(S))]
        &=\Pr[E_S]\sum_{i=1}^k v_i-\sum_{i=1}^k p_i\\
        &\ge(1-\mu(S))\sum_{i=1}^k v_i-\sum_{i=1}^k p_i
          &&\text{(by \cref{lem:muS}, with $t=1$)}\\
        &\ge(1-\mu(S))\sum_{i=1}^k(w_i+p_i)-\sum_{i=1}^k p_i\\
        &=(1-\mu(S))\sum_{i=1}^k w_i-\mu(S)\sum_{i=1}^k p_i\\
        &\ge(1-\mu(S))\sum_{i=1}^k w_i-\alpha\mu(S)\sum_{i=1}^k w_i
          &&\text{(since $p_i\le\alpha w_i$)}\\
        &=\sum_{i=1}^k w_i\bigl(1-(1+\alpha)\mu(S)\bigr)\\
        &\ge\sum_{i=1}^k w_i\left(1-\frac{1+\alpha}{3(1+\alpha)}\right)
          &&\text{(since $\mu(S)\le U$)}\\
        &=\frac23\sum_{i=1}^k w_i.
        \end{align*}
        \end{proof}

        We can now conclude with the main result of this section.
        \begin{proof}[Proof of \Cref{thm:batch-approx}]
        If $\Phi(1;1)=0$, then \Cref{lem:adaptphi} gives $\adapt(I)=0$, and the empty policy suffices. Hence assume $\Phi(1;1)>0$. By \Cref{lem:batch-utility-S} and the singleton guarantee, the branch selected by the algorithm has expected utility at least
        \[
            \max\left\{\frac23\sum_{i\in S}w_i,w_{\max}\right\}.
        \]
        If $S=\mathcal C$, then \Cref{lem:batch-utility-S} immediately gives a lower bound of $\frac23\Phi(1;1)$, which is stronger than the claimed guarantee. Hence assume that candidate $k+1$ exists. Since $\sum_{i=1}^{k+1}\mu_i>U$, $\sum_{i\in\mathcal C}\mu_i\le1$, and the items are ordered by density,
        \[
            \sum_{i=1}^k w_i+w_{\max}
            \ge\sum_{i=1}^{k+1}w_i
            >U\Phi(1;1)=\frac{\Phi(1;1)}{3(1+\alpha)}.
        \]
        It follows that
        \[
            \max\left\{\sum_{i=1}^k w_i,w_{\max}\right\}
            \ge\frac12\left(\sum_{i=1}^k w_i+w_{\max}\right)
            \ge\frac{\Phi(1;1)}{6(1+\alpha)}.
        \]
        Therefore,
        \begin{align*}
            \E[\batchpk]
            &\ge\max\left\{\frac23\sum_{i=1}^k w_i,w_{\max}\right\}\\
            &\ge\frac23\max\left\{\sum_{i=1}^k w_i,w_{\max}\right\}\\
            &\ge\frac{1}{9(1+\alpha)}\Phi(1;1)
             \ge\frac{1}{18(1+\alpha)}\adapt(I),
        \end{align*}
        where the last inequality follows from \Cref{lem:adaptphi}, which gives
        \[
            \adapt(I)=\adapt(1)\le\Phi(2;1)\le2\Phi(1;1).
        \]
        \end{proof}

\section{Return-on-Investment in the Classic Pandora's Box Model}
\label{sec:pandora}

In this section, we provide strong evidence for the importance of the return-on-investment parameter in stochastic optimization with costly information acquisition. We revisit one of the most seminal such problems, Weitzman's Pandora's box problem~\cite{weitzman1978optimal}.\footnote{We thank an anonymous reviewer for the suggestion to explore the Pandora's box model through the simplicity-optimality lens.}

\paragraph{Pandora's Box Problem. }

A decision maker is faced with $n$ boxes. Each box $i\in[n]$ has a known cost $c_i$, and an unknown value $v_i$, whose value comes from an independent, non-negative and known value distribution $F_i$. Let $X_i$ be the random variable representing the value of box $i$. When the decision maker opens box $i$, they incur a cost $c_i$, and value $v_i$ is sampled from distribution $F_i$ and revealed to the decision maker. The decision maker can only keep the highest revealed prize, but incurs the costs of all opened boxes. Let $I_i$ be an indicator random variable that determines whether box $i$ was inspected; the policy maker's goal is to devise a policy that (approximately) maximizes

\[
\E[\max_{i}I_i v_i - \sum_{i} I_i c_i],
\]
where the expectation is over the values sampled from the distributions. We will use policies that randomly choose which boxes get opened in order to simplify our analysis, in which case, the expectation is taken over the randomness of the policy as well.\footnote{Every guarantee obtained by randomized policies can be obtained by a deterministic one by considering the instantiation that gets the best expected utility.} Without loss of generality, we assume $\E[v_i]\ge c_i$ for every $i$, as otherwise, a box is always non-profitable. 

\paragraph{Policy types.} 

In general, a policy can adaptively decide whether to stop and collect the current highest prize or to open another box, and which box to open next. The seminal result of~Weitzman~\cite{weitzman1978optimal}, described below, shows that there exists a simple optimal policy that chooses a predefined non-adaptive order in which to open the boxes, and stops once the reward is high enough such that the expected marginal contribution by any future box is eclipsed by its cost. This is analogous to the Stopping-Adaptive policies defined in~\Cref{sec:model}.

In this section, we inspect the power of non-adaptive policies in approximating Weitzman's optimal policy. Since there are no feasibility constraints, a non-adaptive policy simply chooses a set of boxes in advance, opens all selected boxes, incurs their total cost, and keeps the maximum realized prize. This is analogous to the batch policy described in~\Cref{sec:model}. 
For a fixed Pandora's box instance \(I\), let \(\adapt(I)\) denote the expected utility of Weitzman's optimal policy, and let \(\batch(I)\) denote the maximum expected utility of any batch policy. 

\paragraph{Weitzman's optimal policy.} 

For a box $i$, let $\sigma_i$ be $i$'s \textit{fair cap}, defined by $\E[(v_i-\sigma_i)^+]=c_i$. Such a cap exists whenever $c_i\le \E[v_i]$, as we assume. Weitzman's optimal policy sorts (and renames) boxes such that $\sigma_1\ge \sigma_2\ge\ldots\ge \sigma_n$, and we set $\sigma_{n+1}=0$. The policy opens boxes according to this order; after opening box $i$, it stops if the highest value obtained so far is at least $\sigma_{i+1}$. Intuitively, the cap represents the potential value of the box, after discarding the cost.

The adaptivity gap for the Pandora's box problem was discussed in~Singla~\cite{singla2018price} and Boodaghians, Fusco, Lazos, and Leonardi~\cite{boodaghians2020pandora}, where both works showed that this gap can be unbounded. \cite{singla2018price} used this as a justification to focus on adaptive policies, while \cite{boodaghians2020pandora} considered a relaxed objective, where the optimal policy gets $1/2$ of the obtained reward, but pays the full cost. We discuss the relationship between these results, fixed-subset benchmarks, and our ROI-based perspective in \Cref{app:pandora-related-work}.  

\paragraph{Tail IOR.}

The standard IOR parameter from Pandora's Knapsack does not directly capture the relevant return-on-investment in Pandora's Box. In Pandora's Knapsack, values add across feasible jobs; in Pandora's Box, the decision maker keeps only the maximum realized value. Thus, the relevant contribution of a box is not its total expected value, but the additional value it can provide beyond the current maximum. This motivates measuring a box only through the part of its distribution above its Weitzman fair cap. As illustrated in \Cref{sec:bad-roi}, not adapting the IOR to the Pandora's box objective can make the previously considered IOR artificially small even when the adaptivity gap is large. 

Formally, let
\begin{eqnarray}
\tilde{v}_i  = v_i\cdot \indic{v_i\ge \sigma_i}, \qquad \tilde{w}_i = \E[\tilde{v}_i]-c_i. \label{eq:tail-profit}
\end{eqnarray}
We define the Tail ROI and Tail Inverse ROI (or T-IOR) by
\begin{eqnarray}
\widetilde{\mathrm{ROI}}(i)=\tilde{w}_i/c_i,
\qquad
\tilde{\alpha} = \max_i 1/\widetilde{\mathrm{ROI}}(i). \label{eq:tail-roi-ior}
\end{eqnarray}

\textbf{Remark:} One can consider a special case of the Pandora's box problem and stochastic probing problem~\cite{gupta2016algorithms,gupta2017adaptivity}, where the value distributions are Bernoulli, and the feasibility set is a single box. In this case, the tail IOR parameter coincides with the $\alpha$ IOR parameter introduced for the Pandora's knapsack problem. 
Our main result in this section is the following.

\begin{theorem} \label{thm:pandora-tight.}
The adaptivity gap of Pandora's box is \(\Theta(\tilde{\alpha})\): for every Pandora's box instance \(I\),
\[
    \frac{\adapt(I)}{\batch(I)} = O(\tilde{\alpha}),
\]
and for every \(\tilde{\alpha}\), there exists a Pandora's box instance \(I\) with tail IOR \(\tilde{\alpha}\) such that \(\adapt(I)/\batch(I) = \Omega(\tilde{\alpha})\).
\end{theorem}

We proceed by introducing a randomized non-adaptive policy that $O(\tilde{\alpha})$-approximates the optimal adaptive policy, and then giving an instance showing this is essentially tight.

\subsection{\texorpdfstring{$O(\tilde{\alpha})$ Batch Policy for Pandora's Box}{O(tail-IOR) Batch Policy for Pandora's Box}}

For a Pandora's box instance \(I\), let $p_i$ be the probability that Weitzman's optimal policy opens box $i$.\footnote{$p_i$ can be efficiently computed given $\{\sigma_j\}_{j\in [n]}$ and oracle access to the CDFs of the value distributions.} For $\beta=1/(1+\tilde\alpha)$, where \(\beta \in[0,1]\), we define the randomized non-adaptive policy \(\pbatch\) as follows: it picks each box independently with probability \(\beta p_i\), and then simultaneously opens all boxes it picked. That is, the policy is likely to pick boxes with a good tail return-on-investment (which is proportional to $\beta$) that are likely to get picked by Weitzman's optimal policy.
Let \(\pbatch(I)\) denote its expected utility on instance \(I\).

\begin{theorem}
For every Pandora's box instance \(I\),
\[
    \frac{\adapt(I)}{\pbatch(I)} = O(\tilde{\alpha}).
\]

\end{theorem}
\begin{proof}
Let $I_i$ be the indicator random variable indicating whether box $i$ was opened. Recall that

\begin{eqnarray}
\adapt(I)
&=&
\E\left[\max_i I_i v_i - \sum_i I_i c_i\right] \nonumber\\
&=&
\E\left[\max_i I_i v_i\right] - \sum_i \E[I_i]c_i \nonumber\\
&=&
\underbrace{\E\left[\max_i I_i v_i\right]}_{R}
-
\underbrace{\sum_i p_i c_i}_{C}.
\label{eq:adapt_bound}
\end{eqnarray}

The proof follows from the two lemmas proved below.

\begin{restatable}{lemma}{rbound}\label{lem:r_bound}
$$R\le (1+\tilde{\alpha})\adapt(I).$$
\end{restatable}

\begin{restatable}{lemma}{pbatchprize}\label{lem:pbatch_prize}
Let $S$ be the random set chosen by $\pbatch$. We have that 
$$\E\left[\max_{i\in S} v_i\right]\ge  (\beta-\beta^2/2)R.$$
\end{restatable}

Using these two lemmas, we obtain

\begin{align*}
\pbatch(I)
&=
\E_S\left[\max_{i\in S} v_i-\sum_{i\in S}c_i\right] \\
&\ge
\left(\beta-\frac{\beta^2}{2}\right)R-\sum_i \beta p_i c_i
&& (\text{by \Cref{lem:pbatch_prize}}) \\
&=
\left(\beta-\frac{\beta^2}{2}\right)R-\beta C \\
&=
\left(\beta-\frac{\beta^2}{2}\right)R-\beta(R-\adapt(I)) \\
&=
\beta\adapt(I)-\frac{\beta^2}{2}R \\
&\ge
\beta\adapt(I)-\frac{(1+\tilde{\alpha})\beta^2}{2}\adapt(I)
&& (\text{by~\Cref{lem:r_bound}}) \\
&=
\frac{1}{1+\tilde{\alpha}}\adapt(I)-\frac{1}{2(1+\tilde{\alpha})}\adapt(I)
&& (\beta=1/(1+\tilde{\alpha})) \\
&=
\frac{1}{2(1+\tilde{\alpha})}\adapt(I) .
\end{align*}
\end{proof}

\begin{proof}[Proof of \Cref{lem:r_bound}]
Since Weitzman's optimal policy stops after opening box $i$ whenever the highest obtained value is at least $\sigma_{i+1}$, and since $\sigma_i\ge \sigma_{i+1}$, at most one opened box can satisfy $v_i\ge\sigma_i$. Thus,
$$\sum_i I_i\indic{v_i\ge \sigma_i}\le 1.$$ Therefore,
\begin{eqnarray}
R = \E[\max_i I_i v_i]\ge \E[\sum_i I_i v_i\indic{v_i\ge \sigma_i}] = \sum_i p_i\E[\tilde{v}_i], \label{eq:reward_bound}
\end{eqnarray}
where the last equality follows from the independence of the value of box $i$ and opening box $i$.

By the definition of $\tilde{\alpha}$, we have
\begin{eqnarray}
C = \sum_i p_i c_i
&\le& \tilde{\alpha} \sum_i p_i \tilde{w}_i \nonumber\\
&=& \tilde{\alpha} \left(\sum_i p_i \E[\tilde{v}_i] - \sum_i p_i c_i \right) \nonumber\\
&\le& \tilde{\alpha}(R-C)=\tilde{\alpha}\adapt(I), \label{eq:c_bound}
\end{eqnarray}
where the last inequality follows from Equations~\eqref{eq:adapt_bound} and~\eqref{eq:reward_bound}.

We get
\begin{eqnarray}
R=\adapt(I)+C\le (1+\tilde{\alpha})\adapt(I). \label{eq:r_bound}
\end{eqnarray}
\end{proof}

\begin{proof}[Proof of \Cref{lem:pbatch_prize}]
For \(t\ge0\), let \(q_i(t)=\Pr[v_i\ge t]\). Since \(\pbatch\) chooses each box independently with probability \(\beta p_i\), we have
\begin{eqnarray}
\Pr\left[\max_{i\in S} v_i\ge t\right]
=1-\prod_i(1-\beta p_iq_i(t))
\ge 1-e^{-\beta\sum_i p_iq_i(t)} . \label{eq:max_batch_bound}
\end{eqnarray}

Moreover, for \(t>0\), the event that Weitzman's policy opens box \(i\) is independent of \(v_i\), since the policy decides whether to open box \(i\) before observing its value. Therefore,
\begin{eqnarray*}
\Pr[\max_i I_i v_i\ge t] & \le & \Pr[\exists i \ : \ I_i v_i\ge t]\\
& \le&\sum_i \Pr[I_i v_i\ge t] \\
& = & \sum_i \Pr[I_i] \Pr[ v_i\ge t] \\
& =& \sum_i p_i q_i(t).
\end{eqnarray*}

Thus, by Eq.~\eqref{eq:max_batch_bound},
\begin{eqnarray*}
\Pr\left[\max_{i\in S} v_i\ge t\right]  & \ge & 1-e^{-\beta \Pr[\max_i I_i v_i\ge t]} \\
&\ge& \beta \Pr[\max_i I_i v_i\ge t] - \beta^2 \Pr[\max_i I_i v_i\ge t]^2/2\\
& \ge&  \Pr[\max_i I_i v_i\ge t](\beta-\beta^2/2),
\end{eqnarray*}
where the second inequality follows from \(1-e^{-x}\ge x-x^2/2\) for \(x\ge0\), and the last uses \(\Pr[\max_i I_i v_i\ge t]\le1\).

Integrating the above over \(t\ge0\) gives 
\begin{eqnarray}
\E\left[\max_{i\in S} v_i\right]\ge (\beta-\beta^2/2)\E[\max_i I_i v_i] = (\beta-\beta^2/2)R.\label{eq:pbatch-value}
\end{eqnarray}
\end{proof} 

The matching \(\Omega(\tilde{\alpha})\) lower bound is proved in \Cref{app:pandora-lower-bound}.

\subsection{\texorpdfstring{$\Omega(\tilde{\alpha})$ Adaptivity Gap for Pandora's Box}{Omega Tail-IOR Adaptivity Gap for Pandora's Box}}
\label{app:pandora-lower-bound}

In this section, we give an instance in which the adaptivity gap is at least \(\Omega(\tilde{\alpha})\). In our instance, all boxes are identical: the random values of the boxes are identical Bernoulli variables, and all costs are identical. Therefore, as in the above remark, the tail IOR and IOR parameters coincide. 
\begin{theorem}
For every \(\tilde{\alpha}\), there exists a Pandora's box instance \(I\) with tail IOR \(\tilde{\alpha}\) such that
\[
    \frac{\adapt(I)}{\batch(I)} = \Omega(\tilde{\alpha}).
\]
\end{theorem}
\begin{proof}

Consider $n$ identical boxes with cost $1$ and value $v = n+1$ with probability $1/n$ and $0$ with the remaining probability. For this case, 
\begin{eqnarray}
\tilde{\alpha} = \alpha = 1/(1+1/n-1)= n. \label{eq:alpha_pbatch_lb}
\end{eqnarray}

Consider the utility of Weitzman's optimal policy. Since all boxes are identical, order does not play a role, and the policy simply opens boxes until there is a box with a high value $v=n+1$, if there is such a box. Therefore, its expected reward is 
\begin{eqnarray*}
\adapt(I) & = & (n+1)\Pr[\exists i\ : \ v_i=n+1] - 1\E[\mbox{\# boxes opened}] \\
& =& (n+1)\Pr[\exists i\ : \ v_i=n+1]- \sum_{i=1}^{n} \Pr[\ge i\mbox{ boxes opened}]\\
& = & (n+1)(1-(1-1/n)^n) - \sum_{i=1}^{n}(1-1/n)^{i-1} \\
& = & (n+1)(1-(1-1/n)^n) - n\cdot(1-(1-1/n)^n) \\ 
& \approx & 1-1/e. 
\end{eqnarray*}

Let $B(k)$ be the reward of a batch policy that opens exactly $k$ boxes. We have that 
\begin{eqnarray}
B(k) = (n+1)(1-(1-1/n)^k)-k.
\end{eqnarray}
One can verify that $B'(k)\le 0$ for $k\ge 2$. Therefore, it suffices to consider $k=1$ and $k=2$. We get that $B(1)=1/n>1/n-1/n^2=B(2)$. Therefore, $\batch(I)= 1/n$ for this instance. We get that $\adapt(I)/\batch(I)\approx(1-1/e)n= \Omega(\tilde \alpha).$ Therefore, for every $\tilde\alpha$, there exists such an instance.
\end{proof}


    \section{Knapsack Contracts: An Application to Contract Design}
    \label{sec:contract-application}

    In this section we show the formal connection between \mpk\ and a natural multi-agent contract setting, in which a principal sequentially delegates jobs to agents up to a deadline. Intuitively, this setting is very similar to \mpk\, since the principal has a value for each completed job, there are multiple choices of effort levels at which each job can be executed, and incentivizing different choices incurs different costs on the principal (the cost of paying the agent enough to incentivize the chosen effort level; this cost is well-understood based on existing contract theory).
    There is, however, one main difference: in the contract setting (unlike \mpk), the principal's costs are \emph{random variables}, since the principal pays each agent according to the \emph{realized} size of each job. In \cref{sub:contract-setup} we set up the contract problem, and in \cref{sub:reduction} we show how to reduce it to \mpk\ despite the aforementioned difference.

    \subsection{Setup} 
    \label{sub:contract-setup}
    
    An instance $I$ of the \emph{Knapsack Contracts} problem consists of a principal, $n$ agents, and a knapsack of size $S$. Without loss of generality, let $S=1$, where $S$ represents a budget of some limited resource. As our running example, we assume $S$ is the time remaining until a set deadline. Each agent $i$ has a job which they can carry out to the benefit of the principal. 
    Agent $i$ is represented by the following: 

    \begin{itemize}
        \item $v_{i}$: The value which the principal obtains if agent $i$ completes their job by the deadline.
        \item $\{(F_{i}^{j},c_{i}^{j})\}_{j\in[m]}$: The possible \emph{effort levels} of agent $i$.
            \footnote{All agents have the same number of effort levels without loss of generality, since we can add dummy effort levels that would never be incentivized by the principal.} Each effort level $j\in [m]$ determines a distribution $F_i^j$ over the time it takes the agent to complete the job, as well as an agent-cost $c_{i}^{j}\ge 0$. If the agent chooses to exert effort level $j$, they incur agent-cost $c_i^j$, and complete their job within time $s_{i}$ drawn (independently) from distribution $F_{i}^{j}$. We assume non-decreasing agent-costs $c_{i}^{1}\le\dots\le c_{i}^{m}$. Without loss of generality, effort level $1$ is a \emph{null} effort level, modeling the agent's option to decline: $c_i^1 = 0$, and the job is not carried out, consuming no time and yielding zero value to the principal. This ensures individual rationality for the agent.
            Let $\bF_{i}=\left(F_{i}^{1},\ldots, F_{i}^{m}\right)$. Let $\supp(F_{i}^{j})$ denote the support of $F_{i}^{j}$, and $\supp(\bF_{i})=\cup_{j\in [m]}\supp(F_{i}^{j}).$ We assume the support of each distribution is finite and is explicitly given.        
    \end{itemize}

    The principal knows $I=(v_{i}, \{F_{i}^{j},c_{i}^{j}\}_{j\in[m]})_{i\in[n]}$ in advance. After an agent completes a job, the principal observes the time it took the agent to complete it, but they \textit{do not} observe the effort level chosen by the agent. The principal offers each agent $i$ a contract $t_{i}:\mathbb{R}_{\ge 0}\rightarrow \mathbb{R}_{\ge 0}$, mapping $i$'s observed processing time to a transfer paid by the principal to the agent. Given contract $t_{i}$, agent $i$ chooses an effort level
    \[
        j \in \arg\max_{\ell}\E_{s_i\sim F_{i}^{\ell}}[t_{i}(s_{i})] - c_{i}^{\ell}.
    \]
    As is standard, ties are broken in favor of the principal.
    \begin{definition}[Implementable effort level] \label{def:implementable}
        For agent $i$, an effort level $j$ is \emph{implementable} if there exists a contract $t_{i}$ for which this action maximizes the agent's expected utility among all actions given the contract. In this case, we say contract $t_{i}$ incentivizes action $j$. 
        
        Any implementable effort level $j$ for agent $i$ has an optimal contract that incentivizes it, which can be obtained by solving the following LP:
    \begin{align}
        \min \quad       & \sum_{s \in \supp(F_i^j)}t_i(s)\cdot F_{i}^{j}(s)                                                                                  &  & \nonumber                                                        \\
        \text{s.t.}\quad & \sum_{s \in \supp(F_i^j)}t_i(s)\cdot F_{i}^{j}(s) - c_{i}^{j}\ge \sum_{s \in \supp(F_i^{j'})}t_i(s)\cdot F_{i}^{j'}(s) - c_{i}^{j'} &  & \forall j' \in [m] \nonumber                                     \\
                         & t_i(s)\ge t_i(s')                                                                                                                   &  & \forall s < s' \in \supp(\mathbf{F}_{i}) \label{eq:monotonicity} \\
                         & t_i(s)\ge 0                                                                                                                        &  & \forall s \in \supp(\mathbf{F}_{i})\nonumber
    \end{align}
    Constraint \eqref{eq:monotonicity} ensures monotonicity of payments. This is required as without it, an agent might have an incentive to hover around after completing the job and get a higher payment.
    \end{definition}

    The principal's problem can then be described as follows: when hiring agent $i$, the principal chooses an effort level $j$ to incentivize, and offers the optimal contract that incentivizes that effort level. The principal's collected value is the sum of the values of the jobs that the agents get to complete with a total processing time at most the budget $S$. The principal aims to maximize their utility: the expected total value minus the expected sum of transfers made to the agents.

    \subsection{Reduction} 
    \label{sub:reduction}
    The following proposition shows that in order to find an approximately optimal policy for the Knapsack Contracts problem, it suffices to find an approximately optimal policy for the \mpk\ problem.

\begin{proposition}[Reduction to Multi-choice Pandora's Knapsack]
    \label{lem:reduction}
    Given a polynomial-time algorithm that finds a policy that $c$-approximates the optimal policy for the \mpk\ problem,
    one can efficiently find a policy that $c$-approximates the optimal policy for the Knapsack Contracts problem.
    \end{proposition}

        \begin{proof}
        Given a Knapsack Contracts instance, for each agent $i \in [n]$, we create a corresponding item $i$ in the \mpk\ instance. For each implementable effort level $j$ of agent $i$, we create a corresponding choice $j$ for item $i$. The parameters are mapped as follows:
        \begin{itemize}[nosep,noitemsep]
            \item The value of item $i$ is set to the principal's value $v_i$.
            \item The size of choice $(i,j)$ is a random variable with the distribution $F_i^j$.
            \item The cost of choice $(i,j)$ is set to $p_i^j=\E_{s\sim F_i^j}[ t_i(s)]$, where $t_i$ is the contract that minimizes the expected payment, as given by the LP in \Cref{def:implementable}.
        \end{itemize}

        Now, consider any policy $\varphi$ for the contracts problem, and denote by $\varphi'$ the corresponding policy for the knapsack problem. $\varphi$ specifies the choice of 
        (agent, effort level) pairs for the contracts problem, and $\varphi'$ specifies the corresponding item-choice pairs for the knapsack problem.

         It is left to show that the expected profits are identical. Denote by $S$ the set of successfully completed jobs in the contracts and knapsack problems, and by $A$ the set of agents/items attempted in the contracts and knapsack problems. Notice that $S$ and $A$ have exactly the same distribution under $\varphi$ and $\varphi'$, as both policies make the same choices under the same randomization.
         
         The utility for policy $\varphi$ is a random variable that depends on the (possibly random) choices of the policy and the random sizes of the attempted jobs. Its expectation is given by 
        \begin{equation}\label{eq:profit_phi_expected}
            \E[\utility(\varphi)] 
            = \E\left[\sum_{(i,j)\in S}{v_i} - \sum_{(i,j)\in A}{t_{i}(s_i)}\right] 
            = \E\left[\sum_{(i,j)\in S}{v_i}\right] - \E\left[\sum_{(i,j)\in A}{t_{i}(s_i)}\right].
        \end{equation}
        Expanding the transfer term, we get:
        \begin{align*}
            \E\left[\sum_{(i,j)\in A}{t_{i}(s_i)}\right] & = \E\left[\sum_{(i,j)}{\mathbbm{1}_{(i,j)\in A} \cdot t_{i}(s_i)}\right] = \sum_{(i,j)}\E\left[{\mathbbm{1}_{(i,j)\in A} \cdot t_{i}(s_i)}\right] \\
            & \stackrel{(\star)}{=}  \sum_{(i,j)}\E\left[\mathbbm{1}_{(i,j)\in A}\right] \cdot \E_{s_i\sim F_i^j}\left[t_{i}(s_i)\right] \stackrel{(\star\star)}{=} \sum_{(i,j)}\E\left[\mathbbm{1}_{(i,j)\in A}\right] \cdot p_i^j \\
            &= \E\left[\sum_{(i,j)}\mathbbm{1}_{(i,j)\in A}\cdot p_i^j\right] = \E\left[\sum_{(i,j)\in A}p_i^j\right],
        \end{align*}
        where equality $(\star)$ is due to the independence between $\mathbbm{1}_{(i,j)\in A}$ and $s_i^j$, and equality $(\star\star)$ is due to the definition of $p_i^j$.
        Plugging this back in \cref{eq:profit_phi_expected} yields:
\[
    \E[\utility(\varphi)] = \E\left[\sum_{(i,j)\in S}{v_i} - \sum_{(i,j)\in A} p_i^j\right] = \E[\utility(\varphi')]~,
\] which implies the desired result.
        
    Thus, given a $\varphi'$ that $c$-approximates the optimal policy in the constructed instance of the \mpk\ problem, the corresponding $\varphi$ $c$-approximates the optimal policy in the original Knapsack Contracts instance.
    \end{proof}

\paragraph{\textbf{Implications of the reduction to the Knapsack Contracts Problem.}} The above implies: $(i)$~there exists a skipping-adaptive policy that gives an $O(1)$ approximation to the principal's optimal utility (by~\Cref{thm:a}); and $(ii)$~one can naturally adapt the ROI parameter to the Knapsack Contracts setting, treating the expected transfer that the principal makes when incentivizing agent $i$ to exert effort level $j$ as $p_i^j$ in~\Cref{def:RoI}. This implies that an instance with $\alpha$ IOR for the Knapsack Contracts problem is reduced to an \mpk\ instance with $\alpha$ IOR. Thus, \Cref{subsec:batch-policy} implies an $O(\alpha)$-approximate batch policy for the Knapsack Contracts problem. As all of our lower bound instances use items with a single choice, the lower bounds also trivially extend to the Knapsack Contracts problem, where each agent has a single effort level, and the principal need only cover the agent-cost exactly. Therefore, all results obtained for the \mpk\ problem, summarized by~\Cref{fig:adaptivity_levels}, also apply to the contract problem.

\section{Conclusion} 
\label{sec:conclusion}

In this work, we study an economically motivated extension of the canonical Stochastic Knapsack problem through the lens of simple-vs-optimal policies. By introducing costs, we show that the adaptivity gap can be unbounded, and that this gap is governed by a natural return-on-investment (ROI) parameter. We also introduce a hierarchy of increasingly adaptive policies, and show that limited adaptivity can still achieve near-optimal utility.

Our results characterize every level of this hierarchy up to constant factors: relative to the optimal fully-adaptive policy, batch and non-adaptive policies lose a $\Theta(\alpha)$ factor, stopping-adaptive policies lose a $\Theta(\sqrt{\alpha})$ factor, and skipping-adaptive policies lose only a constant factor.

A broader direction is to revisit other classic decision-making problems where introducing costs is well motivated, and to study how this changes the optimization landscape. We expect simple-vs-optimal approaches to remain useful in such settings, along with the ROI parameter.

Finally, we suggest re-examining other optimization problems with mixed-sign objectives, and exploring whether our approach can yield new insights there as well. We provide strong evidence for the usefulness of our approach by revisiting the canonical Pandora's box problem, and giving tight adaptivity gap bounds by reasoning about the return-on-investment of the boxes.

\ifanonymous\else
\section*{Acknowledgments} 

A.~Eden is incumbent of the Harry \& Abe Sherman Senior Lectureship at the School of Computer Science and Engineering at the Hebrew University. The work of A.\ Berlin and A.\ Eden was supported by the Israel Science Foundation (grant No.~533/23). The work of I.R.~Cohen was supported by the Israel Science Foundation (grant No.~1737/21). The work of Z.\ Barak, O.\ Porat and I.\ Talgam-Cohen was supported by the European Research Council (ERC) under the European Union’s Horizon Europe research and innovation program (grant No.~101077862, project ALGOCONTRACT), by the Israel Science Foundation (grant No.~3331/24), by the NSF-BSF (grant No.~2021680), and by a Google Research Scholar Award.
\fi

\section*{AI Methodology}
We used ChatGPT and Claude to assist with editing, generating the figure, and rephrasing parts of some of the arguments. Astra 6 was used to optimize the parameters in the lower bound construction of \cref{thm:adapt_stadapt_gap}, improving our lower bound from $\Omega(\alpha^{1/3})$ (which was achieved via a similar construction, without the use of AI) to $\Omega(\sqrt{\alpha})$. All outputs were verified by the authors. We assume responsibility for all content.

    
    \appendix

\section*{Appendix Organization}
\Cref{app:pandora-related-work} includes detailed related work. The pseudo-polytime conversion and continuous-distribution details for the skipping- and stopping-adaptive policies from Sections~\ref{sub:adaptive-const-approx} and~\ref{sub:stopping-ub} appear in \Cref{app:skipping-details}, while the LP formulation and properties used by the batch policy from Section~\ref{subsec:batch-policy} are in \Cref{app:batch-upper-bounds}.
\Cref{sec:bad-roi} shows that the inverse-ROI parameter from Pandora's Knapsack does not control the adaptivity gap in the classic Pandora's Box model discussed in Section~\ref{sec:pandora}.
Model relaxations appear in \Cref{sec:relaxations}. 

\section{Detailed Related Work: Pandora's Box, Fixed Subsets, and ROI}
\label[appendix]{app:pandora-related-work}

Our work is related to the Pandora's Box literature, initiated by the classic work of Weitzman~\cite{weitzman1978optimal}. In Weitzman's model, boxes have inspection costs and random values, and the optimal policy follows a reservation-value order while retaining the ability to stop adaptively. Thus, even this canonical policy is only order-non-adaptive, and is not a fixed-subset/open-all policy. We refer the reader to recent surveys, e.g., Beyhaghi and Cai~\cite{beyhaghi2024recent}, for a broader overview of the many variants of Pandora's Box and their connections to prophet inequalities, constrained exploration, and adaptivity gaps.

The closest Pandora-style result to our batch benchmark is Theorem~3 of Boodaghians, Fusco, Lazos, and Leonardi~\cite{boodaghians2020pandora}. They show that, for Pandora's Box with prefix-closed constraints, every adaptive strategy admits a non-adaptive strategy---equivalently, a feasible set ($S$) opened obliviously---whose expected utility is at least one half of the adaptive reward term minus the full adaptive cost term. In our notation, their guarantee is of the form
\[
\mathbb{E}\left[\max_{i\in S}X_i-\sum_{i\in S}c_i\right]
\ge
\frac12
\mathbb{E}\!\left[\max_{i\in\mathcal{S}}X_i\right]
-
\sum_i p_i c_i,
\]
where \(\mathcal{S}\) denotes the random set of boxes opened by the adaptive policy, and \(p_i=\Pr[i\in\mathcal{S}]\). This is very close in policy structure to a fixed-subset/open-all policy. However, it is not a multiplicative approximation to the adaptive net utility
\[
\mathbb{E}\!\left[\max_{i\in\mathcal{S}}X_i\right]
-
\sum_i p_i c_i.
\]
Indeed, when reward and cost nearly cancel, the right-hand side may be negative. This is precisely the low-ROI regime that drives our lower bounds. Thus, their theorem provides an important fixed-set Pandora analogue, while our results identify ROI as the parameter governing when such fixed-set guarantees can translate into approximations of mixed-sign utility.

The fixed-subset/open-all version of Pandora's Box can also be viewed as the optimization problem
\[
\max_{S\subseteq[n]}
\left\{
\mathbb{E}\left[\max_{i\in S}X_i\right]
-
\sum_{i\in S}c_i
\right\}.
\]
Since \(S\mapsto \mathbb{E}[\max_{i\in S}X_i]\) is monotone submodular, subtracting linear costs yields a generally non-monotone submodular objective. The costless cardinality-constrained analogue was studied by Mehta, Nadav, Psomas, and Rubinstein~\cite{mehta2020hitting}, who give a PTAS for selecting \(k\) independent random variables to maximize the expected maximum. This result is structurally related to fixed-subset selection, but does not capture the mixed-sign, ROI-driven adaptivity gap that arises in our model.

More broadly, constrained Pandora's Box results, such as Singla~\cite{singla2018price} and Boodaghians, Fusco, Lazos, and Leonardi~\cite{boodaghians2020pandora}, compare adaptive policies with simpler non-adaptive or partially adaptive policies. In these works the main obstruction comes from feasibility or order constraints, whereas in our setting the obstruction comes from low return-on-investment. Finally, while Beyhaghi and Kleinberg~\cite{beyhaghi2019pandora} study committing policies for Pandora's problem with nonobligatory inspection, their commitment notion is farther from our model: the policy does not commit to a subset that must all be opened, and it still retains early stopping and selection flexibility.


  \section{Appendix for Sections~\ref{sub:adaptive-const-approx} and~\ref{sub:stopping-ub}: Skipping- and Stopping-Adaptive Policies}
\label[appendix]{app:skipping-details}

\subsection{From a Pseudo-Polynomial-Time to a Polynomial-Time Algorithm} \label[appendix]{sec:polytime_implementation}
    In this section, we show how to turn the $O(1)$ skipping-adaptive algorithm (\Cref{alg:adaptive-skc}) and the $O(\sqrt{\alpha})$ stopping-adaptive algorithm (\Cref{alg:stopping-pk}), which both use the LP $\Psi(T)$, from pseudo-polynomial-time algorithms into polynomial-time algorithms. We do so by giving a compact LP that may be solved in polynomial time yet gives a $4$-approximation to the original LP.
    \label[appendix]{sec:round-lp-for-adaptive-policy}

    We formulate an LP $\Psi_P(b)$---a compact version of $\Psi(b)$ with $O(n m \log(b))$ constraints.
    
    We first show that $\Psi_P(b) \ge \frac{1}{4}\Psi(b)$ (\cref{lem:small-lp-approx-large-lp}), and then give a way to create an approximate solution to $\Psi(b)$ given a solution to $\Psi_P(b)$ (\cref{lem:at-least-half-val}). Thus, it is enough to use $\Psi_P(b)$ in our algorithms. We finally discuss how to modify the algorithms to use the compact LP.
    
    For any $b \in \N$, let $\mathcal{T}_b := \crl*{0} \cup \crl*{2^l : l \in \crl*{0,\ldots,\lfloor\log(b)\rfloor}}$ be the set of breakpoints used by the compact LP.

    Let $\Psi_P(b)$ be the following linear program:
    \begin{align*}
    \Psi_P(b) &:= \max_{x} \quad
      \sum_{i\in[n]} \sum_{j\in[m]} \sum_{\tau \in \mathcal{T}_b} w_{i,j,2\tau} \, x_{i,j,\tau}
      \\[2mm]\notag
    \text{s.t.}\quad
    & \sum_{i\in[n]} \sum_{j\in[m]} \sum_{\tau' \le \tau} x_{i,j,\tau'} \, \mu_{i,j,2\tau} \le 2\tau,
    && \forall\, \tau \in \mathcal{T}_b,
    \numberthis \label{constraint:capacity-p}
    \\[1mm]
    & \sum_{j \in [m]} \sum_{\tau \in \mathcal{T}_b} x_{i,j,\tau} \le 1,
    && \forall\, i \in [n],
    \numberthis \label{constraint:take-each-item-time-at-most-once-p}
    \\[2mm]
    & 0 \le x_{i,j,\tau} \le 1,
    && \forall\, i \in [n],\, j \in [m],\, \tau \in \mathcal{T}_b.
    \numberthis \label{constraint:x-is-prob-p}
    \end{align*}
    Here $w_{i,j,2\tau}$ and $\mu_{i,j,2\tau}$ are given by the same formulas as in \Cref{sub:adaptive-const-approx} also when $2\tau > b$.

    \begin{lemma}\label{lem:small-lp-approx-large-lp}
        $\Psi_P(b) \ge \frac{1}{4} \Psi(b)$.
    \end{lemma}
    \begin{proof}
    Let $x$ be the optimal solution to $\Psi(b)$.
    
    Let $\hat{x}_{i,j,0} = \frac{1}{4} \sum_{t \in \crl*{0,1}} x_{i,j,t}$, and $\hat{x}_{i,j,\tau} = \frac{1}{4} \sum_{t \in [2\tau, 4\tau)} x_{i,j,t}$ for any $\tau \in \mathcal{T}_b \setminus \crl*{0}$, where $x_{i,j,t} := 0$ for $t > b$.

    Let us show that all constraints of $\Psi_P(b)$ hold for $\hat{x}$.

    For any $i \in [n]$, by constraint \ref{constraint:take-each-item-time-at-most-once} of $\Psi(b)$:
    
    \[
        \sum_{j \in [m]} \sum_{\tau \in \mathcal{T}_b} \hat{x}_{i,j,\tau} \le \frac{1}{4} \sum_{j \in [m]} \sum_{t \ge 0} x_{i,j,t}\le \frac{1}{4}.
    \]

    Also, for $\tau = 0$ constraint \ref{constraint:capacity-p} holds trivially since $\mu_{i,j,0} = 0$, and for $\tau \ge 1$ with $4\tau - 1 \le b$,

    \begin{align*}
        \sum_{i\in[n]} \sum_{j\in[m]} \sum_{\tau' \le \tau} \hat{x}_{i,j,\tau'} \, \mu_{i,j,2\tau} & = \sum_{i\in[n]} \sum_{j\in[m]} \sum_{t' = 0}^{4\tau - 1} \frac{1}{4} x_{i,j,t'} \cdot  \mu_{i,j,2\tau} \\
        & \le \frac{1}{4} \sum_{i\in[n]} \sum_{j\in[m]} \sum_{t' = 0}^{4\tau - 1} x_{i,j,t'} \cdot  \mu_{i,j,4\tau - 1} \le \frac{1}{4} \cdot 2 \prn*{4\tau - 1} \le 2\tau,
    \end{align*}
    where the first inequality is due to the fact that $\mu_{i,j,2\tau} \le \mu_{i,j,4\tau-1}$ by the definition of $\mu$, and the second inequality is due to the fact that $x$ is a feasible solution to $\Psi(b)$ and thus constraint \ref{constraint:capacity} holds.
    If $4\tau - 1 > b$, we use constraint \ref{constraint:capacity} at time $b$ instead. Since $\min\crl*{s, ct} \le c \min\crl*{s,t}$ for $c \ge 1$, we have $\mu_{i,j,2\tau} \le \max\crl*{1, \frac{2\tau}{b}} \mu_{i,j,b}$, and thus
    \begin{align*}
        \sum_{i\in[n]} \sum_{j\in[m]} \sum_{\tau' \le \tau} \hat{x}_{i,j,\tau'} \, \mu_{i,j,2\tau} & \le \frac{1}{4} \max\crl*{1, \frac{2\tau}{b}} \sum_{i\in[n]} \sum_{j\in[m]} \sum_{t=0}^{b} x_{i,j,t} \cdot \mu_{i,j,b} \\
        & \le \frac{1}{4} \max\crl*{1, \frac{2\tau}{b}} \cdot 2b = \frac{1}{4} \max\crl*{2b, 4\tau} \le 2\tau,
    \end{align*}
    where the last inequality holds since $b < 4\tau$.

    So all constraints of $\Psi_P(b)$ hold and thus $\hat{x}$ is a feasible solution.
    Thus, the value obtained from this solution is a lower bound for $\Psi_P(b)$ and therefore:
    
    \[
        \Psi_P(b) \ge \sum_{i\in[n]} \sum_{j\in[m]} \sum_{\tau \in \mathcal{T}_b} w_{i,j,2\tau} \, \hat{x}_{i,j,\tau} \ge \frac{1}{4} \sum_{i\in[n]} \sum_{j\in[m]} \sum_{t \in \crl*{0,\ldots,b}} w_{i,j,t} \, x_{i,j,t} = \frac{1}{4} \Psi(b).
    \]
    
    \end{proof}
    
    Let $x^*$ be the optimal solution of $\Psi_P(b)$. For $\tau \in \mathcal{T}_b \setminus \crl*{0}$, let $I_\tau := [\tau, 2\tau) \cap \crl*{0,\ldots,b}$ (so $I_\tau = [\tau, 2\tau)$, except possibly for the largest breakpoint). Let $y$ be defined as follows: $y_{i,j,0} = x^*_{i,j,0}$ and $y_{i,j,t} = \frac{x^*_{i,j,\tau}}{|I_\tau|}$ for any $i,j \in [n] \times [m]$, $\tau \in \mathcal{T}_b \setminus \crl*{0}$ and $t \in I_\tau$. In the following lemma we show that $y$ is a feasible solution of $\Psi(b)$ which yields at least the same value as $x^*$.

    \begin{lemma}\label{lem:at-least-half-val}
        Let $x^*$ be the optimal solution of \ $\Psi_P(b)$, and let $y \in [n] \times [m] \times \crl*{0,\ldots,b}$ be defined by $y_{i,j,0} = x^*_{i,j,0}$ and $y_{i,j,t} = \frac{x^*_{i,j,\tau}}{|I_\tau|}$ for any $i,j \in [n] \times [m]$, $\tau \in \mathcal{T}_b \setminus \crl*{0}$ and $t \in I_\tau$.
        Then $y$ is a feasible solution of \ $\Psi(b)$ yielding a value of at least a quarter of the optimal LP value.
    \end{lemma}
    \begin{proof}
    Since $x^*$ is a feasible solution of $\Psi_P(b)$, constraints \ref{constraint:take-each-item-time-at-most-once}, \ref{constraint:x-is-prob} hold (due to constraints \ref{constraint:take-each-item-time-at-most-once-p}, \ref{constraint:x-is-prob-p} of $\Psi_P(b)$).
    
    For $t = 0$, constraint \ref{constraint:capacity} holds trivially since $\mu_{i,j,0} = 0$. Otherwise, let $t \in I_\tau$ for some $\tau \in \mathcal{T}_b \setminus \crl*{0}$.

    \begin{align*}
        \sum_{i\in[n]} \sum_{j\in[m]} \sum_{t' \le t} y_{i,j,t'} \, \mu_{i,j,t} & \le \sum_{i\in[n]} \sum_{j\in[m]} \sum_{t' \le t} y_{i,j,t'} \, \mu_{i,j,2\tau} \le \sum_{i\in[n]} \sum_{j\in[m]} \sum_{t' < 2\tau} y_{i,j,t'} \, \mu_{i,j,2\tau} \\
        & = \sum_{i\in[n]} \sum_{j\in[m]} \sum_{\tau' \le \tau} x^*_{i,j,\tau'} \, \mu_{i,j,2\tau} \le 2\tau \le 2t,
    \end{align*}
    where the first inequality is due to the monotonicity of $\mu_{i,j,t}$ in $t$, the second inequality is due to summing over more elements, the equality holds since the sets $I_{\tau'}$ for $\tau' \in \mathcal{T}_b \setminus \crl*{0}$, $\tau' \le \tau$, partition $[1, 2\tau) \cap \crl*{0,\ldots,b}$, and the last inequality is due to constraint \ref{constraint:capacity-p} of $\Psi_P(b)$.
    
    We deduce that constraint \ref{constraint:capacity} holds, and thus $y$ is a feasible solution to $\Psi(b)$.
    
    Finally:
    \begin{align*}
        \sum_{i\in[n]} \sum_{j\in[m]} \Bigl( w_{i,j,0} \, x^*_{i,j,0} + \sum_{\tau \in \mathcal{T}_b \setminus \crl*{0}} \sum_{t \in I_\tau} w_{i,j,t} \, \frac{x^*_{i,j,\tau}}{|I_\tau|} \Bigr) & \ge \sum_{i\in[n]} \sum_{j\in[m]} \Bigl( w_{i,j,0} \, x^*_{i,j,0} + \sum_{\tau \in \mathcal{T}_b \setminus \crl*{0}} \sum_{t \in I_\tau} w_{i,j,2\tau} \, \frac{x^*_{i,j,\tau}}{|I_\tau|} \Bigr)  \\
        & = \sum_{i\in[n]} \sum_{j\in[m]} \sum_{\tau \in \mathcal{T}_b} w_{i,j,2\tau} \, x^*_{i,j,\tau} = \Psi_P(b),
    \end{align*}
    where the first inequality holds due to the monotonicity of $w_{i,j,t}$ in $t$, as $t < 2\tau$ for every $t \in I_\tau$.
    We thus get the desired result by \cref{lem:small-lp-approx-large-lp}.
    \end{proof}
    
    The complete polynomial-time algorithms can simply solve the $\Psi_P(T)$ LP, and use its solution $x^*$ such that for each $x^*_{i,j,\tau}$ we add an additional step of choosing $t$ in a uniform way: $t \sim Unif(I_\tau)$ (and $t = 0$ for $\tau = 0$).
    This way, the probability to choose item choice $(i,j)$ with $b_{i,j_i} = t$ is $y_{i,j,t}$ (and so the proof works as before).

    Since this time $y$ is not the optimal solution, but a $4$-approximate solution, we pay an additional constant factor and still get an $O(1)$ approximation ratio for the skipping-adaptive policy. The same holds for the stopping-adaptive policy, whose analysis only uses the feasibility of $x^*$ for $\Psi(T)$, the non-negativity of $w_{i,j,t}$ on its support (which is preserved by discarding the sampled $t$ if $w_{i,j,t} < 0$), and its value. Hence, replacing $\Psi(T)$ by $\Psi_P(T)$ also in Step~2 of \Cref{alg:stopping-pk}, we still get an $O(\sqrt{\alpha})$ approximation ratio; specifically, $\adapt(I)\le72\sqrt{1+\alpha}\cdot\E[\utility(\pi_I)]$.
\subsection{Handling Continuous Distributions} \label[appendix]{sec:continuous-distributions}
        In \Cref{thm:a,thm:stop-sqrt-ub}, which both use the LP $\Psi(T)$, we assume that (similar to \cite{gupta2011approximation}) the size distributions are discrete; namely, that all the items' sizes are integers between $0$ and $T+1$ (or, equivalently, that the sizes are $\left\{0,\frac{1}{T},\frac{2}{T},\cdots,1,1+\frac{1}{T}\right\}$, where the largest size represents any size larger than the knapsack). In this section, we describe how to apply these results to instances with continuous distributions, using discretization and $\eps$-resource augmentation. For continuous distributions, the policies evaluate their decision rules using the discretized observed sizes.

        Fix $\eps > 0$. We set $T = \lceil n/\eps \rceil$ and the discretization parameter $\delta = 1/T$, so that $n\delta \le \eps$.
        Given an instance $I$ with continuous size distributions, we construct a rounded instance $\hat{I}$ where all item sizes $s_i^j$ are rounded down to the nearest multiple of $\delta$: $\hat{s}_i^j = \delta \lfloor s_i^j / \delta \rfloor$, where rounded sizes larger than $1$ are replaced by $1+\delta$.
        
        Let $OPT(I)$ and $OPT(\hat{I})$ be the optimal expected utilities for the original and rounded instances, respectively, with capacity 1. Since reducing item sizes can only increase the probability of successfully fitting any set of items, the optimal value for the rounded problem is an upper bound on the optimal value for the original problem. Thus, $OPT(\hat{I}) \ge OPT(I)$.
        
        Let $\varphi$ be any policy for $\hat{I}$; for example, the skipping-adaptive policy of \Cref{thm:a} or the stopping-adaptive policy of \Cref{thm:stop-sqrt-ub}, computed for $\hat{I}$.

        We simulate $\varphi$ on the rounded instance $\hat{I}$ and mirror its execution, skipping, and stopping decisions on the original instance $I$. Whenever a true size $s_i^j$ is observed, we feed the rounded size $\hat{s}_i^j$ to the simulated policy. The real execution terminates whenever the simulated execution terminates, including after a simulated overflow. Thus the order, choices, and decision rules remain fixed in advance, and both executions incur the same costs.

        We claim that given an augmented capacity of $1 + \eps$, the expected utility of $\varphi$ on $I$ is at least the expected utility of $\varphi$ running on $\hat{I}$.

        For any execution path (realization) of $\varphi$, let $\mathcal{S}$ be the set of item-choice pairs successfully inserted by $\varphi$ in $\hat{I}$. By the knapsack constraint: $\sum_{(i,j)\in \mathcal{S}} \hat{s}_{i}^{j} \le 1$. 

        For every $(i,j)\in\mathcal{S}$, we have $\hat{s}_i^j\le1$, so overflow capping was not applied. Hence its true size is $s_i^j = \hat{s}_i^j + \eta_i^j$, where $0 \le \eta_i^j < \delta$. Since $\mathcal{S}$ contains at most $n$ items, the total rounding error is bounded by $n \delta \le \eps$. Thus, the total size in the original instance is:
        \[ \sum_{(i,j)\in\mathcal{S}} s_{i}^{j} < \sum_{(i,j)\in\mathcal{S}} \hat{s}_{i}^{j} + \eps \le 1 + \eps. \]

        Hence every pair in $\mathcal{S}$ also fits in $I$ with capacity $1+\eps$. If $\varphi$ overflows in $\hat{I}$, the overflow pair yields utility $-p_i^j$ in $\hat{I}$, and at least $-p_i^j$ in $I$.
        We deduce that the same policy $\varphi$ applied to instance $I$ given an additional capacity of $\eps$ gets at least the same utility as $\varphi$ applied to instance $\hat{I}$ with no additional capacity. Since $OPT(\hat{I}) \ge OPT(I)$, every approximation guarantee of $\varphi$ on $\hat{I}$ carries over to $I$ with capacity $1+\eps$.


    \section{Appendix for Section~\ref{subsec:batch-policy}: LP for the Batch Policy}
    \label[appendix]{app:batch-upper-bounds}
    \subsection{LP Formulation and Properties}
    \label[appendix]{sec:alg}

    We first define an LP that both guides our solution and gives an upper bound on the performance of every adaptive policy. For every actual item-choice pair $(i,j)$, let
    \[
        w_{ij}:=v_i\Pr_{s_i^j\sim F_i^j}[s_i^j\le1]-p_i^j,
        \qquad
        \mu_{ij}(\tau):=\E_{s_i^j\sim F_i^j}[\min\{s_i^j,\tau\}].
    \]
    The first quantity is the expected utility of attempting the choice in an empty unit-capacity knapsack; the second is its expected size truncated at $\tau$. The objective coefficients remain those of the original unit-capacity instance, whereas the truncation level is specified separately from the load budget. This LP relaxation generalizes the LP used by Dean, Goemans, and Vondr\'ak~\cite{2004DeanGoemansVon} to our model with costs and multiple choices.

    For a load budget $B>0$ and a truncation level $\tau>0$, define
    \begin{equation}\label{eq:phi_t}
        \Phi(B;\tau):=
        \max\left\{\sum_{i,j}w_{ij}x_{ij}:
        \sum_{i,j}\mu_{ij}(\tau)x_{ij}\le B,\quad
        \forall i,\ \sum_jx_{ij}=1,\quad x_{ij}\ge0\right\}.
    \end{equation}
    When the parameters coincide, write $\Phi(t):=\Phi(t;t)$. In particular, the LP used in \Cref{sec:batch-ub} is $\Phi(1)=\Phi(1;1)$.

    Every item includes a dummy null choice $j_\bot$ with zero value, cost, and size; its coefficients are $w_{ij_\bot}=\mu_{ij_\bot}(\tau)=0$. Selecting it represents not attempting the item. Actual choices with $w_{ij}\le0$ may be replaced by the null choice. We make this replacement before solving the LP, so every non-null candidate returned below has positive weight.

    We use an optimal LP solution to guide our algorithm. First, we characterize its structure. A closely related structural result for the LP relaxation of the standard multiple-choice knapsack problem was shown by Sinha and Zoltners~\cite{MultipleChoiceKnapsack1979}.

\begin{lemma}
  \label{lem:opt_solution}
    There exists an optimal solution to~\(\Phi(B;\tau)\) satisfying the following properties:
  \begin{enumerate}[nosep,noitemsep]
    \item At most one item~\(i^{*}\) has exactly two choices~\(j,j'\) such that \(x_{i^* j}, x_{i^* j'} > 0\).
    \item Every other item~\(i \neq i^{*}\) has at most one choice~\(j\) with \(x_{ij} = 1\).
  \end{enumerate}
  Moreover, such a solution can be computed in polynomial time.

\end{lemma}
\begin{proof}[Proof of Lemma~\ref{lem:opt_solution}] 
Note that any basic feasible solution to the LP satisfies these properties; the basic solution has at most $n+1$ positive variables, and the constraints ensure that for each item $i$, there is at least one choice $j$ with $x_{ij} > 0$, hence at most one item has two non-zero choices.

Such a solution can be obtained by first computing optimal primal and dual solutions of \(\Phi(B;\tau)\) in polynomial time and then applying the strongly polynomial procedure of Megiddo~\cite{megiddo_finding_1991} to recover an optimal basis. The basic feasible solution associated with this basis has the required structure. Moreover, \cite{dyer84} show that the multiple-choice knapsack linear program can be solved in linear time, obtaining an optimal solution with the desired structure directly in strongly polynomial time.
\end{proof}

    We now present a helpful subroutine, \textbf{Virtual-Items}, which converts an \mpk\ instance into a candidate set containing at most one choice per original item. If the optimal solution uses two choices of one item, we replace them by one virtual choice, represented as their convex combination. The virtual choice samples one of its underlying choices according to the LP coefficients, independently of all other items. This sampling may be performed before executing the policy, or equivalently when the virtual choice is attempted. Null choices are omitted from the candidate set, but may be one of the two underlying choices of a virtual candidate.

    \begin{algorithm}[H]
    \caption{Virtual-Items (of Pandora's Knapsack)}
    \label{alg:virtual-items}
    \KwIn{Instance $I$, load budget $B>0$, and truncation level $\tau>0$}
    \KwOut{Candidate set $\mathcal C$ with at most one possibly virtual choice per original item}
    Replace each non-null choice with $w_{ij}\le0$ by the null choice\;
    Solve $\Phi(B;\tau)$ to obtain an optimal solution $x$ with the structure from \Cref{lem:opt_solution}\;
    $\mathcal C\gets\{(i,j):j\ne j_\bot,\ x_{ij}=1\}$\;
    \If{an item $i^*$ has two choices $j,j'$ with $x_{i^*,j},x_{i^*,j'}>0$}{
        Create a virtual choice $(i^*,j^*)$ that samples $j$ and $j'$ with probabilities $x_{i^*,j}$ and $x_{i^*,j'}$, respectively\;
        $w_{i^*,j^*}\gets x_{i^*,j}w_{i^*,j}+x_{i^*,j'}w_{i^*,j'}$\;
        $\mu_{i^*,j^*}(\tau)\gets x_{i^*,j}\mu_{i^*,j}(\tau)+x_{i^*,j'}\mu_{i^*,j'}(\tau)$\;
        Add $(i^*,j^*)$ to $\mathcal C$\;
    }
    \Return $\mathcal C$\;
    \end{algorithm}

    For the single possibly virtual item, every coefficient is understood as the same convex combination of the corresponding coefficients of its underlying choices, including a possible null choice. When the candidate set is empty, the calling algorithm returns the empty policy, and we set $w_{\max}:=0$. A positive-weight candidate with $\mu_i(\tau)=0$ is assigned density $+\infty$.

    The batch policy in \Cref{sec:batch-ub} uses $B=\tau=1$ and the prefix threshold $U=1/[3(1+\alpha)]$. In its density order, ties are broken by placing all integral candidates before the virtual item. If the virtual item mixes a non-null choice with the null choice, it is never included in the prefix $S$. To see this, let $\mu_{\mathrm{virt}}$ be its truncated mass. If $\mu_{\mathrm{virt}}>U$, it cannot belong to $S$. Otherwise, the LP load constraint is tight: if it were slack, increasing the coefficient of the positive-weight non-null choice would improve the objective. The integral candidates have total truncated mass $1-\mu_{\mathrm{virt}}\ge1-U>U$ and density at least that of the virtual item, by complementary slackness. Thus the prefix terminates before the virtual item. The virtual item remains eligible for the singleton branch. In particular, any virtual item included in $S$ mixes two non-null choices and retains the deterministic value $v_i$.

    For a set of item choices $\mathcal S$, write
    \[
        \operatorname{size}(\mathcal S):=\sum_{(i,j)\in\mathcal S}s_i^j,
        \qquad
        \mu_\tau(\mathcal S):=\sum_{(i,j)\in\mathcal S}\mu_{ij}(\tau).
    \]
    The following lemmas extend the corresponding arguments of Dean, Goemans, and Vondr\'ak~\cite{dean2008approximating} to multiple choices, costs, and a general capacity $t\in(0,1]$.

    \begin{lemma}\label{lem:muS}
    For every set of item choices $\mathcal S$ and every $t\in(0,1]$,
    \[
        \Pr[\operatorname{size}(\mathcal S)<t]\ge1-\frac1t\mu_t(\mathcal S).
    \]
    \end{lemma}
    \begin{proof}
    By Markov's inequality,
    \begin{align*}
        \Pr[\operatorname{size}(\mathcal S)\ge t]
        &\le\frac1t\E[\min\{\operatorname{size}(\mathcal S),t\}]\\
        &\le\frac1t\E\left[\sum_{(i,j)\in\mathcal S}\min\{s_i^j,t\}\right]
         =\frac1t\mu_t(\mathcal S).
    \end{align*}
    \end{proof}

    \begin{lemma}\label{lem:E_le2}
    For any adaptive policy for the \mpk\ problem with capacity $t\in(0,1]$, let $\mathcal S$ be the random set of item choices attempted by the policy. Then
    \[
        \E[\mu_t(\mathcal S)]\le2t.
    \]
    \end{lemma}
    \begin{proof}
    Let $I_{ij}$ indicate that $(i,j)$ is attempted. Since the attempt decision is made before $s_i^j$ is observed, $I_{ij}$ is independent of $s_i^j$, and hence
    \[
        \E[\mu_t(\mathcal S)]
        =\E\left[\sum_{i,j}I_{ij}\min\{s_i^j,t\}\right].
    \]
    On every realization, the total size preceding the final attempted pair is at most $t$, and the truncated size of the final pair is at most $t$. Thus the sum inside the expectation is at most $2t$; it is zero when no pair is attempted.
    \end{proof}

    \begin{lemma}\label{lem:adaptphi}
    For every $t\in(0,1]$ and $c\ge2$,
    \[
        \adapt(t)\le\Phi(ct;t)\le c\Phi(t),
    \]
    where $\adapt(t)$ is the optimal expected utility of an adaptive policy with capacity $t$. Moreover, for every $B,\tau>0$ and $c\ge1$,
    \[
        \Phi(cB;\tau)\le c\Phi(B;\tau),
    \]
    and, whenever $0<\tau\le\tau'$, one has
    \[
        \Phi(B;\tau')\le\Phi(B;\tau).
    \]
    \end{lemma}
    \begin{proof}
    Fix an adaptive policy for capacity $t$. For every positive-weight actual choice, let $x_{ij}$ be its attempt probability, and assign all remaining probability mass of item $i$ to its null choice. This includes both the probability of not attempting the item and the attempt probabilities of choices with $w_{ij}\le0$. Moving nonpositive choices to the null choice only increases the objective upper bound and decreases the load. By \Cref{lem:E_le2},
    $\sum_{i,j}\mu_{ij}(t)x_{ij}\le2t$.
    Conditioned on attempting $(i,j)$, the remaining capacity is at most one and the size is still unobserved, so its expected contribution is at most $w_{ij}$. Consequently,
    \[
        \adapt(t)\le\Phi(2t;t)\le\Phi(ct;t).
    \]
    For scaling, let $x$ be optimal for $\Phi(cB;\tau)$, divide every non-null coordinate by $c$, and assign the remaining mass to the null choice. The resulting solution is feasible for $\Phi(B;\tau)$ and has $1/c$ of the objective value. Hence
    $\Phi(cB;\tau)\le c\Phi(B;\tau)$, which also gives
    $\Phi(ct;t)\le c\Phi(t;t)=c\Phi(t)$.
    Finally, $\mu_{ij}(\tau)\le\mu_{ij}(\tau')$ for $\tau\le\tau'$, so every solution feasible for $\Phi(B;\tau')$ is feasible for $\Phi(B;\tau)$.
    \end{proof}

        \subsection{LP Overconfidence: An \texorpdfstring{$\Omega(\alpha )$}{Omega(alpha)} Overestimation of the True Utility} \label[appendix]{sec:LP_gap}

    In this section, we provide an instance of the \mpk\ problem where the value of the LP introduced in Eq.~\eqref{eq:phi_t} is larger than the obtainable value of any policy by a factor of $\Omega(\alpha)$. This is in stark contrast to the case without costs, where~Dean, Goemans, and Vondr\'ak~\cite{dean2008approximating} show a non-adaptive policy whose utility is a constant factor away from the value of the analogous LP. As our batch policy in \Cref{sec:batch-ub} gives an $O(\alpha)$-approximation to the value of the LP in Eq.~\eqref{eq:phi_t}, this approximation is the best possible with respect to the value of the LP.

    Fix $\epsilon\in(0,1)$. In our instance, we have $n=\lceil1/(2\epsilon)\rceil+1$ identical items with value $1$ and cost $1-\epsilon$. The size distribution of the items is  $$s=\begin{cases}\epsilon^{n}, & \text{w.p. }1-\epsilon,\\ 1, & \text{w.p. }\epsilon;\end{cases}.$$

    By \Cref{def:RoI}, the value of the IOR parameter is $\alpha = \frac{1-\epsilon}{\epsilon}=O(1/\epsilon)$. For every item $i$, we have $\mu_i=(1-\epsilon)\cdot \epsilon^n+\epsilon\cdot 1=\epsilon^n+\epsilon-\epsilon^{n+1}\le 2\epsilon$ and $w_i=\epsilon$. Setting $x_i=1$ for $\ell=\lfloor\frac{1}{2\epsilon}\rfloor$ items, and for an additional item $i'$ set $x_{i'}=\frac{1}{2\epsilon}-\ell$, implies that $$\sum_i x_i\cdot \mu_i\le2\epsilon\cdot \sum_i x_i = 2\epsilon \cdot\frac{1}{2\epsilon}=1,$$ which is a feasible assignment to the LP. Therefore, $\Phi(1)\ge \frac{1}{2\epsilon}\cdot \epsilon=1/2.$ We next bound the utility of any policy for this instance. 
    
    For any policy, after inserting the first item, we have that any item fits with probability at most $1-\epsilon$; thus, we have that the benefit from trying to insert any item after the first one is at most $(1-\epsilon)v-p=0.$ Therefore, the policy that only inserts one item is optimal, and it obtains a utility of $\epsilon\le 2\epsilon\cdot  \Phi(1)=\Phi(1)/\Omega(\alpha).$

    This shows that we cannot hope for a tighter approximation to the LP's value.
    

        \section{Appendix for Section~\ref{sec:pandora}: ROI Counterexample for Pandora's Box}
        \label[appendix]{sec:bad-roi}
        
        We now show that the IOR parameter defined for Pandora's knapsack does not
        capture the adaptivity gap in the Pandora's box problem. Recall that, in that
        definition, the inverse ROI is
        \[
        \alpha=\max_i \frac{c_i}{\E[v_i]-c_i}.
        \]
        This definition measures the profitability of a box if no other boxes are present. In
        Pandora's box, this does not capture the return-on-investment, since the decision maker obtains the
        maximum value, and not the sum of values. Therefore, a value that is obtained
        almost surely can make the ROI look high, even though it contributes
        little to the marginal value of opening additional boxes.
        
        We show that this issue can make the adaptivity gap unbounded even when
        $\alpha=1$.
        
        \begin{theorem}
        \label{thm:empty-roi-does-not-control-pandora}
        For every integer $a\ge 2$, there exists an instance of Pandora's box with
        $n=a^2$ boxes and
        \[
        \max_i \frac{c_i}{\E[v_i]-c_i}=1
        \]
        such that
        \[
        \frac{\adapt}{\batch}=\Omega(a)=\Omega(\sqrt n).
        \]
        \end{theorem}
        
        \begin{proof}
        Fix an integer $a\ge 2$, and let $n=a^2$. All boxes are identical. Each box has
        cost $1$, and value
        \[
        v=
        \begin{cases}
        M+V, & \text{with probability } q,\\
        M, & \text{otherwise,}
        \end{cases}
        \]
        where
        \[
        q=\frac{1}{a^2},
        \qquad
        M=1-\frac1a,
        \qquad
        V=a^2+a.
        \]
        For every box,
        \[
        \E[v]
        =
        M+qV
        =
        1-\frac1a+\frac{1}{a^2}(a^2+a)
        =
        2.
        \]
        Since the cost is $1$, we get
        \[
        \frac{c}{\E[v]-c}=1.
        \]
        Therefore, $\alpha=1$.
        
        We first lower-bound the value of Weitzman's policy. Consider the adaptive
        policy that opens boxes sequentially until either a high value $M+V$ is
        observed, or all boxes have been opened. This is a feasible adaptive policy, and
        therefore gives a lower bound on $\adapt$. Let
        \[
        P=1-(1-q)^n
        \]
        be the probability that at least one high value appears. The number of opened
        boxes is a truncated geometric random variable, and hence
        \[
        \E[\mbox{\# boxes opened}]
        =
        \sum_{t=0}^{n-1}(1-q)^t
        =
        \frac{1-(1-q)^n}{q}
        =
        \frac{P}{q}.
        \]
        The final value is $M+V$ if some high value appears, and $M$ otherwise. Thus the
        expected utility of this policy is
        \[
        M+VP-\frac{P}{q}
        =
        M+\left(V-\frac1q\right)P.
        \]
        By the choice of parameters,
        \[
        V-\frac1q
        =
        a^2+a-a^2
        =
        a.
        \]
        Since $n=a^2$ and $q=1/a^2$,
        \[
        P
        =
        1-\left(1-\frac{1}{a^2}\right)^{a^2}
        \ge
        1-e^{-1}.
        \]
        Therefore,
        \[
        \adapt
        \ge
        M+aP
        \ge
        a(1-e^{-1}).
        \]
        
        It remains to upper-bound the value of every non-adaptive policy. Since all
        boxes are identical, a deterministic batch policy is determined only by the
        number $k$ of boxes it opens. If $k=0$, the value is $0$. If $k\ge 1$, its
        expected utility is
        \[
        B(k)
        =
        M+V\left(1-(1-q)^k\right)-k.
        \]
        We show that $B(k)<2$ for every $k$.
        
        It is enough to prove that
        \[
        V\left(1-(1-q)^k\right)-k\le 1.
        \]
        We use the inequality
        \[
        1-x\ge \exp\left(-\frac{x}{1-x}\right),
        \qquad 0\le x<1.
        \]
        Applying it with $x=q$, we get
        \[
        1-(1-q)^k
        \le
        1-\exp\left(-\frac{kq}{1-q}\right).
        \]
        Let
        \[
        y=\frac{kq}{1-q}.
        \]
        Since $q=1/a^2$, we have $k=(a^2-1)y$. Therefore,
        \[
        V\left(1-(1-q)^k\right)-k
        \le
        (a^2+a)(1-e^{-y})-(a^2-1)y.
        \]
        Define
        \[
        h(y)=(a^2+a)(1-e^{-y})-(a^2-1)y.
        \]
        The function $h$ is concave. Its maximizer satisfies
        \[
        e^{-y}=\frac{a^2-1}{a^2+a}=\frac{a-1}{a},
        \]
        and therefore
        \[
        y^*=\ln\left(\frac{a}{a-1}\right).
        \]
        At this point,
        \[
        h(y^*)
        =
        a+1-(a^2-1)\ln\left(1+\frac{1}{a-1}\right).
        \]
        Using
        \[
        \ln(1+x)\ge \frac{2x}{2+x},
        \qquad x\ge 0,
        \]
        with $x=1/(a-1)$, we get
        \[
        \ln\left(1+\frac{1}{a-1}\right)
        \ge
        \frac{2}{2a-1}.
        \]
        Therefore,
        \[
        h(y^*)
        \le
        a+1-\frac{2(a^2-1)}{2a-1}
        =
        \frac{a+1}{2a-1}
        \le
        1,
        \]
        where the last inequality follows since $a\ge 2$. Hence
        \[
        V\left(1-(1-q)^k\right)-k\le 1
        \]
        for every $k$, and so every batch policy has value at most
        \[
        M+1<2.
        \]
        Randomization cannot improve this bound, since any randomized batch policy is a
        distribution over deterministic batch policies. Thus
        \[
        \batch\le 2.
        \]
        
        Combining the two bounds gives
        \[
        \frac{\adapt}{\batch}
        \ge
        \frac{a(1-e^{-1})}{2}
        =
        \Omega(a)
        =
        \Omega(\sqrt n).
        \]
        \end{proof}

    \section{Relaxations}
    \label[appendix]{sec:relaxations}

    In this section, we consider several relaxations of the original model. In particular, we focus on the setting where items are size-bounded with high probability. To facilitate the analysis, we first study a relaxed variant in which there is no explicit bound on item sizes. In this variant, the policy receives the value of each item inserted before the capacity is met, including the item that causes the capacity to be met or exceeded. We then transform this policy into one evaluated under the original \mpk\ objective, where no value is collected from an overflowing item.

    \subsection{Start-by-Deadline Model}\label[appendix]{sec:overflow}

    For $t\in(0,1]$, define $\pkof(t)$ as follows. It calls
    $\text{\textbf{Virtual-Items}}(I;t,t)$ (\Cref{alg:virtual-items}) and obtains a candidate set $\mathcal C=\{1,\ldots,r\}$, sorted in non-increasing order of density $w_i/\mu_i(t)$. It returns the empty policy when $\mathcal C=\emptyset$. Otherwise, let $w_{\max}=\max_{i\in\mathcal C}w_i$. If $w_{\max}\ge\Phi(t)/4$, the policy inserts only a maximizing candidate $i_{\max}$. Otherwise, it considers the candidates in density order, attempting an item whenever the accumulated size is below $t$, and stops once this size reaches or exceeds $t$. The policy earns the value of every attempted item, including the final item, and pays its cost. Write $\pkof:=\pkof(1)$.

    \begin{lemma}\label{lem:skc_of_bound}
    For every constant $c\ge1$ and every $t\in(0,1]$,
    \[
        \E[\pkof(t)]\ge\frac1c\,\frac{\Phi(ct;t)}4.
    \]
    \end{lemma}
    \begin{proof}
    If $\Phi(t)=0$, then $\Phi(ct;t)=0$ by \Cref{lem:adaptphi}, so the claim is immediate. Hence assume $\Phi(t)>0$, so that $\mathcal C\ne\emptyset$ and $i_{\max}$ is well defined.

    If $w_{\max}\ge\Phi(t)/4$, we take only $i_{\max}$ and obtain
    \[
        v_{i_{\max}}-p_{i_{\max}}
        \ge w_{\max}\ge\frac{\Phi(t)}4
        \ge\frac1c\,\frac{\Phi(ct;t)}4,
    \]
    where, for a virtual candidate, the first expression denotes its expected utility, and the last inequality follows from \Cref{lem:adaptphi}. Hence assume $w_{\max}<\Phi(t)/4$.

    Let $\pkof_k$ denote the utility of candidate $k$. This item starts whenever the total size of preceding items is below $t$. By \Cref{lem:muS},
    \begin{equation}\label{eq:pkof-item-start}
    \begin{aligned}
        \E[\pkof_k]
        &\ge(v_k-p_k)\left(1-\frac1t\sum_{j<k}\mu_j(t)\right)\\
        &\ge w_k\left(1-\frac1t\sum_{j\le k}\mu_j(t)\right).
    \end{aligned}
    \end{equation}
    For a virtual candidate, $v_k-p_k$ denotes the convex combination of the net values of its underlying choices. Summing over $k\in[r]$ gives
    \begin{equation}\label{eq:skc_of_bound}
    \begin{aligned}
        \E[\pkof(t)]
        &\ge\Phi(t)-\frac1t\sum_{1\le j\le k\le r}w_k\mu_j(t)\\
        &\ge\Phi(t)-\frac1{2t}\sum_{k,j=1}^r w_k\mu_j(t)
                         -\frac1t\sum_{k=1}^r w_k\mu_k(t)\\
        &\ge\frac{\Phi(t)}2-w_{\max}
         \ge\frac{\Phi(t)}4
         \ge\frac1c\,\frac{\Phi(ct;t)}4.
    \end{aligned}
    \end{equation}
    For the second inequality, the density order gives
    $w_k\mu_j(t)\le w_j\mu_k(t)$ whenever $j<k$. The remaining inequalities use
    $\sum_k\mu_k(t)\le t$, $\sum_kw_k=\Phi(t)$,
    $w_{\max}<\Phi(t)/4$, and \Cref{lem:adaptphi}.
    \end{proof}

    If the policy receives the value of every item inserted while some capacity remains, even if it overflows, the adaptivity gap is constant. To show this, given an instance $I$ of the start-by-deadline model, let $\bar I$ be obtained by replacing every size $s_i^j$ by $\bar s_i^j:=\min\{s_i^j,1\}$. These instances are pathwise equivalent for the unit-capacity start-by-deadline objective: any attempted item whose size is changed terminates the process in both instances and earns its value in both. In the remainder of this subsection, $\bar\Phi$, $\bar w_{ij}$, $\bar w_{\max}$, and $\pkof$ refer to $\bar I$; in particular, choices are evaluated and the LP is solved on the truncated instance.

    \begin{lemma}\label{thm:ADAPT_le_phi2}
    For every instance $I$ of the start-by-deadline model,
    \[
        \adapt_{\mathrm{OF}}(I)=\adapt_{\mathrm{OF}}(\bar I)
        \le2\bar\Phi(1).
    \]
    \end{lemma}
    \begin{proof}
    Since $\bar s_i^j\le1$ almost surely, the LP coefficient for every actual choice of $\bar I$ is
    \[
        \bar w_{ij}=v_i\Pr[\bar s_i^j\le1]-p_i^j=v_i-p_i^j.
    \]
    Let $\mathcal S$ be the random set of choices attempted by an optimal policy for $\bar I$, and set $x_{ij}:=\Pr[(i,j)\in\mathcal S]$, completing every item with its null choice. Nonpositive-weight choices can be omitted. The pathwise argument of \Cref{lem:E_le2} also applies to the start-by-deadline stopping rule, so $x$ is feasible for $\bar\Phi(2;1)$. Therefore,
    \begin{equation}\label{eq:adapt_of_bound}
    \begin{aligned}
        \adapt_{\mathrm{OF}}(I)
        &=\adapt_{\mathrm{OF}}(\bar I)
         =\sum_{i,j}\bar w_{ij}x_{ij}\\
        &\le\bar\Phi(2;1)
         \le2\bar\Phi(1;1)
         =2\bar\Phi(1).
    \end{aligned}
    \end{equation}
    The scaling inequality follows from \Cref{lem:adaptphi}.
    \end{proof}

    \begin{theorem}\label{lem:of_approximation}
    The policy $\pkof$ computed for $\bar I$ satisfies
    \[
        \frac{\adapt_{\mathrm{OF}}(I)}{\E[\pkof]} = O(1).
    \]
    More precisely, the ratio is at most $8$.
    \end{theorem}
    \begin{proof}
    If $\bar\Phi(1)=0$, then $\bar w_{\max}=0$ and \Cref{thm:ADAPT_le_phi2} gives $\adapt_{\mathrm{OF}}(I)=0$, so the claim is immediate. Hence assume $\bar\Phi(1)>0$. By \Cref{thm:ADAPT_le_phi2,lem:skc_of_bound}, applied to $\bar I$,
    \[
        \frac{\adapt_{\mathrm{OF}}(I)}{\E[\pkof]}
        \le\frac{2\bar\Phi(1)}{\bar\Phi(1)/4}.
    \]
    If $\bar w_{\max}\ge\bar\Phi(1)/4$, then $\E[\pkof]=\bar w_{\max}$, and
    \[
        \frac{\adapt_{\mathrm{OF}}(I)}{\E[\pkof]}
        \le\frac{8\bar w_{\max}}{\bar w_{\max}}=8.
    \]
    Otherwise,
    \[
        \frac{\adapt_{\mathrm{OF}}(I)}{\E[\pkof]}
        \le\frac{2\bar\Phi(1)}{\bar\Phi(1)/4}=8.
    \]
    \end{proof}

    \subsection{Bounded-Size Items}\label[appendix]{sec:bounded}

    We now return to the original \mpk\ objective and the untruncated input instance; all weights and LP values in this subsection refer to that instance. Fix $\delta\in(0,1)$. For $\alpha>0$, we call the variant satisfying
    \[
        \Pr_{s_i^j\sim F_i^j}[s_i^j>1-\delta]\le\frac1{2n\alpha}
        \qquad\text{for every actual choice }(i,j)
    \]
    the $\mpk$-BOUND problem. When $\alpha=0$, the constant-factor guarantee of \Cref{thm:batch-approx} already proves the desired result without a bounded-size assumption.

    For $\alpha\ge1$, our algorithm $\pkbound$ runs $\pkof(\delta)$ on the original instance, stopping once the total realized size reaches or exceeds $\delta$, but its utility is evaluated under the original \mpk\ objective. For $0\le\alpha<1$, it instead executes the batch returned by \Cref{thm:batch-approx} in any fixed order, stopping on overflow or when that sequence ends. This sequential execution pointwise weakly dominates the batch policy and gives an $18(1+\alpha)<36$ approximation. Hence the proof below may assume $\alpha\ge1$.

    \begin{theorem}\label{thm:bounded_delta}
    The policy $\pkbound$ satisfies
    \[
        \frac{\adapt(1)}{\E[\pkbound]} = O(1/\delta).
    \]
    More precisely, the ratio is at most $8/\delta$ for $\alpha\ge1$ and at most $18(1+\alpha)<36$ for $0\le\alpha<1$.
    \end{theorem}
    \begin{proof}
    Let $W:=\Phi(\delta)=\Phi(\delta;\delta)$. If $W=0$, the final LP upper bound below implies $\adapt(1)=0$, so assume $W>0$. If $w_{\max}\ge W/4$, then $\pkof(\delta)$ selects only a maximum-weight candidate. Since it is attempted in an empty unit-capacity knapsack,
    \[
        \E[\pkbound]=w_{\max}\ge W/4.
    \]
    Now assume $w_{\max}<W/4$. Let $A_k$ be the event that candidate $k$ is attempted. Using the preceding-items bound from the first line of \eqref{eq:pkof-item-start}, together with the same density argument as in \Cref{lem:skc_of_bound}, gives
    \begin{equation}\label{eq:pkbound-prefix}
        \sum_{k=1}^r w_k\Pr[A_k]
        \ge W-\frac1\delta\sum_{1\le j<k\le r}w_k\mu_j(\delta)
        \ge\frac W2.
    \end{equation}
    Immediately before an attempted item, the accumulated size is below $\delta$; hence it certainly fits whenever its size is at most $1-\delta$. For a possibly virtual candidate $k$, with underlying choices $a$ and convex coefficients $\lambda_{ka}$, an expression of the form $v_k\Pr[s_k>x]$ denotes the linear quantity
    \[
        \sum_a\lambda_{ka}v_a\Pr[s_a>x],
    \]
    with the null choice contributing zero. Using independence of the current choice and its size from $A_k$, we obtain
    \begin{equation}\label{eq:pkbound-direct-loss}
    \begin{aligned}
        \E[\pkbound]
        &\ge\sum_{k=1}^r\Pr[A_k]
          \bigl(w_k-v_k\Pr[s_k>1-\delta]\bigr)\\
        &\ge\frac W2-\sum_{k=1}^r v_k\Pr[s_k>1-\delta].
    \end{aligned}
    \end{equation}
    For an actual positive-weight pair $(i,j)$, let
    $q_{ij}:=\Pr[s_i^j\le1]$ and $\rho_{ij}:=\Pr[s_i^j>1-\delta]$.
    Since $w_{ij}=v_iq_{ij}-p_i^j>0$, $p_i^j\le\alpha w_{ij}$, and $q_{ij}\ge1-\rho_{ij}$, the bounded-size assumption gives
    \[
    \begin{aligned}
        v_i\rho_{ij}
        &=\frac{\rho_{ij}}{q_{ij}}(w_{ij}+p_i^j)\\
        &\le(1+\alpha)\frac{\rho_{ij}}{1-\rho_{ij}}w_{ij}
         \le\frac{1+\alpha}{2n\alpha-1}w_{ij}
         \le\frac{2}{2n-1}w_{ij},
    \end{aligned}
    \]
    where the last inequality uses $\alpha\ge1$. This inequality also holds for the possibly virtual candidate by taking the same convex combination. Every candidate has positive weight, their weights sum to $W$, and $w_{\max}<W/4$, so there are at least five candidates; in particular, $n\ge5$. Consequently,
    \[
        \sum_{k=1}^r v_k\Pr[s_k>1-\delta]
        \le\frac{2}{2n-1}\sum_{k=1}^r w_k
        \le\frac{2W}{9}.
    \]
    Substituting into \eqref{eq:pkbound-direct-loss} yields
    \[
        \E[\pkbound]\ge\frac W2-\frac{2W}{9}
        =\frac{5W}{18}\ge\frac W4.
    \]
    Thus, in both branches,
    \begin{equation}\label{eq:pkbound-lower}
        \E[\pkbound]\ge\frac14\Phi(\delta;\delta).
    \end{equation}
    Finally, \Cref{lem:adaptphi} gives
    \[
        \adapt(1)\le\Phi(2;1)
        \le\frac2\delta\Phi(\delta;1)
        \le\frac2\delta\Phi(\delta;\delta).
    \]
    The second inequality scales the first parameter while fixing the truncation level at $1$; the third uses monotonicity in the second parameter. Combining this bound with \eqref{eq:pkbound-lower} proves
    \[
        \E[\pkbound]\ge\frac\delta8\adapt(1).
    \]
    Together with the $18(1+\alpha)<36$ guarantee for $\alpha<1$ and the fact that $\delta<1$, this proves the $O(1/\delta)$ claim in all cases.
    \end{proof}

    The policy $\pkbound$ fixes the order of item choices in advance. For $\alpha\ge1$, it uses the adaptive stopping rule that terminates when the total realized size first reaches or exceeds $\delta$; for $\alpha<1$, it follows a fixed sequence until overflow or its end. In \Cref{sec:adaptivity_gap_bounded}, we show that even for bounded-size items, a non-adaptive policy may be a factor $\Omega(\sqrt\alpha)$ worse than the optimal policy (\Cref{thm:bounded-size-adaptivity-gap}).

    \subsection{\texorpdfstring{$\Omega(\sqrt \alpha)$}{Omega(sqrt(alpha))} Adaptivity Gap of Bounded-Size Items} 
    \label[appendix]{sec:adaptivity_gap_bounded}

    \begin{theorem}\label{thm:bounded-size-adaptivity-gap}
        For every $\delta\in(0,1/2]$ and every $\alpha$, there exists an instance $I$ of $\mpk$-BOUND with IOR $\alpha$ such that
        \[
            \frac{\stadpt(I)}{\nonadpt(I)} = \Omega(\sqrt{\alpha}).
        \]
    \end{theorem}
    
    \begin{proof}
    For $0\le\alpha<7$, a single size-zero item of value $1+\alpha$ and cost $\alpha$ suffices. Otherwise, fix $\epsilon\in(0,\frac{1}{2}]$, and consider $n=\lceil 1/\epsilon\rceil$ identical items with the following parameters:
    \begin{center}
            \begin{tabular}{l@{\quad}l}
                \toprule Property                 & Item                                                                                                                                                                                                \\
                \midrule Value $v$           & $2$                                                                                                                                                                                                                                \\
                Cost $p$                      & $2 - \epsilon^2$                                                                                                                                                                                                           \\
                \cmidrule{1-2} Size distribution s & $ \begin{cases}\epsilon^{n}, & \text{w.p. }1-\epsilon,\\ \frac{1}{2}, & \text{w.p. }\epsilon;\end{cases}$  \\
                \bottomrule
            \end{tabular}
        \end{center}

        All sizes are at most $1/2$, so the instance satisfies the $\mpk$-BOUND condition for $\delta\le1/2$. Note that $n\epsilon^n\le\frac{1}{2}$, so the items of size $\epsilon^n$ have total size at most $\frac{1}{2}$. We consider the case where the remaining capacity $S'$ satisfies $S'<\frac{1}{2}$, and compute the expected utility per item.
        $$\E[w|S'<\frac{1}{2}]=v\cdot\Pr[s\le S']-p\le2\cdot(1-\epsilon)-(2-\epsilon^2)=2-2\epsilon-2+\epsilon^2=-2\epsilon+\epsilon^2<0$$
        In expectation, the adaptive policy can include $\frac{1-(1-\epsilon)^n}{\epsilon}\ge\frac{1-e^{-1}}{\epsilon}$ items before reaching $S' < 1/2$, yielding expected total utility at least
        \[
        \frac{1-e^{-1}}{\epsilon} \cdot \epsilon^2 = (1-e^{-1})\epsilon.
        \]

        For a non-adaptive policy, the number of attempted insertions is fixed in advance. If the policy inserts only one item, its expected utility is $v - p = \epsilon^2$. Let $\utility(i)$ denote the expected utility from the $i$th item. For $i > 2$:
        \begin{eqnarray*}
            \utility(i) &\le & \E[\utility(i)|s_j=\epsilon^n\ \forall j<i]\cdot\Pr[s_j=\epsilon^n\quad \forall j<i] \\ 
            & &+ \E[\utility(i)|\exists!j<i:\ s_j>\epsilon^n]\cdot\Pr[\exists!j<i:\ s_j>\epsilon^n] \\ 
            & = & (v-p)\cdot (1-\epsilon)^{i-1} + ((v-p)\cdot(1-\epsilon) -p\cdot \epsilon)\cdot\Pr[\exists!j<i:\ s_j>\epsilon^n]\\
            &= & \epsilon^2\cdot (1-\epsilon)^{i-1} + (\epsilon^2-2\epsilon)\cdot\Pr[\exists!j<i:\ s_j>\epsilon^n]\\
            &\le& \epsilon^2\cdot (1-\epsilon)^{i-1} + (\epsilon^2-2\epsilon) \cdot\Pr[s_{i-1}>\epsilon^n \ \wedge\ s_j=\epsilon^n \quad \forall j<i-1]\\
            & = & \epsilon\cdot (1-\epsilon)^{i-2}(\epsilon\cdot (1-\epsilon) +\epsilon^2-2\epsilon)\\
            & \le & \epsilon\cdot (1-\epsilon)^{i-2}(\epsilon^2-\epsilon) <  0.
        \end{eqnarray*}
        In the above, the first inequality holds since if at least two of the first $i-1$ items have size $\frac{1}{2}$, then item $i$ is either not reached or overflows, so its utility is non-positive. The second inequality follows from the fact that $\epsilon^2 - 2\epsilon$ is negative, and that the probability that exactly one of the first $i-1$ items has size greater than $\epsilon^n$ is larger than the probability that only item $i-1$ has size greater than $\epsilon^n$. By \Cref{def:RoI}, the value of the IOR parameter is $\alpha = \frac{2-\epsilon^2}{\epsilon^2}=\Theta(1/\epsilon^2)$.

        Therefore, we have that a non-adaptive policy only loses from having a sequence longer than 2, and the optimal non-adaptive policy has a utility of $2\epsilon^2$. We conclude that for this instance, we have $$\frac{\stadpt(I)}{\nonadpt(I)}\ge\frac{(1-e^{-1})\epsilon}{2\epsilon^2}=\frac{1-e^{-1}}{2\epsilon}=\Omega(\sqrt{\alpha}).$$
        \end{proof}

\bibliographystyle{alphaurl}
\bibliography{references}

\end{document}